\documentclass[aip,pof,reprint,amsmath,amssymb,onecolumn]{revtex4-1}
\pdfoutput=1
\usepackage{dcolumn}
\usepackage{bm}
\usepackage{algorithm}
\usepackage{xcolor}
\usepackage{graphicx}
\usepackage{subfig}
\usepackage{soul}

\graphicspath{{./}}  

\definecolor{blue}{rgb}{0.098,0.357,0.675}
\definecolor{green}{rgb}{0.5,0.75,0.0}
\usepackage[
    pdftex,
    pdftitle={Lift and wakes of flying snakes},
    pdfauthor={Anush Krishnan, John J.~Socha and Pavlos P. Vlachos, Lorena A. Barba},
    pdfpagemode={UseOutlines},
    bookmarks, bookmarksopen,bookmarksnumbered={True},
    colorlinks, linkcolor={blue},citecolor={blue},urlcolor={blue}
    ]{hyperref}

\newcommand{\aoa}{\textsc{aoa}}

\newcommand{\ibm}{\textsc{ibm}}

\newcommand\ignore[1]{} 

\newcommand{\comment}[1]{\textcolor{green}{ #1 }}

\renewcommand{\comment}[1]{}

\begin{document}

\preprint{AIP/123-QED}

\title[Lift and wakes of flying snakes]{Lift and wakes of flying snakes}

\author{Anush Krishnan}
\affiliation{Mechanical Engineering, Boston University, Boston, MA, 02215, USA}
\author{John J. Socha}
\affiliation{Engineering Science and Mechanics, Virginia Tech, Blacksburg, VA, 24061}
\author{Pavlos P. Vlachos}
\thanks{New address: Mechanical Engineering, Purdue University, West Lafayette, IN, 47907}
\affiliation{Mechanical Engineering, Virginia Tech, Blacksburg, VA, 24061, USA}
\author{L. A. Barba}
\thanks{Corresponding Author: labarba@gwu.edu. New address: Mechanical and Aerospace Engineering, George Washington University, Washington, DC, 20052}
\affiliation{Mechanical Engineering, Boston University, Boston, MA, 02215, USA}

\date{\today}
             
\begin{abstract}
Flying snakes use a unique method of aerial locomotion: they jump from tree branches, flatten their bodies and undulate through the air to produce a glide. The shape of their body cross-section during the glide plays an important role in generating lift.  This paper presents a computational investigation of the aerodynamics of the cross-sectional shape.
Two-dimensional simulations of incompressible flow past the anatomically correct cross-section of the species \emph{Chrysopelea paradisi} show that a significant enhancement in lift appears at a $35^\circ$ angle of attack, above Reynolds numbers 2000. Previous experiments on physical models also obtained an increased lift, at the same angle of attack.  The flow is inherently three-dimensional in physical experiments, due to fluid instabilities, and it is thus intriguing that the enhanced lift also appears in the two-dimensional simulations. The simulations point to the lift enhancement arising from the early separation of the boundary layer on the dorsal surface of the snake profile, without stall. The separated shear layer rolls up and interacts with secondary vorticity in the near-wake, inducing the primary vortex to remain closer to the body and thus cause enhanced suction, resulting in higher lift. 
\end{abstract}


\maketitle

\section{Introduction}

Interest in small robotic devices has soared in the last two decades, leading engineers to turn towards nature for inspiration and spurring the fields of biomimetics and bioinspired design. Nature has evolved diverse solutions to animal locomotion in the forms of flapping flight, swimming, walking, slithering, jumping, and gliding.  At least thirty independent animal lineages have evolved gliding flight,\cite{Dudley2007, Norberg1990} but only one animal glides without any appendages: the flying snake. Three species of snakes in the genus \emph{Chrysopelea} are known to glide.\cite{Socha2011}
They inhabit lowland tropical forests in Southeast and South Asia and have a peculiar behavior:  they jump from tree branches to start a glide to the ground or other vegetation, possibly as a way to escape a threat or to travel more efficiently. Their gliding ability is surprisingly good, and one species of flying snake, \emph{Chrysopelea paradisi} (the paradise tree snake), has also been observed to turn in mid-air.\cite{Socha2002, SochaETal2005, SochaETal2010}

Like all snakes, \emph{Chrysopelea paradisi} has a cylindrical body with roughly circular cross-section. But when it glides, this snake reconfigures its body to assume a flatter profile. During the glide, the snake undulates laterally and the parts of the body that are perpendicular to the direction of motion act as lift-generating `wings'.\cite{MiklaszETal2010, Holden2011} As in conventional wings, the cross-sectional shape of the snake's body must play an important role. Early studies indicated that it may even outperform other airfoil shapes in the Reynolds number regime at which the snakes glide ($Re\sim$5000--15,000).\cite{Socha2011}

The first investigations into the aerodynamic characteristics of the flying snake's body profile were made by Miklasz et al.\cite{MiklaszETal2010} They tested sections of tubing that approximated the snake's cross-sectional geometry as filled circular arcs, and encountered some unexpected aerodynamics. The lift measured on the sections increased with angle of attack ({\aoa}) up to $30^{\circ}$,  then decreased gently without experiencing catastrophic stall, and the maximum lift coefficient ($C_{L}=1.5$) appeared as a noticeable spike at the stall angle (defined as the angle of attack at which lift is maximum). Holden et al.,\cite{Holden2011} followed with the first study of a model using an anatomically accurate cross-section. They observed a spike in the lift curve at {\aoa} $35^\circ$ for flows with Reynolds numbers 9000 and higher. The maximum lift coefficient was 1.9, which is unexpectedly high, given the unconventional shape and Reynolds number range.\cite{Shyy2008} The counter-intuitively high lift observed in both studies suggests a lift-enhancement mechanism. Holden et al.\ inferred that the extra lift on the body was due to suction by vortices on the dorsal side of the airfoil. However, the flow mechanism responsible for the enhanced lift by the flying snake's profile has yet to be identified. 

In the present study, we aim to answer the question: what is the mechanism of lift enhancement created by the flying snake's cross-sectional shape?  To address this, we ran two-dimensional simulations of flow over the anatomical cross-section of the snake. We computed the flow at various angles of attack at low Reynolds numbers starting from $Re=500$ and increasing. We found that a marked peak in the lift coefficient does appear at {\aoa} $35^{\circ}$ for $Re=2000$ and above. Hence, our simulations were able to capture some of the unique lift characteristics observed in the physical experiments of Holden et al.
Aiming to explain the lift mechanism, we analyzed vorticity visualizations of the wake, time-averaged pressure fields, surface pressure distributions, swirling strength, and wake-vortex trajectories. 

Although we performed two-dimensional simulations of flow over the cross-section of the snake, we recognize that three-dimensional effects are present at this Reynolds-number regime\cite{Williamson1996}. Further 3-D and unsteady mechanisms due to the motion of the snake likely play a role in real gliding (we discuss these effects in Section \ref{discussion}).
Despite the simplification of the gliding system, the two-dimensional simulations in this study provide insight into flows at Reynolds numbers $10^3$--$10^4$ (sometimes called `ultra-low' Reynolds numbers in the aeronautics literature\cite{KunzKroo2001, Sunada2002a}) where studies \cite{AlamETal2010,Shyy2008,VargasETal2008} are scarce compared to traditional applications in aeronautics.

\section{Background material}
\subsection{Description of the gliding behavior and kinematics of flying snakes}

\begin{figure*}
\centering
	\includegraphics[width=0.7\textwidth]{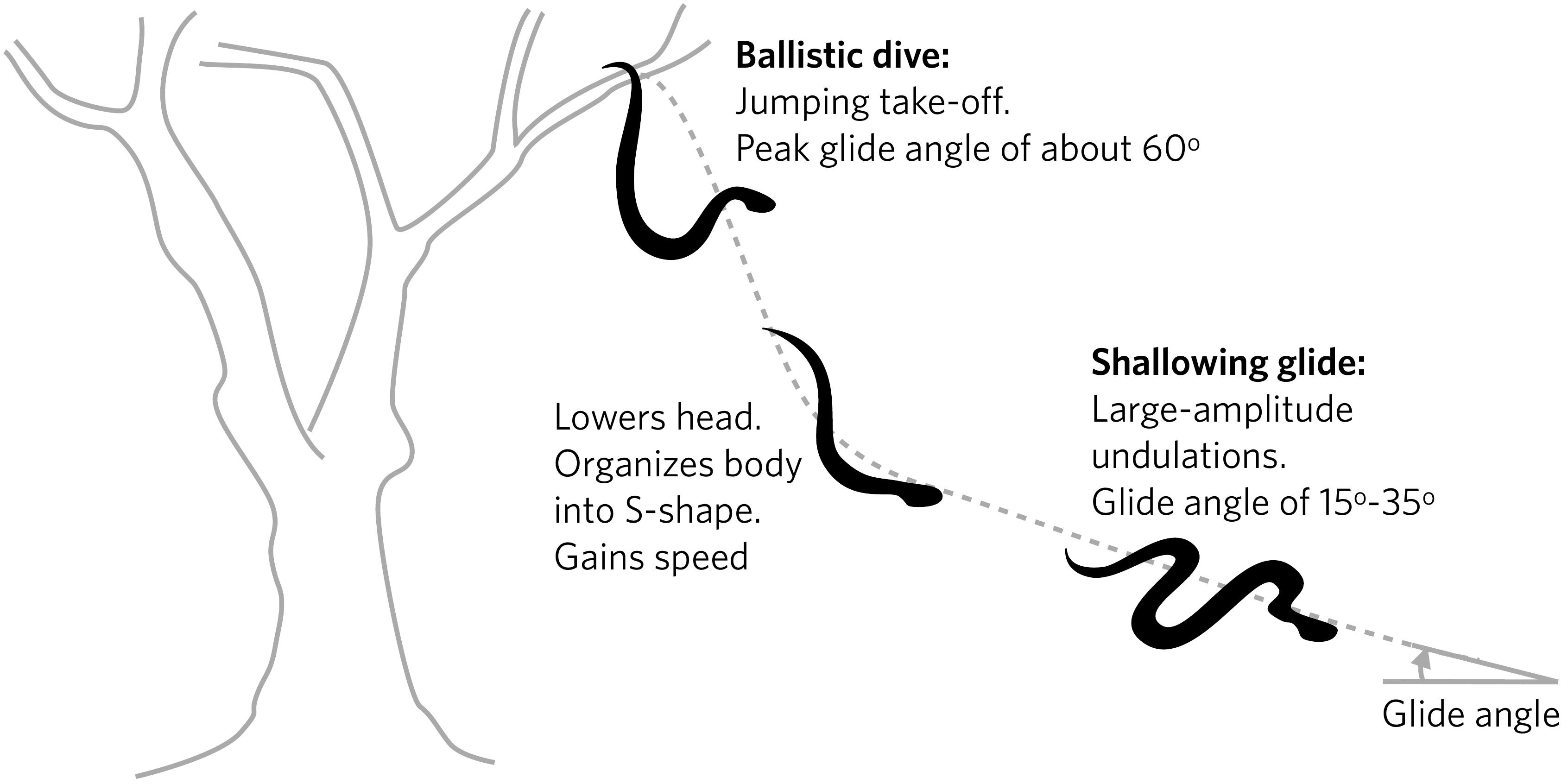}
	\caption{\small Different stages in a typical glide trajectory of the paradise tree snake.}
	\label{fig:snakeDive}
\end{figure*}

\begin{figure}
\centering
	\subfloat[]{ \includegraphics[width=0.40\columnwidth]{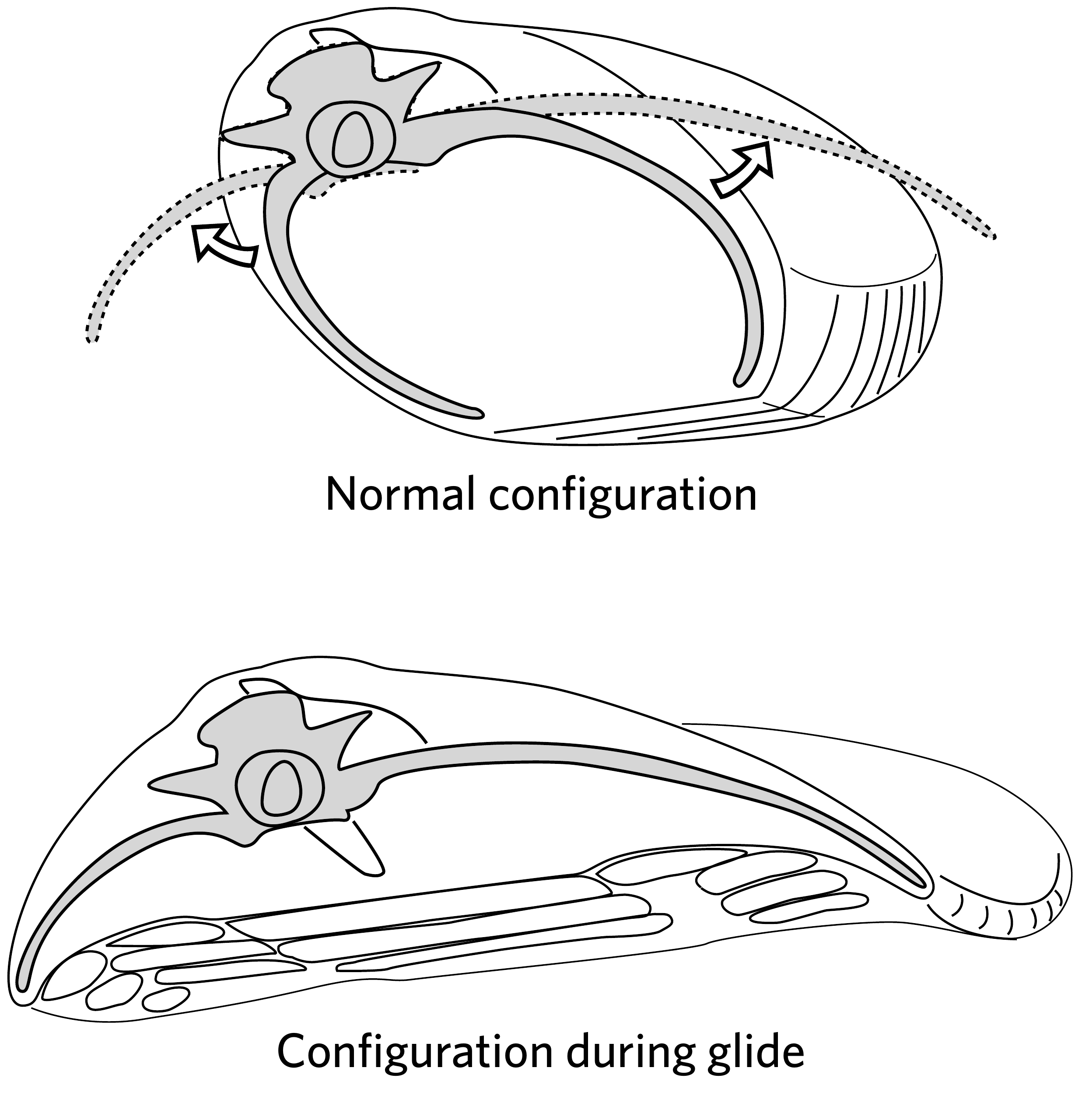} \label{fig:snakeRibs}} \qquad
	\subfloat[]{ \includegraphics[width=0.30\columnwidth]{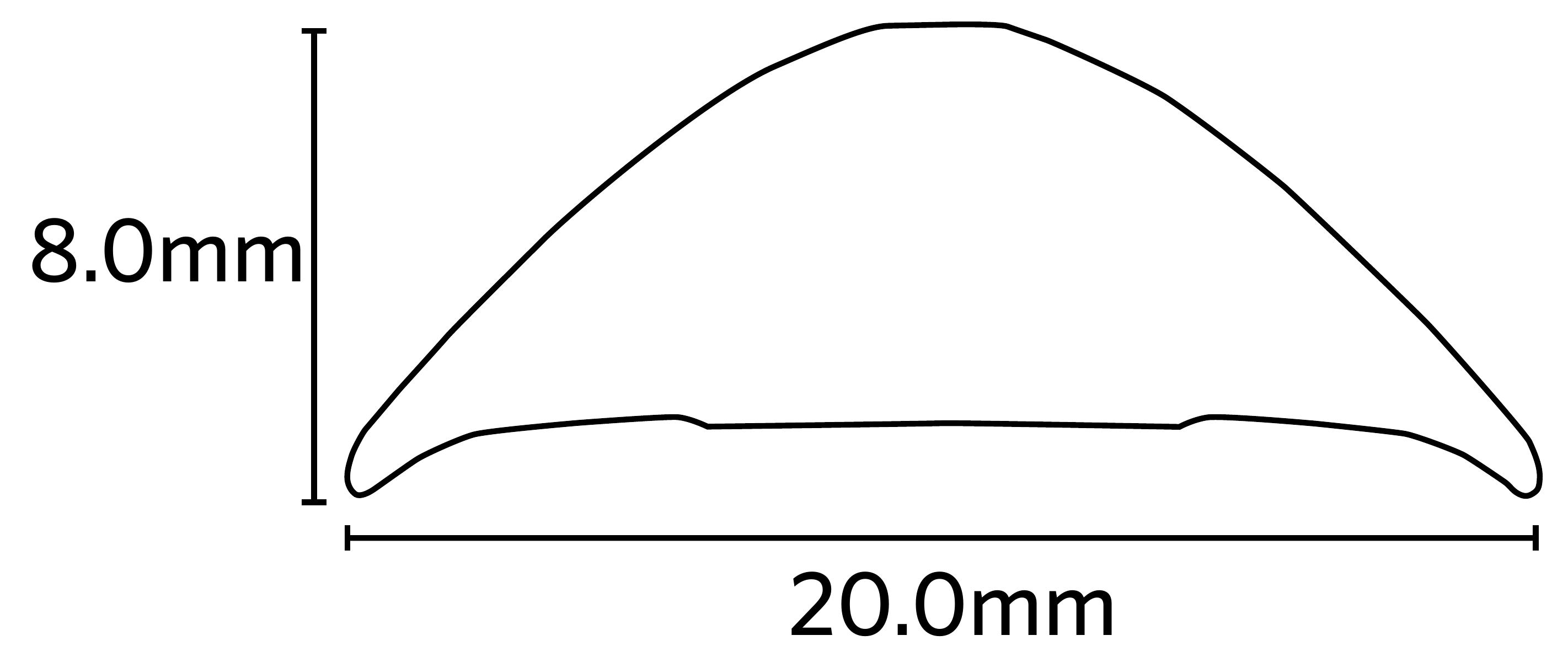} \label{fig:snakeCS}}
\caption{\small (a) Artistic impression of the mechanism that changes the snake's body cross-section for gliding. The rib movement is hypothesized to be directed both anteriorly (toward the head) and dorsally (toward the spine).\cite{Socha2011} The sketch is not to scale and was adapted from an illustration by Tara Dalton Bensen in collaboration with Jake Socha. (b) Cross-section of a typical adult \emph{C.\ paradisi} during glide, modified from a previous study. \cite{Socha2011} Geometry data and plot available under CC-BY license.\cite{KrishnanETal-share705877}}
\label{fig:cs}
\end{figure}

\emph{C.\ paradisi} and other gliding snakes make use of unique kinematics, illustrated in Figure \ref{fig:snakeDive}. The glide begins with a ballistic dive from a tree branch: the snake launches itself with a horizontal velocity and falls with a relatively straight posture. Immediately after the jump, it spreads its ribs apart and nearly doubles the width of its body, changing its cross-section from the cylindrical shape to a flattened shape\cite{Socha2002} (see sketch in Figure \ref{fig:snakeRibs}). The profile can be described as a rounded triangle with fore-aft symmetry and overhanging lips at the leading and trailing edges.\cite{Socha2011} Figure \ref{fig:snakeCS} shows the anatomically accurate cross-section geometry of the paradise tree snake in the airborne configuration.\cite{Socha2011,KrishnanETal-share705877} 
The glide angle during the ballistic dive can reach a maximum of about $60$ degrees,\cite{SochaETal2005} while the snake gains speed, changes its posture to an $S$-shape and begins undulating laterally. It thus uses its body as a continuously reconfiguring wing that generates lift. The speed of descent (i.e.,~the rate at which the snake is sinking vertically) peaks and then decreases, while the trajectory transitions to the shallowing glide phase of the flight. Equilibrium glides have rarely been observed under experimental conditions,\cite{Socha2002, SochaETal2005, SochaETal2010} and the glide angle usually keeps decreasing without attaining a constant value before the end of the descent. In field observations, the snakes cover a horizontal distance of about 10 m, on average, when jumping from a height of about 9 m. Glide angles as low as $13^\circ$ have been observed towards the end of the descent.\cite{SochaETal2005}

\begin{figure*}
\centering
	\subfloat{\includegraphics[width=0.65\textwidth]{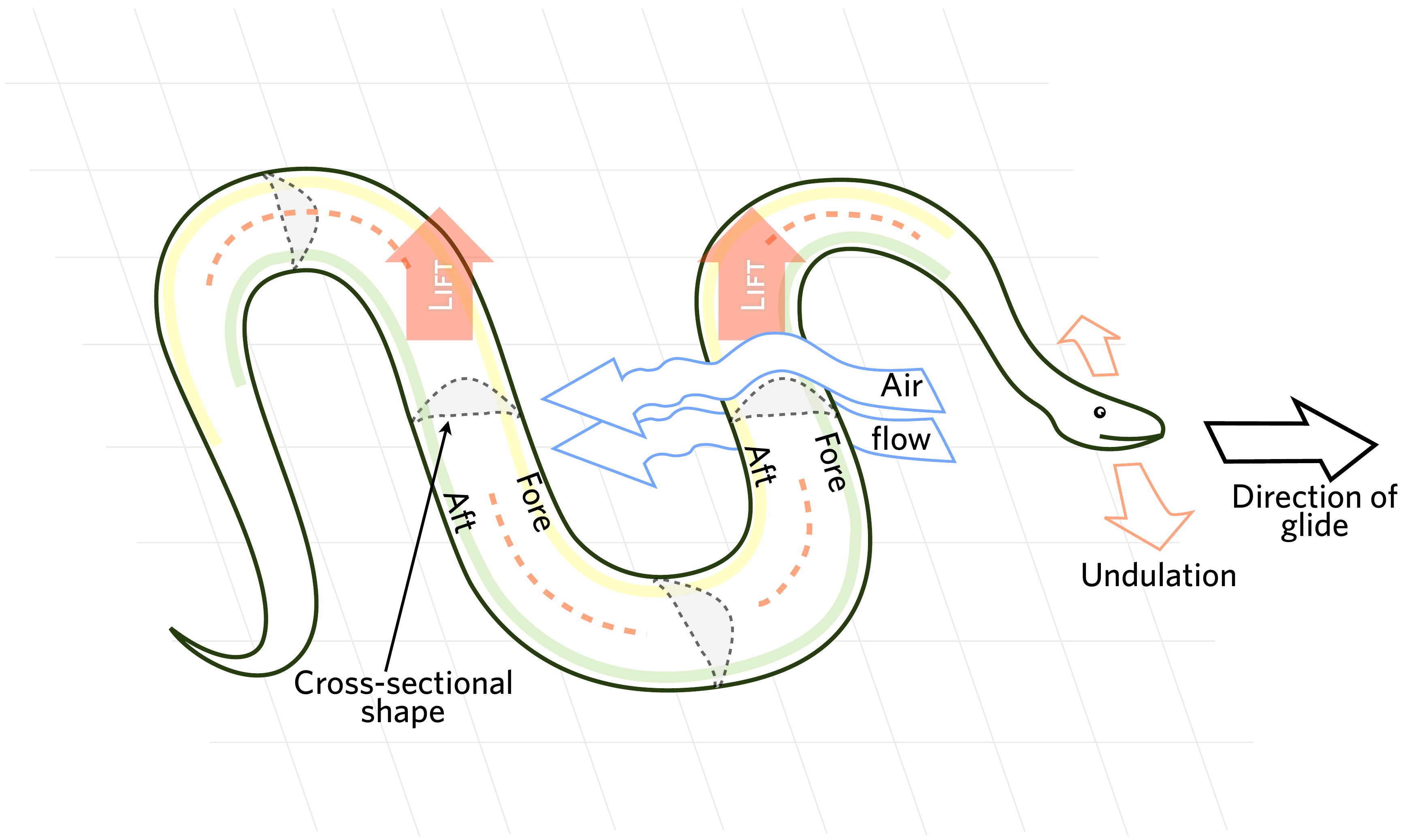}}
	\subfloat{\includegraphics[width=0.3\textwidth]{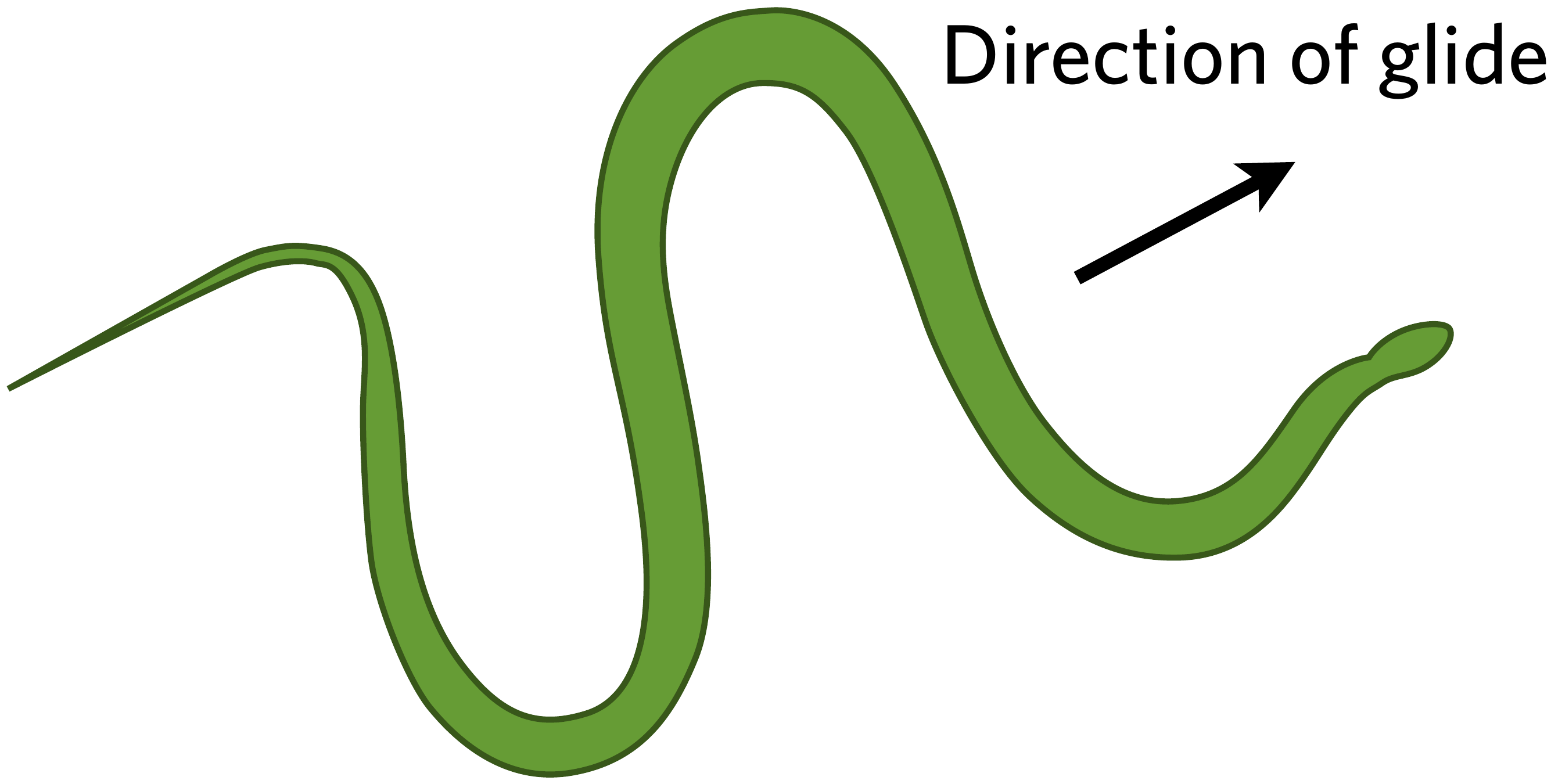}}
	\caption{\small  As it moves through the air, the snake undulates laterally, sending traveling waves down the body. The body dimensions have been exaggerated for the sake of clarity (see sketch on the right for a proportionally more accurate plan-view of the snake in flight).}
	\label{fig:snakeMotion}
\end{figure*}

During a glide, the snake moves its head side-to-side, sending traveling waves down the body, as illustrated in Figure \ref{fig:snakeMotion} (not to scale). With the body forming a wide $S$-shape at an angle with respect to the glide trajectory, long sections of the flattened body that are perpendicular to the direction of motion generate lift. Compared to terrestrial motion,\cite{Socha2002} the undulation frequency in air (1--2 Hz) is lower and the amplitude (10--17\% snout-vent length) is higher.\cite{SochaETal2005} As the body moves forward along the glide path, the fore and aft edges of the body repeatedly swap, which is not a problem aerodynamically thanks to the fore-aft symmetry. All portions of the body move in the vertical axis as well, with the most prominent motion occurring in the posterior end.

\subsection{Previous experimental studies with cross-section models of the snake}

The kinematics of gliding in the paradise tree snake are complex and various factors may contribute to the generation of lift.  It is difficult to study the unsteady aerodynamics of the snake by incorporating simultaneously all of the elements of its natural glide.  But as with any airfoil, we expect the cross-sectional shape to play an important role in the aerodynamics. This cross-section is thick compared to conventional airfoils, and is better described as a lifting bluff body.\cite{Holden2011}

Miklasz et al.\cite{MiklaszETal2010} were the first to investigate the role of the profile, using wind-tunnel testing of stationary models resembling the snake's cross-section. The models consisted of segments perpendicular to the flow, with simple geometrical shapes meant to approximate a snake's straight portions of the body (circular pipes cut to make a semi-circular, filled, half-filled, or empty cylindrical section).
Their experiments provided measurements of lift and drag at various angles of attack ({\aoa}) in a flow with Reynolds number 15,000. The maximum lift ($C_L=1.5$) occurred at an angle of attack of $30^\circ$ and the drag remained approximately the same between $25^\circ$ and $30^\circ$, thus causing a spike in the polar plot (where the lift coefficient is plotted against the drag coefficient). Near-maximal lift was also observed in a wide range of angles of attack ($10^\circ$--$30^\circ$). Beyond {\aoa} $30^\circ$, in the post-stall region,  the lift drops gradually while the drag increases rapidly. This region is characterized by flow separating at the leading edge, resulting in some of the flow being deflected upwards and consequently generating a wide wake. Miklasz et al.\ also tested tandem models finding that the downstream section experienced higher lift when placed below the wake of the upstream model, at horizontal separations up to five chord lengths.

A subsequent study by Holden et al.\cite{Holden2011} was the first to use models with an anatomically accurate cross-section based on \emph{C.\ paradisi}.  They tested the models in a water tunnel at Reynolds numbers in the range 3000--15,000 and used time-resolved digital particle image velocimetry (\textsc{trdpiv}).\cite{Willert1991} Plotting the experimental lift curves, they show a peak in lift coefficient at {\aoa} $35^{\circ}$, with a maximum $C_{L}$ between 1.1 and 1.9, increasing with Reynolds number. For Reynolds numbers 9000 and above, the maximum $C_{L}$ also appeared as a noticeable spike---about 30\% higher than at any other angle of attack. In the case of drag, at Reynolds 7000 and below $C_D$ grows gradually until $35^\circ$ {\aoa} and then increases steeply in the post-stall region. But for Reynolds 9000 and higher, the drag coefficients at $30^\circ$ and $35^\circ$ {\aoa} are the same, similar to the result by Miklasz et al. High values of lift coefficient were maintained in the range $20^\circ$--$60^\circ$. The peak in lift at $35^{\circ}$ is an unexpected feature of the snake's cross-section and the value $C_{L}= 1.9$ is considered high for an airfoil at this Reynolds number.\cite{Shyy2008} But gliding and flying animals have surprised researchers before with their natural abilities and performance.\cite{BahlmanETal2013,Bishop2007}

\section{Methods}

\subsection{Numerical approach: Immersed boundary method}

We solved the Navier-Stokes equations using an immersed boundary method ({\ibm}). Our implementation uses a finite-difference discretization of the system of equations in a projection-based formulation proposed by Taira \& Colonius.\cite{Taira2007} For the convection terms, we used an explicit Adams-Bashforth time-stepping scheme, with a Crank-Nicolson scheme for the diffusion terms. The spatial derivatives for the convection terms were calculated using a second-order symmetric conservative scheme \cite{Morinishi1998} and the diffusion terms were calculated using central differences. The formal spatial order of accuracy of the method is second order everywhere except near the solid boundaries, where it is first order. The temporal order of accuracy is first order, by virtue of the projection method used to step forward in time. Throughout this study, we performed direct numerical simulations of two-dimensional, unsteady, incompressible flow over the cross-section of the snake.

\subsection{Set-up of the computational experiments}

\begin{figure}
\centering
	\includegraphics[width=0.5\columnwidth]{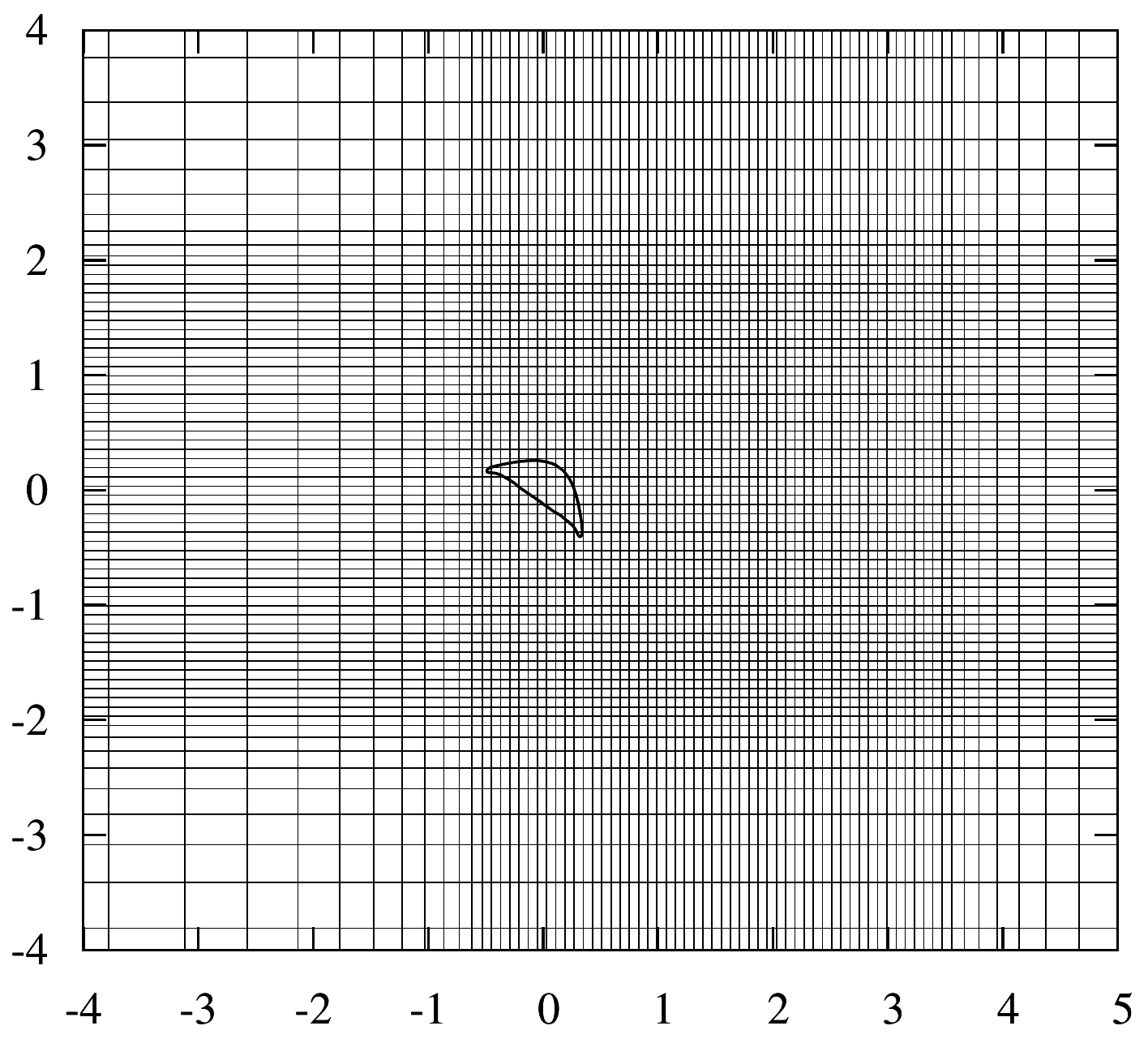}
	\caption{\small A portion of the grid used to simulate two-dimensional flow past the cross-section of the snake, showing only one out of every $20$ grid points, for clarity.}
	\label{fig:grid}
\end{figure}

We used an accurate geometry for the cross-section of the snake, as determined by Socha \cite{Socha2011} and used in a previous study with physical models. \cite{Holden2011} We discretized the flow domain using a stretching Cartesian grid, as shown in Figure \ref{fig:grid}.
The cross-section of the snake was scaled such that the chord length $c$ (defined, as in previous studies, as the maximum width of the profile) was equal to $1$. The body was placed at the center of a domain that spanned the region $[-15,15]\times[-15,15]$, a size chosen so that the velocity of the fluid at the boundaries could be fixed to the uniform velocity of the free stream ($u_\infty$) without affecting the solution. The value of $u_\infty$ was set to $1$ in the $+x$-direction, and the velocities at the inlet, top and bottom boundaries were fixed at this value. A convective boundary condition, $\frac{\partial{u}}{\partial{t}}+u_\infty\frac{\partial{u}}{\partial{x}}=0$, was applied at the outlet.

The grid is uniform with cell width $0.004$ in the region $[-0.52,3.48]\times[-2,2]$, and exponentially stretched in the remaining region with a stretching ratio of 1.01. The total number of cells in the entire domain is $1704{\times}1704$, providing nearly 3 million cells, with the majority located in the area near the body. 

The Reynolds number in the simulations varied in the range 500--3000, in increments of 500, and the angles of attack were varied in the range of $0^\circ$--$45^\circ$ in steps of $5^\circ$. Flows were impulsively started, and run for a sufficiently long period of time until periodic vortex shedding was obtained.
We simulated 80 time units of the unsteady flow in each run. With a time step of $0.0004$, this required 200,000 time steps for each run. We computed the flow over an impulsively started cylinder at $Re=3000$ with these same simulation parameters and grid, and the results matched well against past simulations. \cite{KrishnanBarba-share92789} 
We also conducted a grid-independence study by computing the flow over the  cross-section at $Re=2000$  and angles of attack $30^\circ$ and $35^\circ$, using two different grids and calculating the average lift and drag coefficients in each case. The primary measure of grid refinement used is the width $h$ of the square cells near the body, which are the smallest cells in the grid. The grids were generated such that the aspect-ratios of the cells near the domain boundary are nearly the same in both cases. As shown in Table \ref{table:gridIndependence},  when $h$ is reduced from 0.006 to 0.004, we obtain a change of 0.07\% in the lift coefficient at ${\aoa}=30^\circ$, and a change of 2.3\% in the lift coefficient at ${\aoa}=35^\circ$. This confirms that a grid with $h=0.004$ in our simulations  produces sufficiently accurate solutions.

\begin{table}
	\centering
	\begin{tabular}{ c c c c c c}
		${\aoa}=30^\circ$: & & & & & \\
		~Mesh size~ & ~~~~~$h$~~~~~ & ~Avg. $C_d$~ & $\Delta{C_d}/C_d$ & ~Avg. $C_l$~ & $\Delta{C_l}/C_l$ \\
		\hline
		$1293\times 1293$ & 0.006 & 0.964 & & 1.533 & \\
		$1704\times 1704$ & 0.004 & 0.967 & 0.3\% & 1.532 & 0.07\% \\
		\hline
		\\
		${\aoa}=35^\circ$: & & & & & \\
		~Mesh size~ & ~~~~~$h$~~~~~ & ~Avg. $C_d$~ & $\Delta{C_d}/C_d$ & ~Avg. $C_l$~ & $\Delta{C_l}/C_l$ \\
		\hline
		$1293\times 1293$ & 0.006 & 1.280 & & 2.098 & \\
		$1704\times 1704$ & 0.004 & 1.316 & 2.7\% & 2.147 & 2.3\% \\
		\hline
	\end{tabular}
	\caption{Grid-independence: Average lift and drag coefficients for flow over the cross-section of the snake at $Re=2000$ with ${\aoa}=30^\circ$ and $35^\circ$, using two different grids; $h$ is the width of the smallest cells in the domain, in the uniform region near the immersed boundary.}
	\label{table:gridIndependence}
\end{table}

\subsection{Post-processing of numerical solutions}

\begin{figure*}
\centering
	\includegraphics[width=0.65\textwidth]{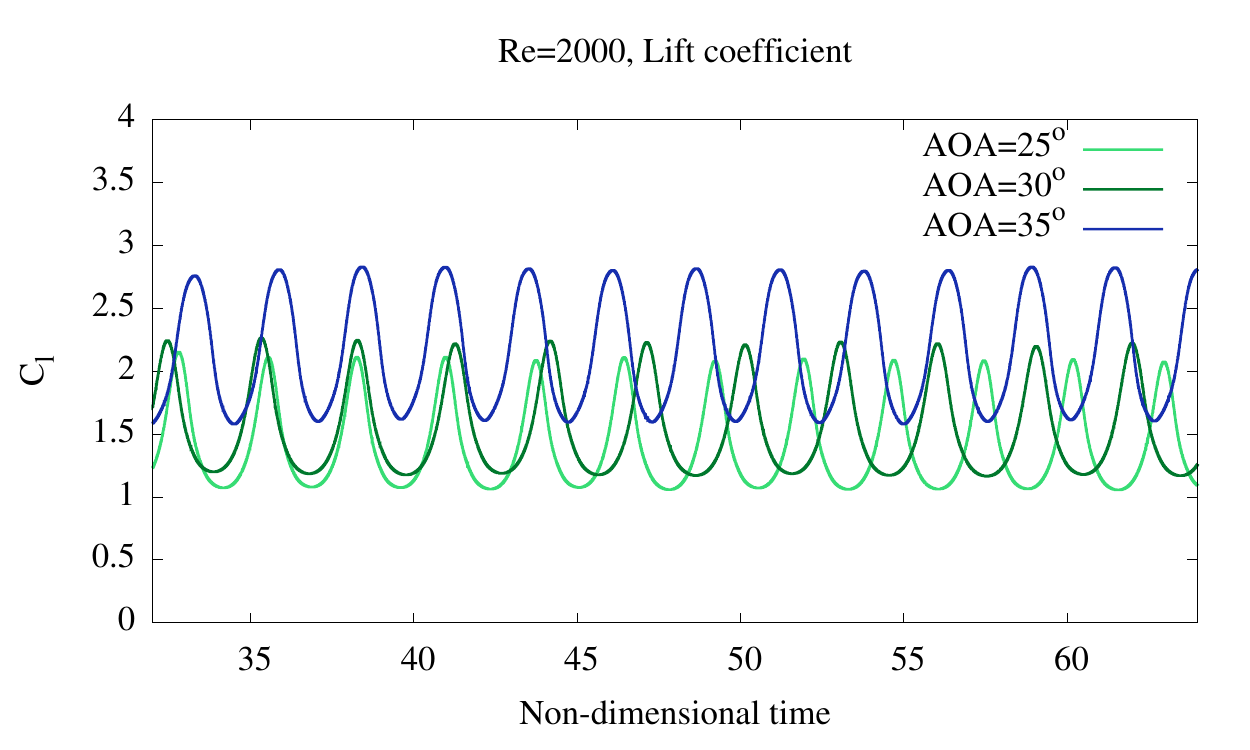}
	\caption{\small Unsteady lift coefficients for $Re=2000$ flow past the cross-section of the snake at {\aoa}=$25^\circ$, $30^\circ$ and $35^\circ$. }
	\label{fig:unsteadyLift}
\end{figure*}

\begin{figure}[H]
\centering
	\includegraphics[width=0.4\columnwidth]{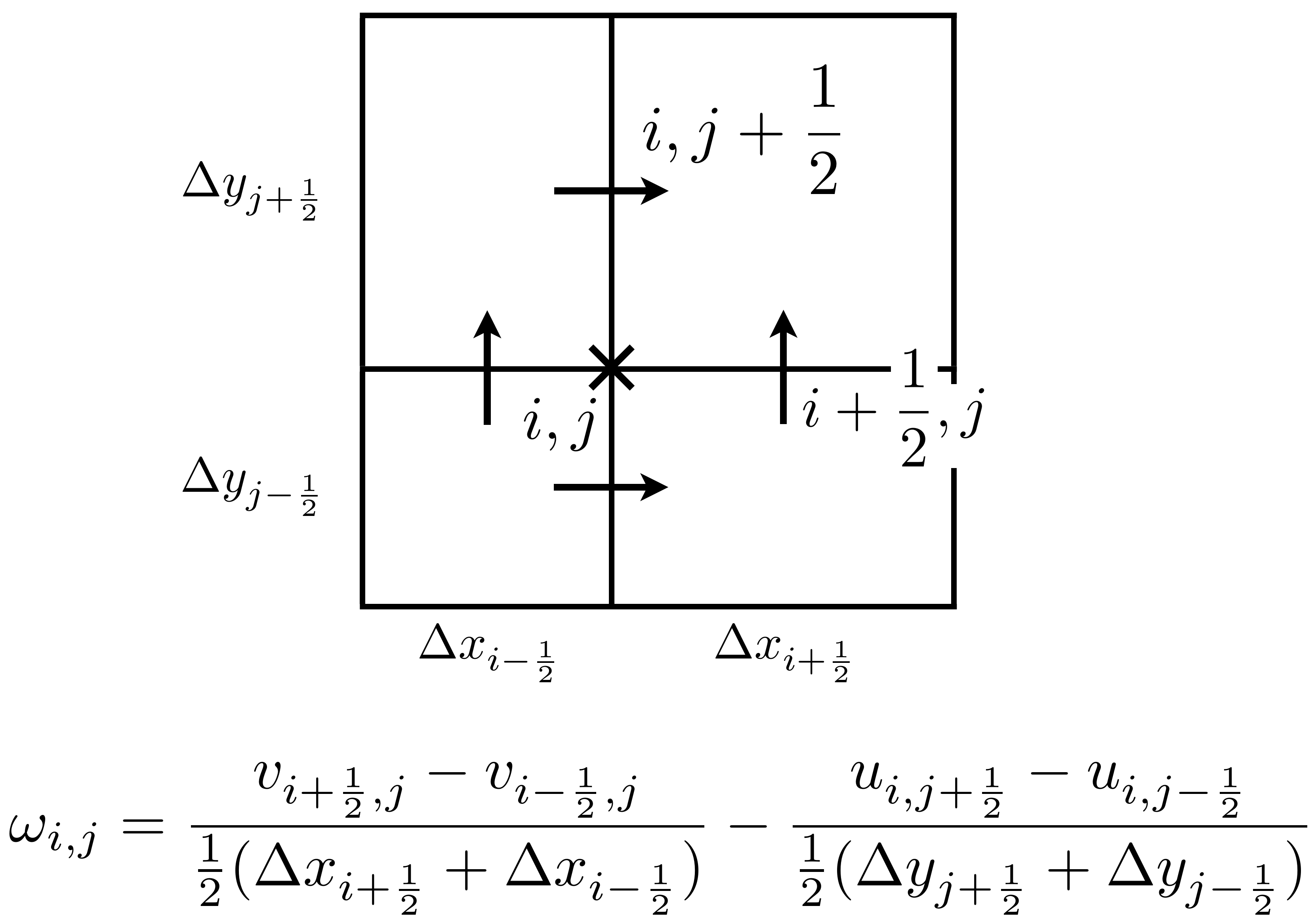}
	\caption{\small Numerical calculation of the discrete vorticity. }
	\label{fig:vortCalc}
\end{figure}

The numerical solutions for the flow field were post-processed to obtain various physical quantities and time-dependent flow visualizations, which we used to analyze the aerodynamic characteristics of the snake's cross-section. Here, we briefly describe the post-processing procedures:
\paragraph{Lift and drag coefficients.}
The lift and drag forces per unit length were obtained by integrating the immersed-boundary-method force distribution along the body. Forces were normalized using the fluid density $\rho$, freestream velocity $u_\infty$, and chord length $c$---all having a numerical value of 1 in the current study---to obtain the lift and drag coefficients:
\begin{equation}
C_l=\frac{L}{\frac{1}{2}\rho {u^2_\infty}c} \qquad
C_d=\frac{D}{\frac{1}{2}\rho {u^2_\infty}c}
\end{equation}
The lift and drag coefficients oscillate due to vortex shedding in the wake (see Figure \ref{fig:unsteadyLift}). To analyze the aerodynamic performance of the snake's cross-section, we calculated the mean lift and drag coefficients by taking the time-average of the unsteady force coefficients in the period of non-dimensional time $t^*=32$--$64$, neglecting the influence of the initial transients ($t^*$ is normalized using the freestream velocity $u_\infty$ and the chord length $c$).
\paragraph{Vorticity.}
In two-dimensional flows, only one component of vorticity exists: the component perpendicular to the plane of the flow, ${\omega}_k=\frac{\partial{v}}{\partial{x}}-\frac{\partial{u}}{\partial{y}}$. On our staggered Cartesian grid, the velocity components are stored at the centers of the cell faces; we calculated vorticity at the node points using a central-difference approximation of the velocity derivatives (see Figure \ref{fig:vortCalc}).
\paragraph{Pressure field.}
Most of the force on a bluff body moving through a fluid arises from differences in surface pressure, with frictional forces playing a minor role. Therefore, we analyzed the pressure field to supplement our observations of the lift and drag coefficients. The time-averaged pressure fields (and surface pressure distributions) were calculated by taking the mean of 125 equally spaced sampling frames for pressure in the period $t^*=32$--$64$.  This sampling rate corresponds to about 12 frames per period of vortex shedding, a choice made on the basis of being sufficient to capture the features of the flow.

The pressure contribution to the lift force is equal to the line integral of the pressure along the surface of the body. This surface distribution is obtained via bilinear interpolation of the pressure from grid points to locations on the body surface. However, the immersed boundary method can produce oscillations in the flow quantities at the surface itself. The width of the mesh cells $h$ near the body is 0.4\% of the chord length, and our discrete delta function for interpolating the velocity field extends to $1.5h$ from the surface in each Cartesian direction. The boundary definition is therefore not sharp, and we avoided spurious artifacts by measuring the surface pressure in the fluid at a distance of 1\% of the chord length normal to the body.

\paragraph{Swirling strength in the vortical wake.}
The vortex dynamics in the wake can provide insight into the mechanism of lift generation. To this end, we attempted to identify the vortices in the flow and plot their evolution. Several methods \cite{ChongETal1990,Hunt1988,jeong+hussain1995} have been developed to objectively define and identify vortices in fluid flows. In the present work, we make use of the swirling strength, \cite{Zhou1999} denoted by $\lambda_{ci}$ and defined as the imaginary part of the complex eigenvalue of the tensor $\mathbf{\nabla{u}}$. In regions of flow where no complex eigenvalues exist, there is no rotating flow and the swirling strength is zero.

\paragraph{Vortex trajectories}

Vortex trajectories were plotted by marking the locations of the centers of vortices at successive equally spaced instants of time. The centers of vortices are assumed to coincide with the minima of the instantaneous pressure field, and the markers were placed at these locations. The set of points obtained for each vortex represents its pathline. Vortex centers can also be identified as maxima in the swirling strength field. In the flows we studied, both methods resulted in the same vortex locations.	

\subsection{Validation \& Verification and Reproducibility}

We previously validated and verified our immersed-boundary-method code (called \texttt{cuIBM}) using analytical solutions, comparisons with published experimental results, and comparisons with other published numerical results.\cite{KrishnanBarba-share92789} The verification tests in two dimensions with an analytical solution used Couette flow between two cylinders, while the other benchmarks consist of lid-driven cavity at $Re=100$ and an impulsively started circular cylinder at different values of $Re$. Comparisons with published numerical results also include the vortex shedding over a circular cylinder and flow over heaving and flapping airfoils. We also computed the temporal and spatial orders of convergence  at several sampling times, using the Couette-flow test.

The Barba research group has a consistent policy of making science codes freely available, in the interest of reproducibility. In line with this, we release the entire code that was used to produce the results of this study. The \texttt{cuIBM} code is made available under the MIT open-source license and we maintain a version-control repository.\cite{Note1} The code repository includes all the files needed for running the numerical experiments reported in this paper: input, configuration, and post-processing. To support our open-science and reproducibility goals, in addition to open-source sharing of the code and the running scripts, several of the plots themselves are also available and usable under a CC-BY license, as indicated in the respective captions.

\section{Results}

\subsection{Lift and Drag coefficients}

\begin{figure}
\begin{center}
	\subfloat[Lift coefficient vs.\ \aoa.]{ \includegraphics[width=0.44\textwidth]{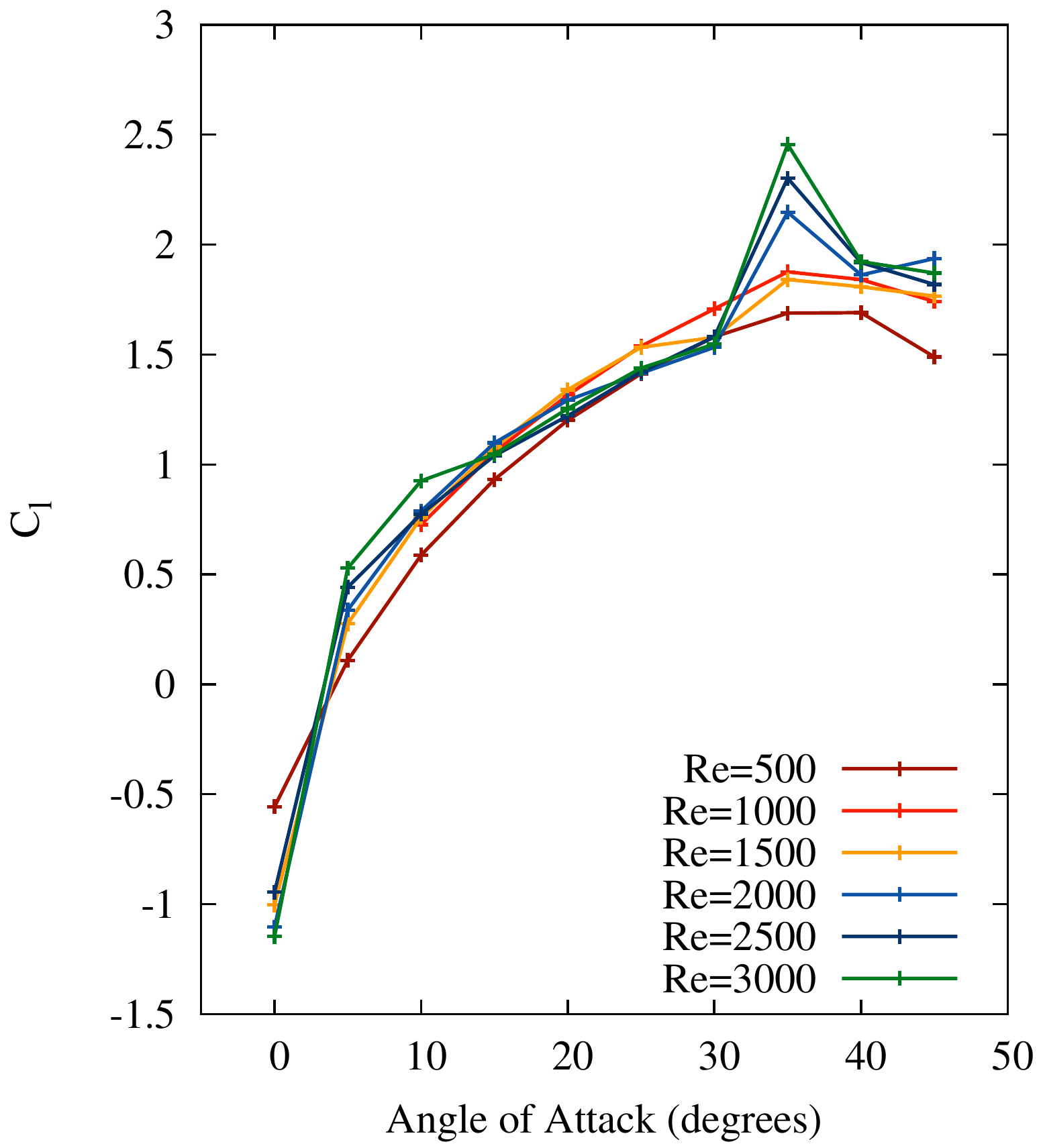} }
	\subfloat[Drag coefficient vs.\ \aoa.]{ \includegraphics[width=0.44\textwidth]{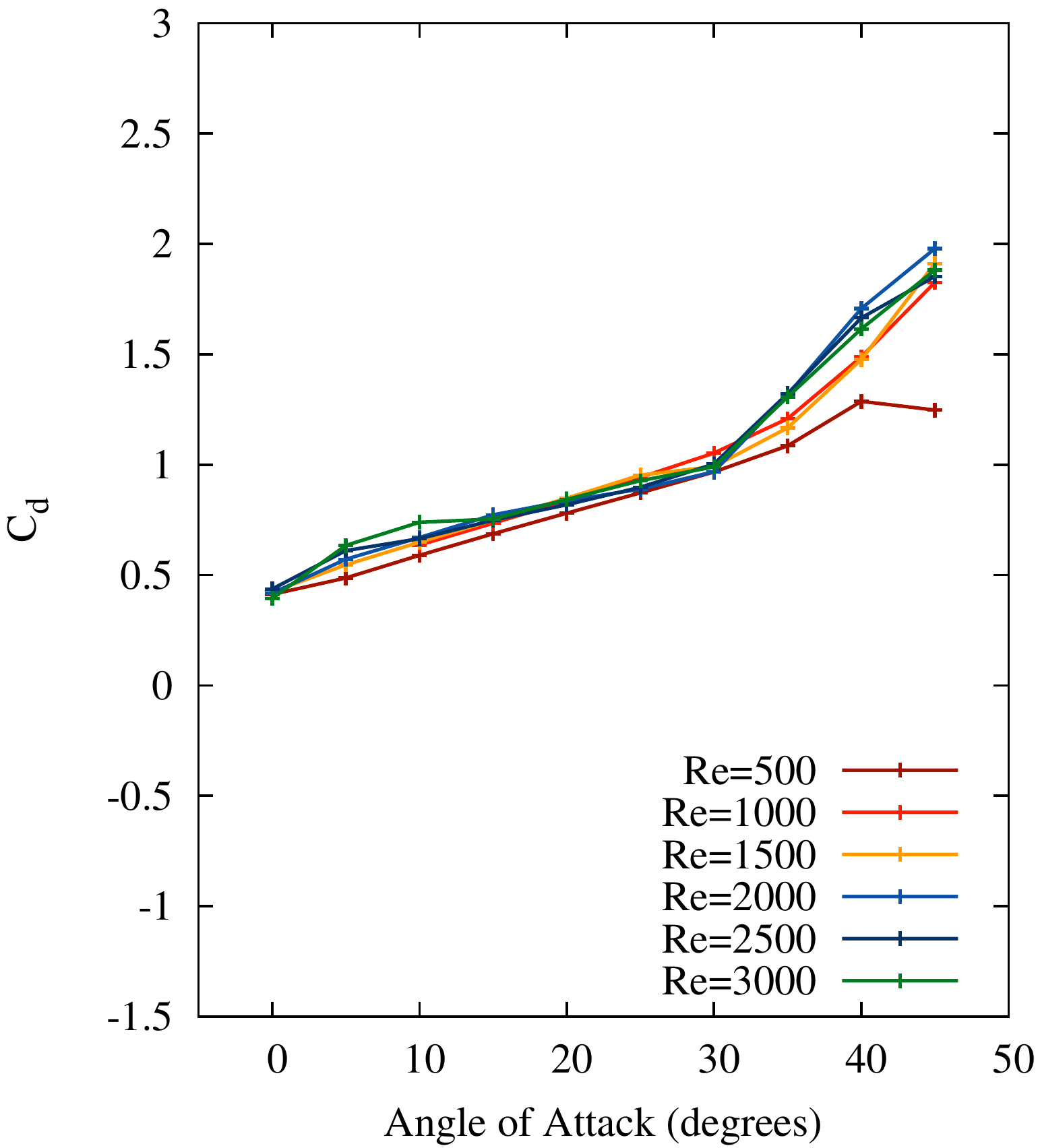} }
\caption{\small Time-averaged force coefficients. Data sets, plotting scripts, and figures available under CC-BY.\cite{KrishnanETal-share705883}}
\label{fig:clcd}
\end{center}
\end{figure}

The first result in this work is the characterization of the lift and drag of the snake profile. 
Figure \ref{fig:clcd} shows curves of lift and drag coefficient versus \aoa\ for Reynolds numbers between 500 and 3000.
The lift coefficient ($C_L$) increases rapidly with \aoa\ at all the Reynolds numbers tested, from a negative lift at $0^{\circ}$ to about $1.5$ at $25^{\circ}$ or $30^{\circ}$.  Beyond this value of \aoa, the lift coefficient increases more slowly, peaks, and then starts falling (stall). For $Re=2000$ and above, the value of $C_L$ jumps to above $2$ at an \aoa$=35^{\circ}$, the stall angle.
The drag coefficient ($C_D$) increases gradually and almost linearly in the range \aoa$= 0$--$30^{\circ}$. Except for the lowest value of the Reynolds number, $Re=500$, the slopes of the $C_D$-versus-{\aoa} curves increase after {\aoa} $30^{\circ}$ or $35^{\circ}$, as the snake profile approaches stall.

In summary, the 2D model of the snake's cross-section exhibits enhanced lift at \aoa\ $35^{\circ}$, just before stall. This \aoa\ coincides with the observations in water-tunnel experiments by Holden and colleagues.\cite{Holden2011} The actual values of the lift coefficient, however, are larger in these simulations than in the experiments.

\subsection{Vorticity of the wake}

\begin{figure*}
\begin{center}
	\subfloat[$Re$=1000, {\aoa}=$30^\circ$]{ \includegraphics[width=0.5\textwidth]{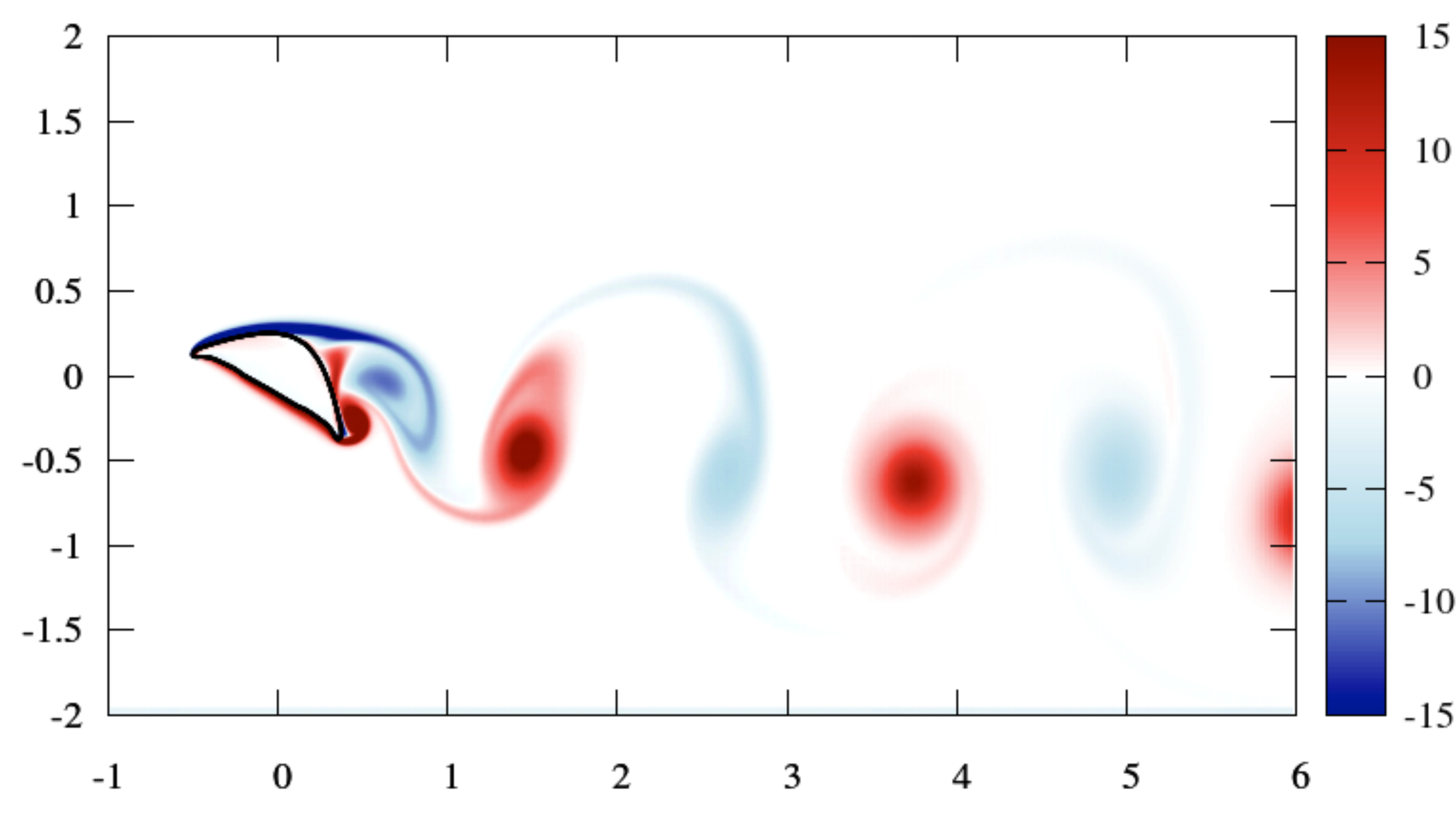} \label{subfig:vort1k30}}
	\subfloat[$Re$=2000, {\aoa}=$30^\circ$]{ \includegraphics[width=0.5\textwidth]{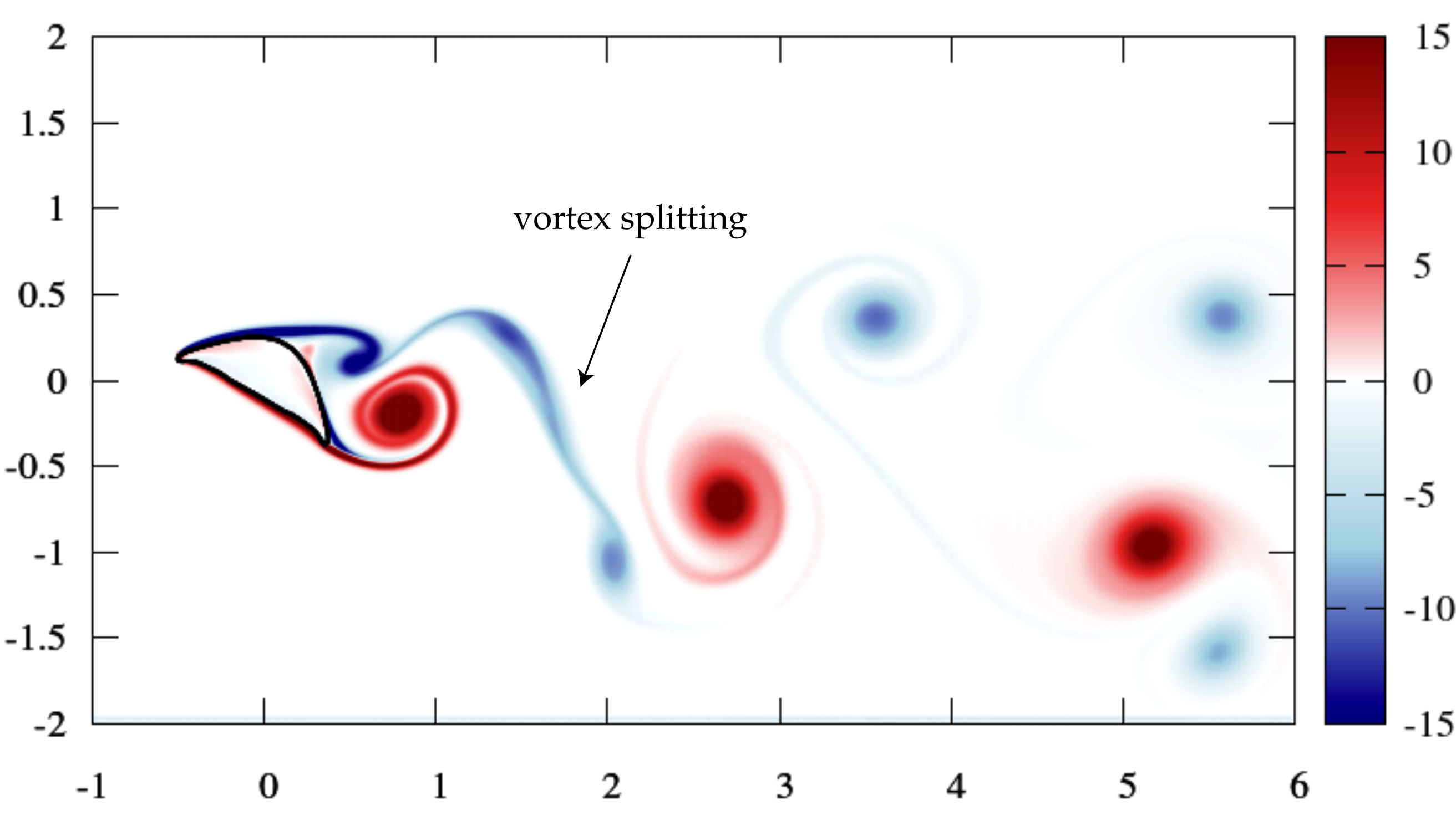} \label{subfig:vort2k30}}  \\
	\subfloat[$Re$=1000, {\aoa}=$35^\circ$]{ \includegraphics[width=0.5\textwidth]{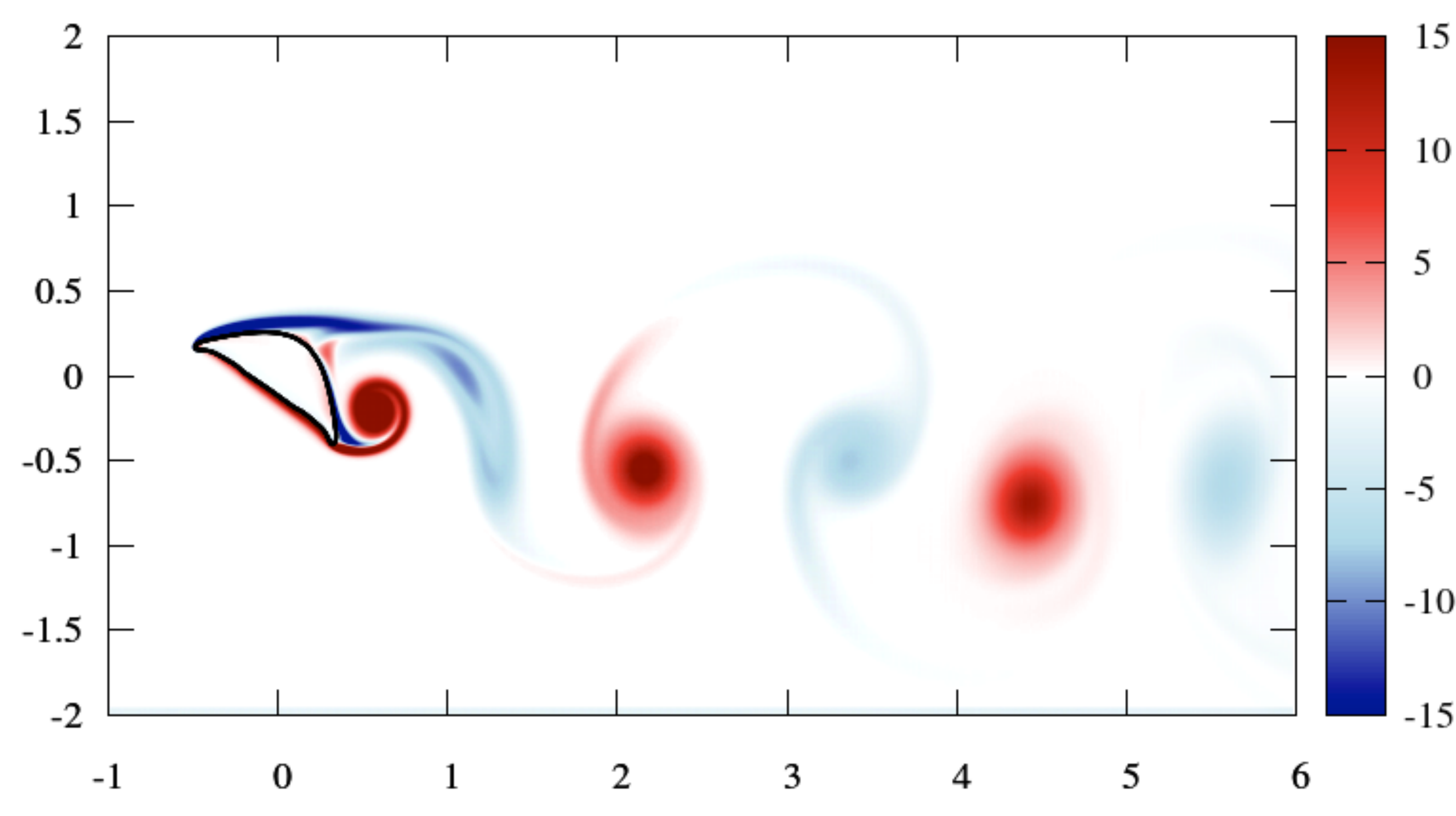} \label{subfig:vort1k35}} 
	\subfloat[$Re$=2000, {\aoa}=$35^\circ$]{ \includegraphics[width=0.5\textwidth]{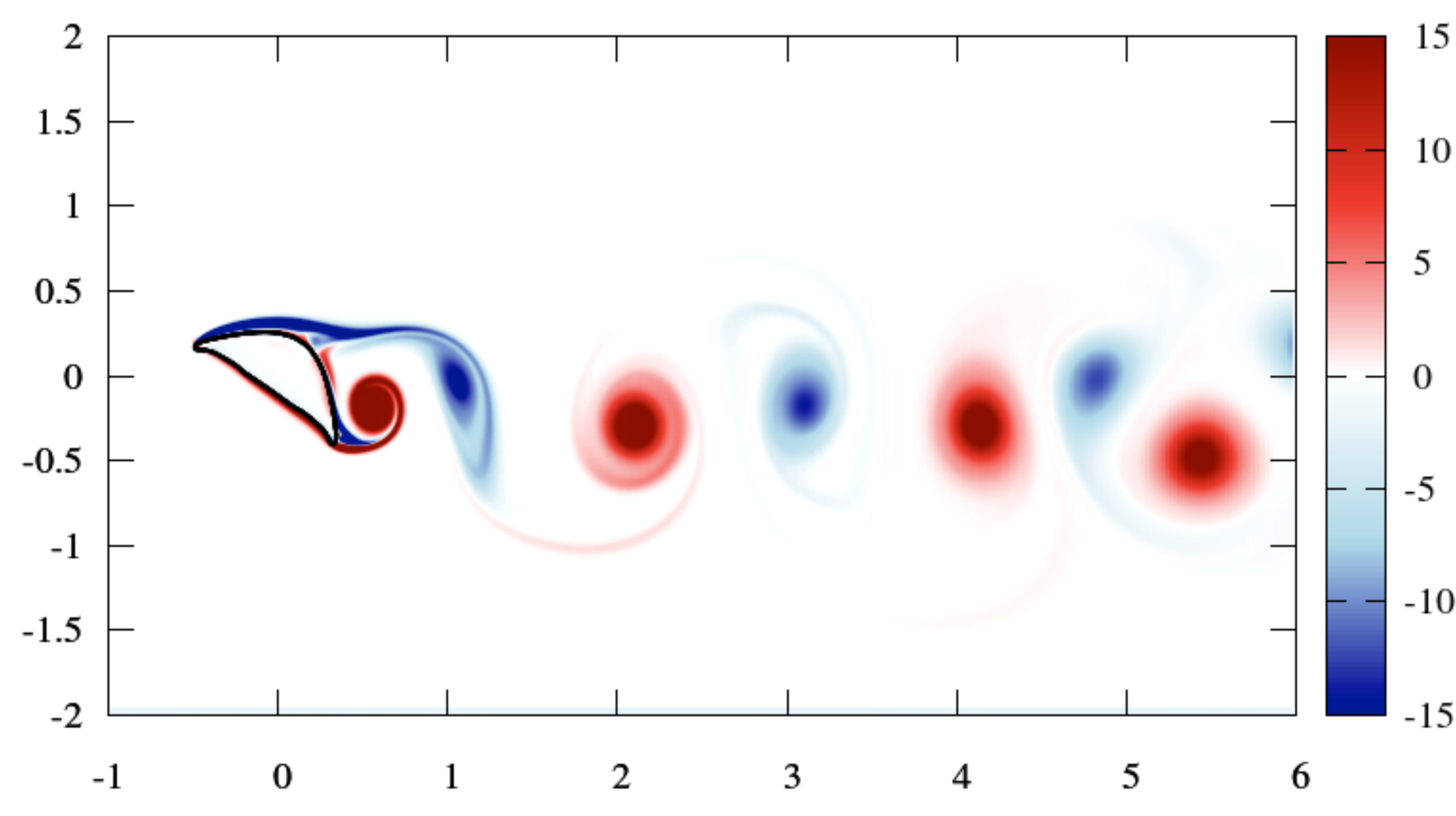} \label{subfig:vort2k35}}  \\
	\subfloat[$Re$=1000, {\aoa}=$40^\circ$]{ \includegraphics[width=0.5\textwidth]{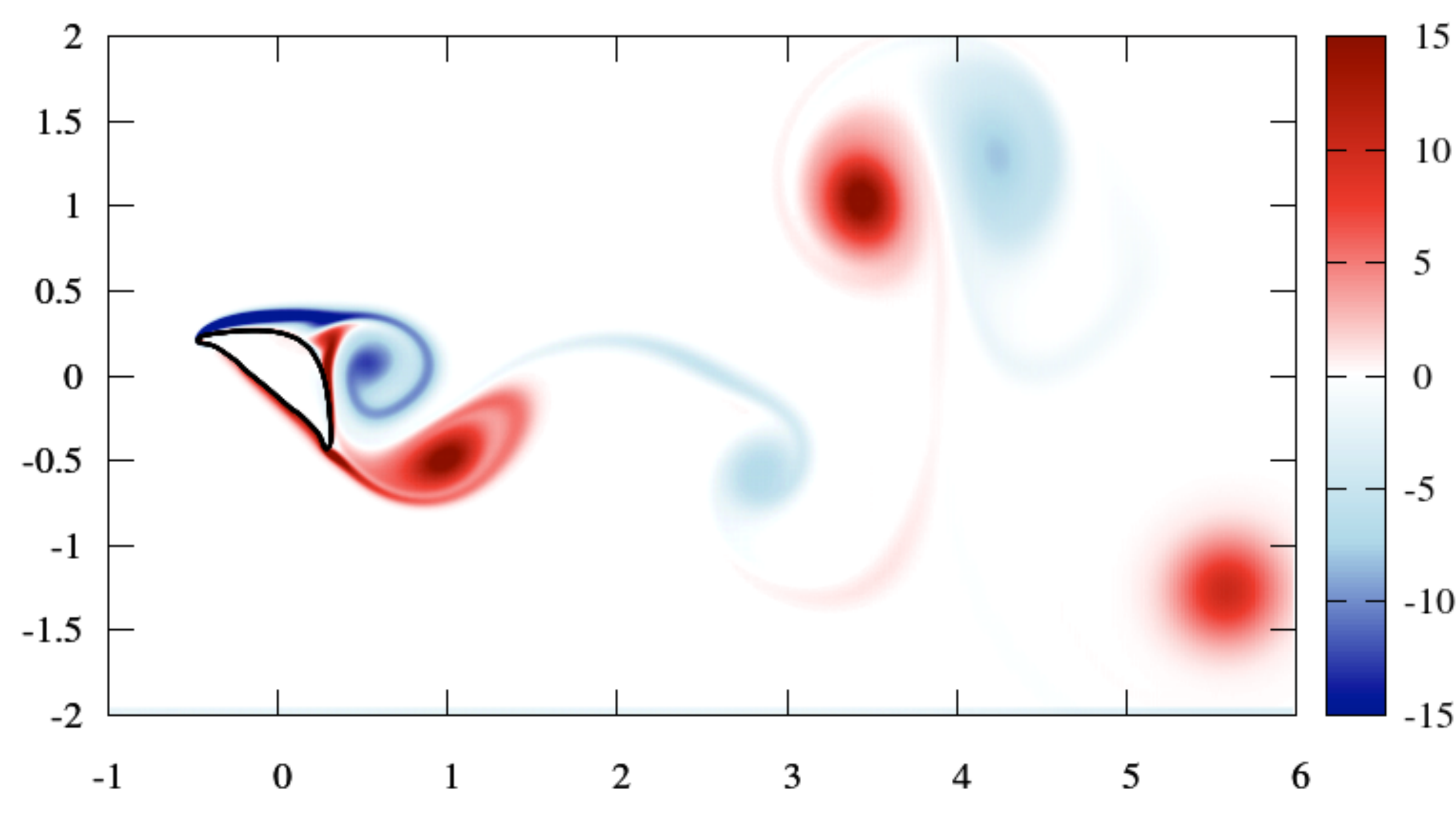} \label{subfig:vort1k40}} 
	\subfloat[$Re$=2000, {\aoa}=$40^\circ$]{ \includegraphics[width=0.5\textwidth]{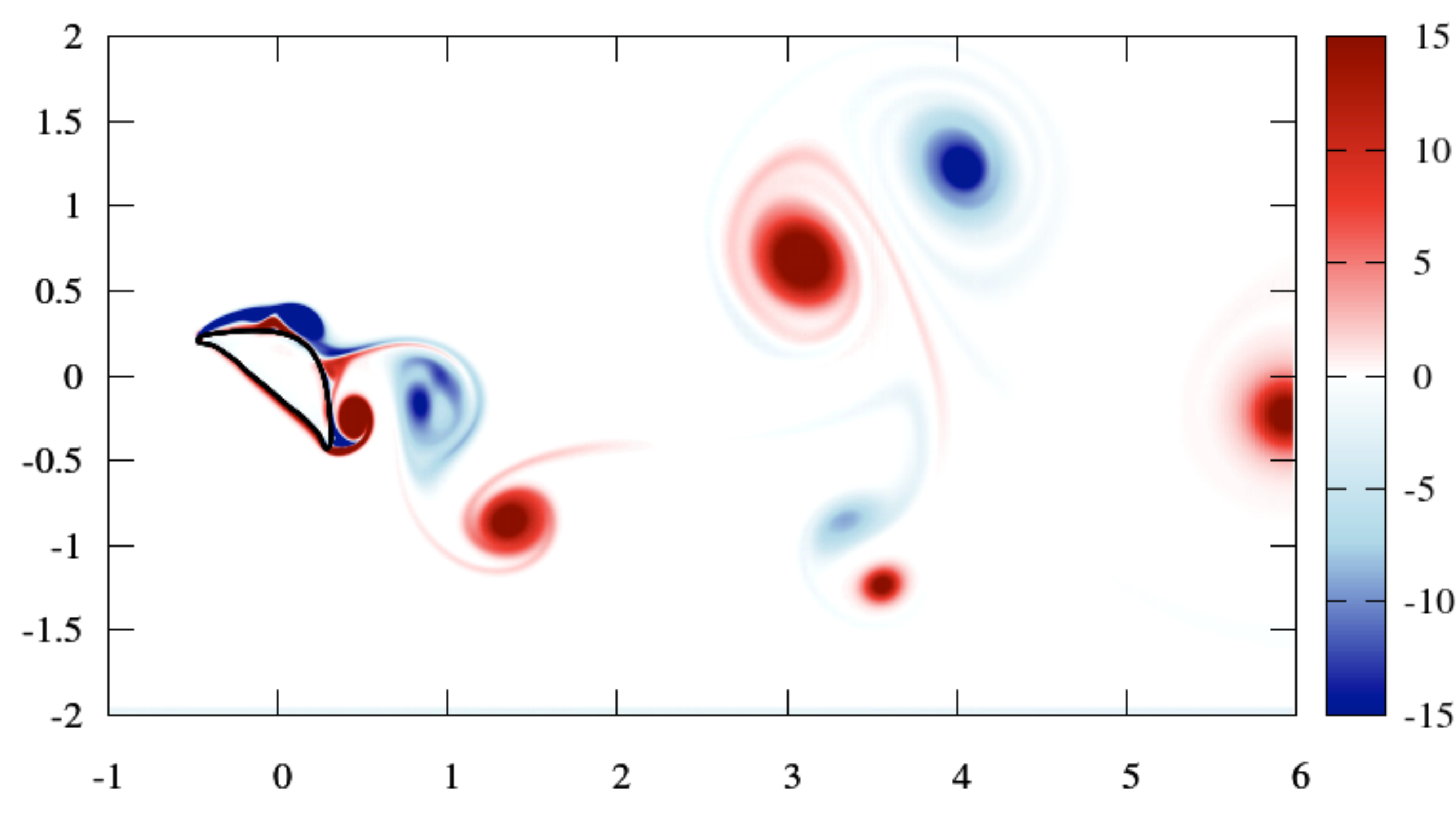} \label{subfig:vort2k40}} 
\caption{\small Vorticity field in the wake of the snake profile at Reynolds numbers 1000 and 2000, for angles of attack $30^\circ$, $35^\circ$ and $40^\circ$. The colors saturate at values of $\pm{15}$. A video showing animations of these wakes is available online.\cite{KrishnanBarba-share157334}}
\label{fig:vorticity}
\end{center}
\end{figure*}

Given that the lift coefficient spikes at \aoa\ $35^{\circ}$ for the simulations with $Re=2000$ and higher, we generated visualizations of the vortical wakes at both $Re=1000$ and $2000$ and searched for any differences.  The results are presented in Figure \ref{fig:vorticity}, where the frames in the left column correspond to $Re=1000$, and the frames in the right column are for $Re=2000$, in both cases for \aoa$=\{30^{\circ}, 35^{\circ}, 40^{\circ}\}$. (Animations of these visualizations are available online as supplementary materials.\cite{KrishnanBarba-share157334})

The wake at $Re=1000$ is a classical von K{\'a}rm{\'a}n vortex street at the lower values of \aoa: alternate clockwise (blue) and counter-clockwise (red) vortices are shed to form a street, which is slightly deflected downwards. At \aoa\ $40^{\circ}$, the wake is different. The separation point of the vortices on the dorsal surface has moved towards the leading edge and larger vortices are formed in the region behind the body, with longer formation times. Some vortices form dipole pairs that are deflected upwards, giving rise to a much wider wake (Figure \ref{subfig:vort1k40}). Associated with this wake behavior, the drag curve shows an increase in slope and the lift coefficient drops (Figure \ref{fig:clcd}).

The vortices produced in the wake at $Re=2000$ are stronger and more compact, as expected. 
In addition, three different wake patterns appear across angles of attack: for \aoa\ $30^\circ$ and below, for $35^\circ$, and for {\aoa} $40^\circ$ and above. At {\aoa} $30^\circ$, the dorsal vortex (blue) interacts with the trailing=edge vortex and is strained to the point that it is split in two (Figure \ref{subfig:vort2k30}).
This gives rise to a wake pattern known as $S+P$, for `single' and `pair': a street of single vortices at the top and dipoles at the bottom (or \emph{vice versa}). 
 At {\aoa} $35^\circ$, the dorsal vortex is stronger and it separates closer to the leading edge. The strain field of the trailing-edge vortex in this case is not strong enough to split the dorsal vortex, and the resulting wake is a classical von K{\'a}rm{\'a}n vortex street. It differs from that of the lower Reynolds number, $Re=1000$, by showing almost no deflection and consisting of stronger vortices that are more tightly close together (Figure \ref{subfig:vort2k35}). Finally, at angles of attack $40^\circ$ and higher, the wake is similar to the $Re=1000$ case, with large vortex pairs deflected upwards and smaller vortices deflected downwards. There are more evident shear-layer instabilities---with higher-frequency vortices appearing in the unstable shear layers---which were not seen at $Re=1000$. 
 
\subsection{Time-averaged pressure field}

\begin{figure*}
\begin{center}
	\subfloat[$Re$=1000, {\aoa}=$30^\circ$]{ \includegraphics[width=0.4\textwidth]{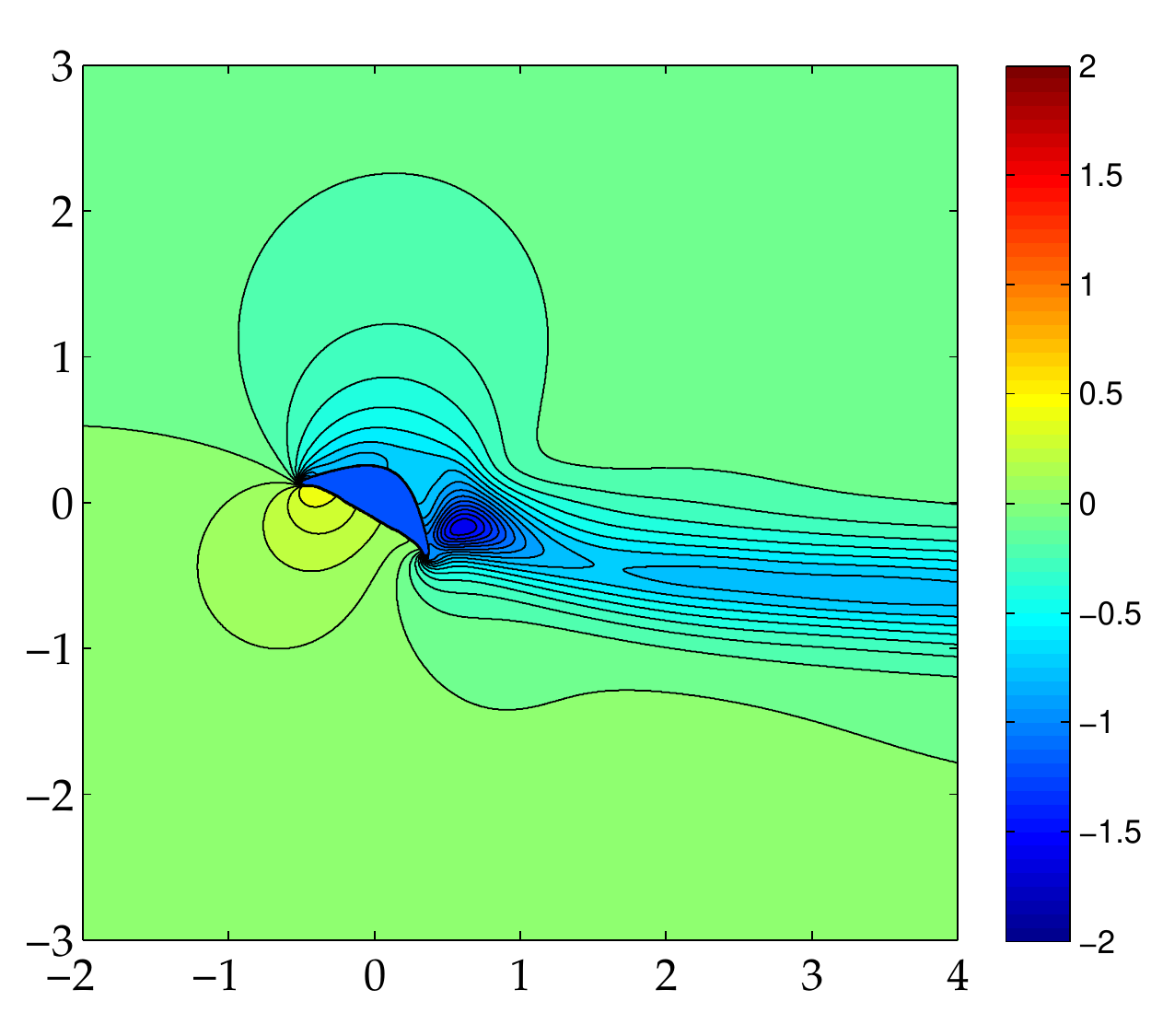} } \qquad \qquad
	\subfloat[$Re$=2000, {\aoa}=$30^\circ$]{ \includegraphics[width=0.4\textwidth]{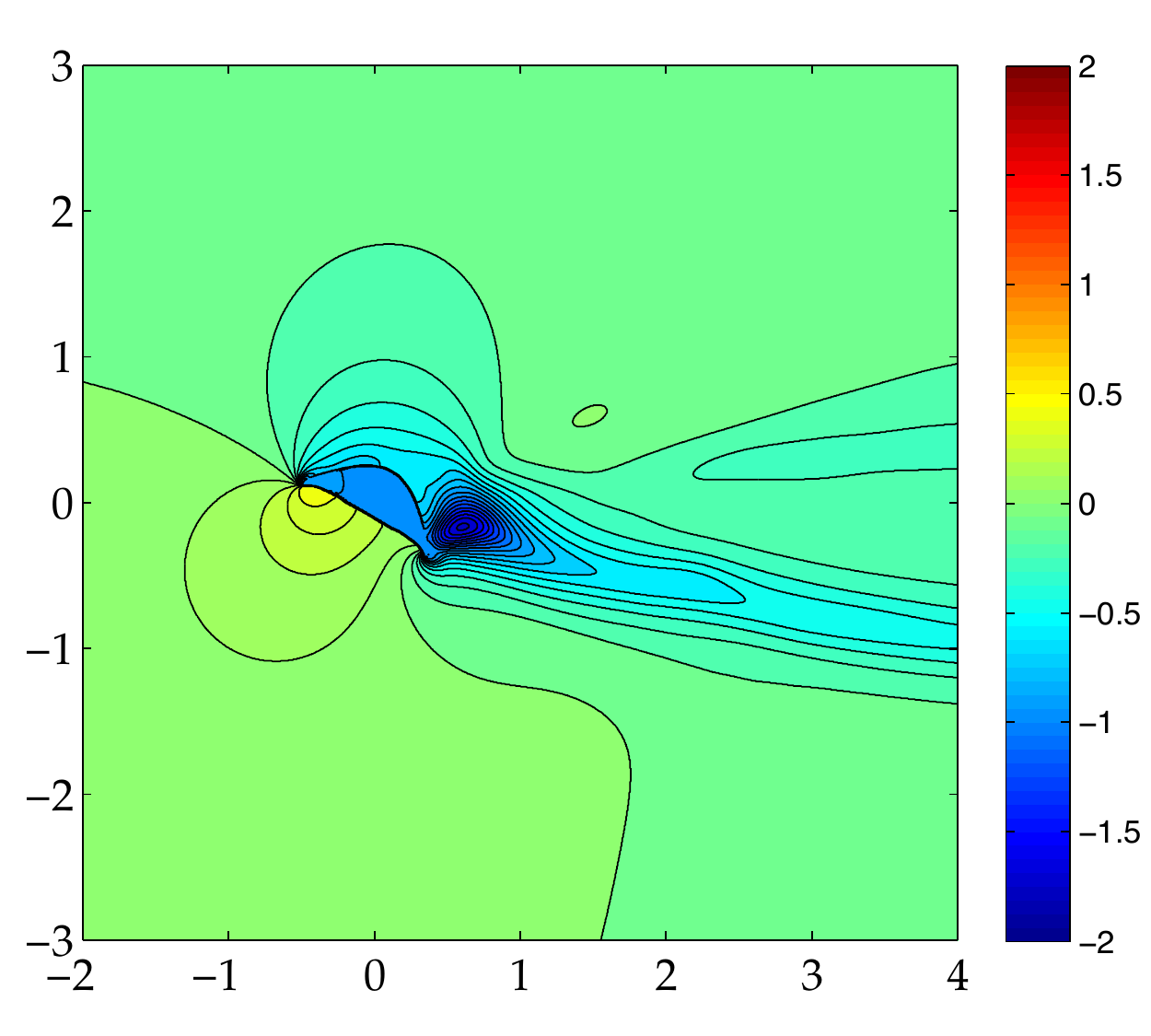} } \\
	\subfloat[$Re$=1000, {\aoa}=$35^\circ$]{ \includegraphics[width=0.4\textwidth]{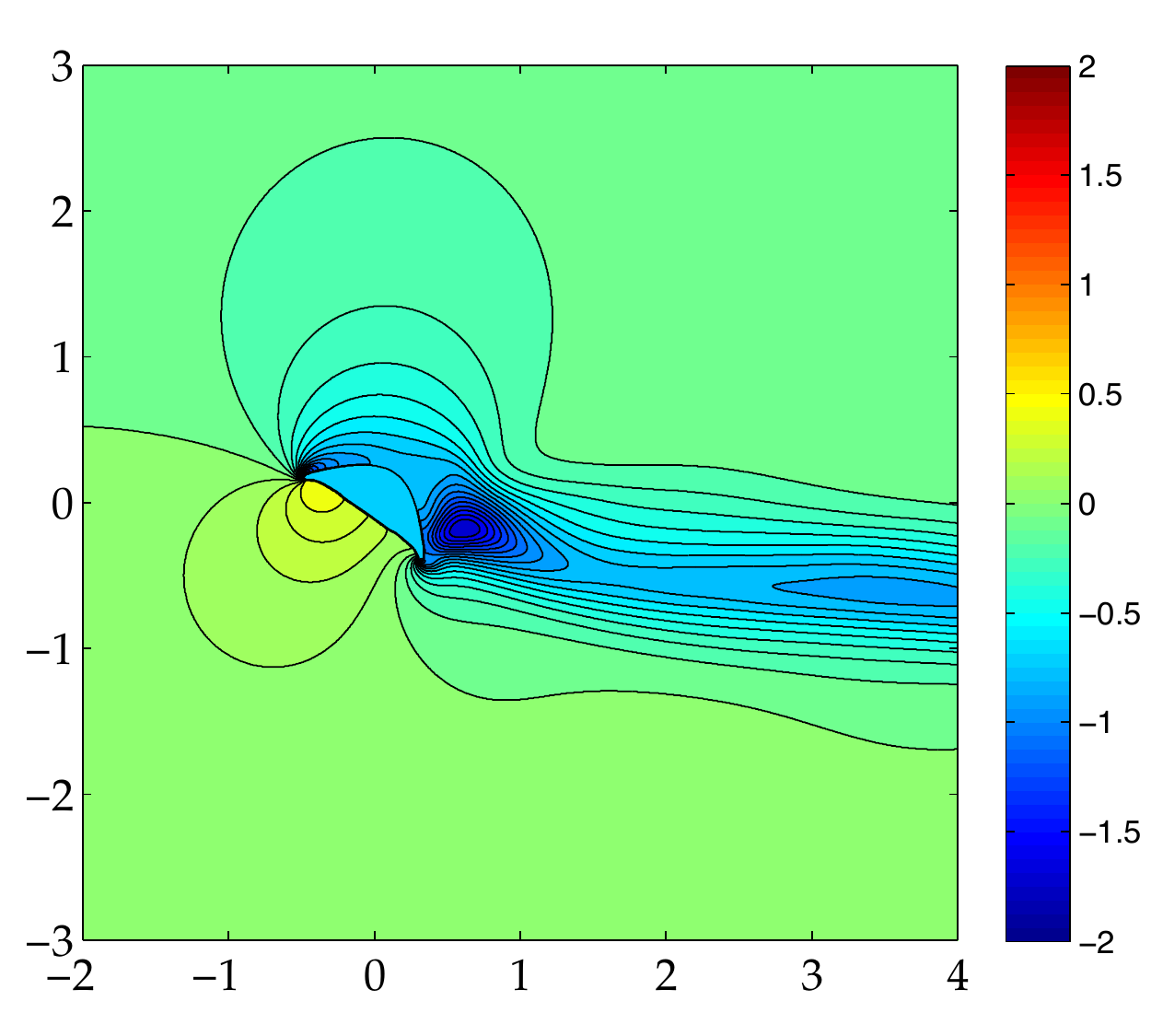} } \qquad \qquad
	\subfloat[$Re$=2000, {\aoa}=$35^\circ$]{ \includegraphics[width=0.4\textwidth]{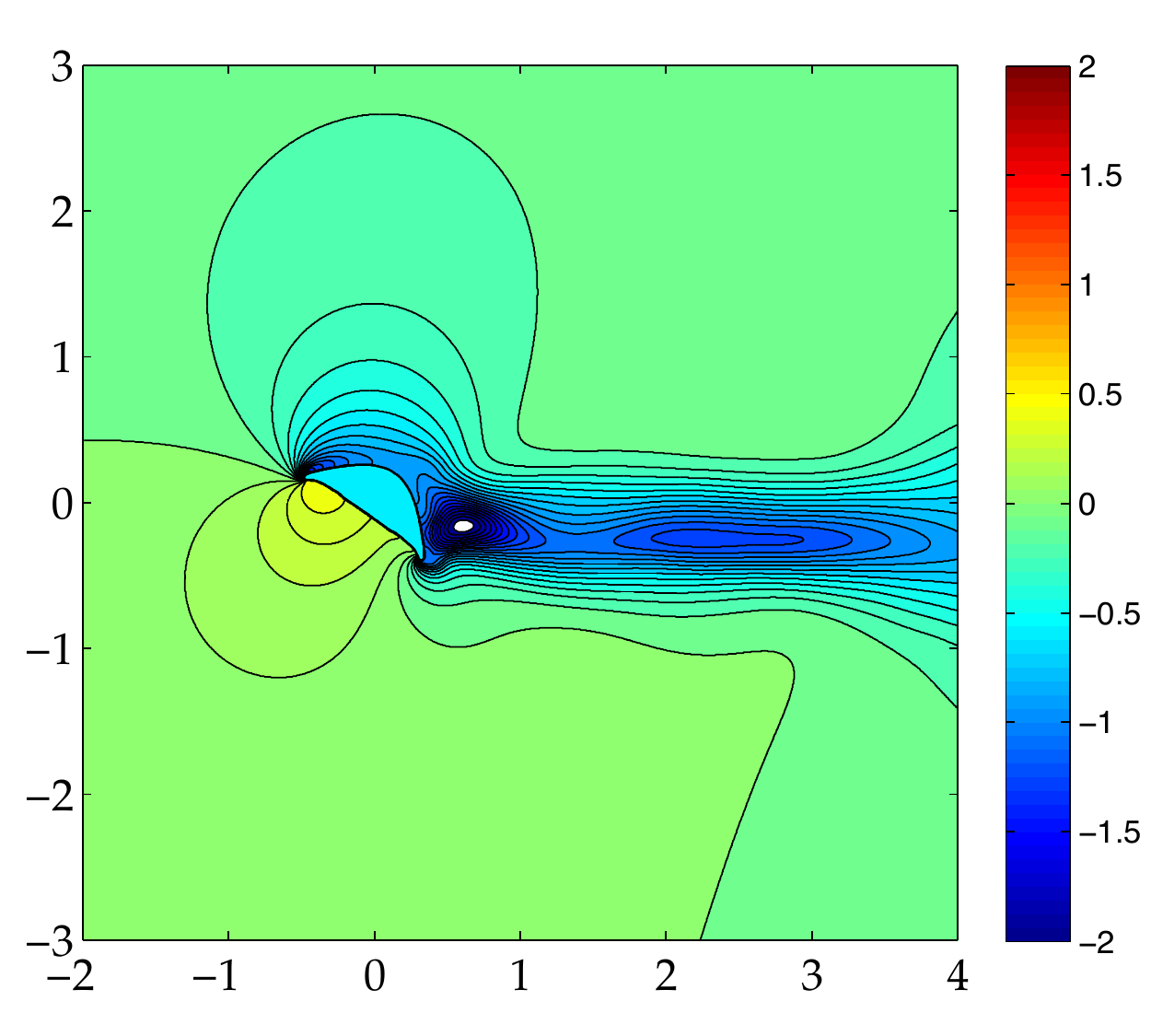} } \\
	\subfloat[$Re$=1000, {\aoa}=$40^\circ$]{ \includegraphics[width=0.4\textwidth]{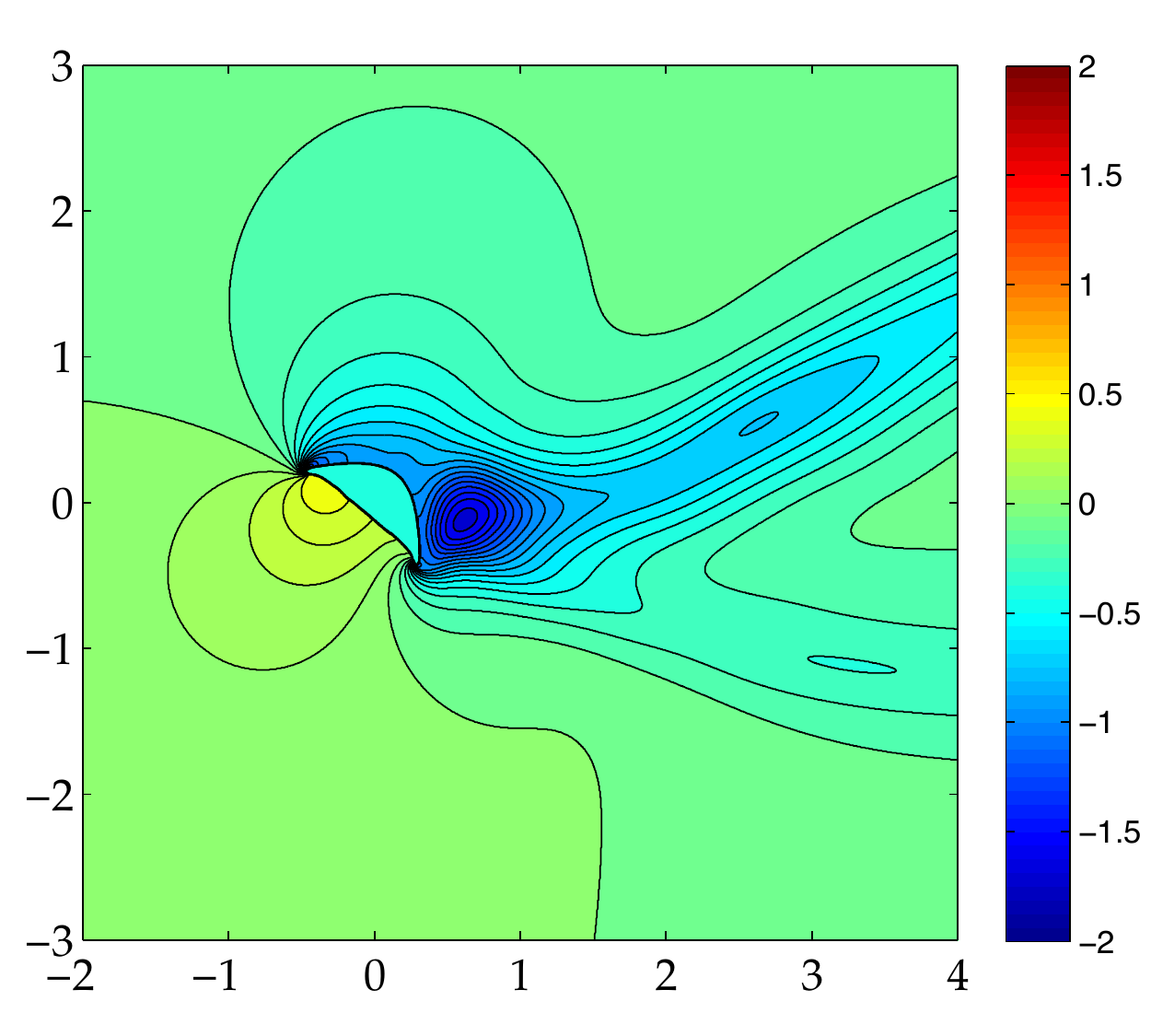} } \qquad \qquad
	\subfloat[$Re$=2000, {\aoa}=$40^\circ$]{ \includegraphics[width=0.4\textwidth]{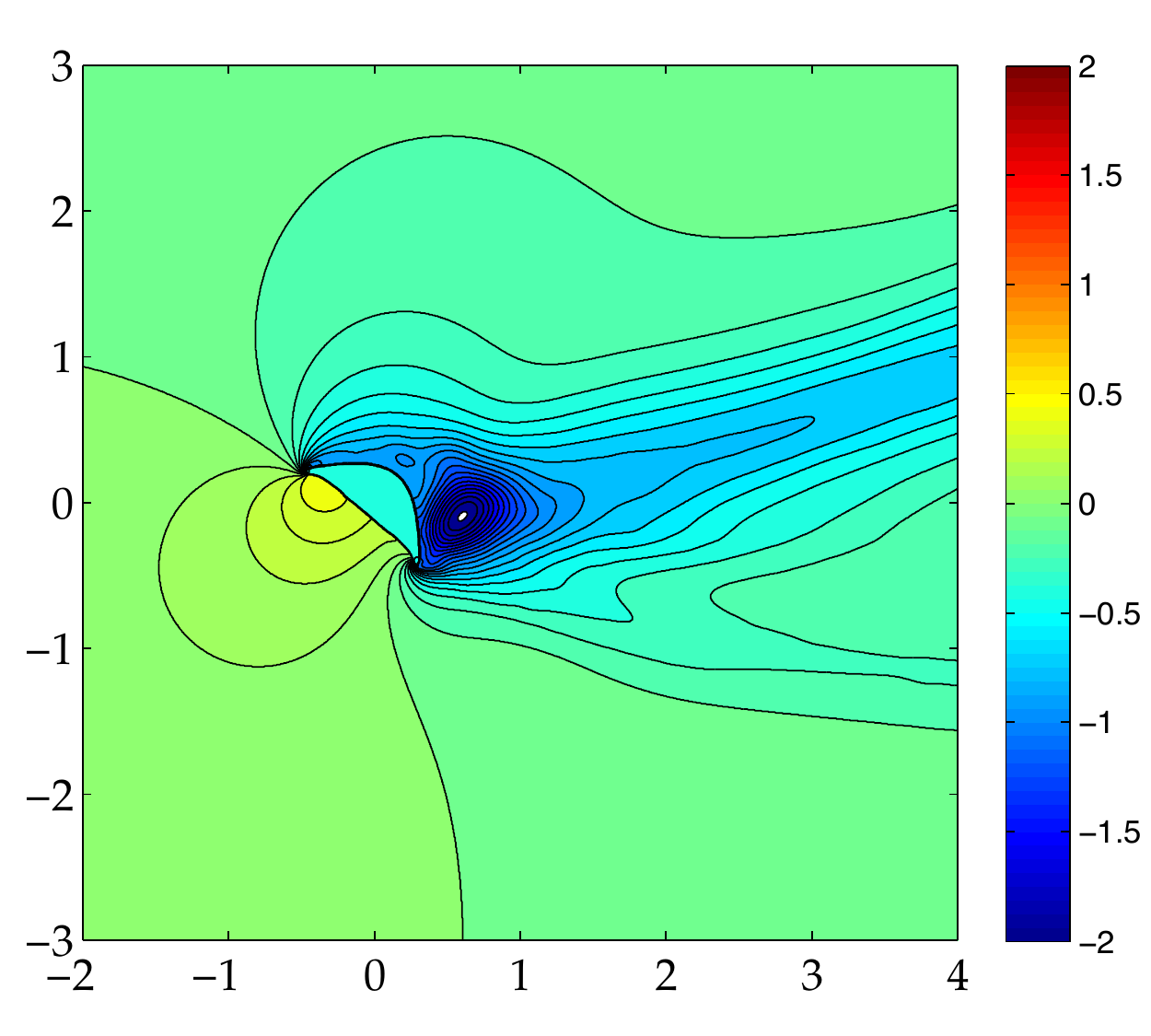} }
\caption{\small Average pressure field of flow over the snake profile at $Re=1000$ and $2000$ for angles of attack $30^\circ$, $35^\circ$ and $40^\circ$. The average was calculated by taking the mean of the pressure fields at 125 equally spaced time frames in the period $t^*=32$--$64$.}
\label{fig:averagePressure}
\end{center}
\end{figure*}

The visualization of time-averaged pressure fields is another tool for characterizing the vortex wakes. Regions of intense vorticity are seen as areas of lower pressure, and thus the main path of the von K{\'a}rm{\'a}n vortices becomes visible. Figure \ref{fig:averagePressure} shows the average pressure field for the flows at $Re= 1000$ and 2000. The slight downwards deflection of the flow at the cases with lower Reynolds number and lower \aoa\ is evident, and the wake at $Re=2000$ and \aoa$=35^{\circ}$ tracks a tight and straight path downstream.
The wakes at {\aoa} $40^\circ$ show two low-pressure tracks in the average pressure field: one corresponding to the large dipoles that are deflected upwards and the other corresponding to the smaller vortices deflected downwards. 

\subsection{Trajectories of vortex cores}

\begin{figure*}
\begin{center}
	\subfloat[$Re=1000$, {\aoa} $30^\circ$]{  
	\includegraphics[width=0.4\textwidth]{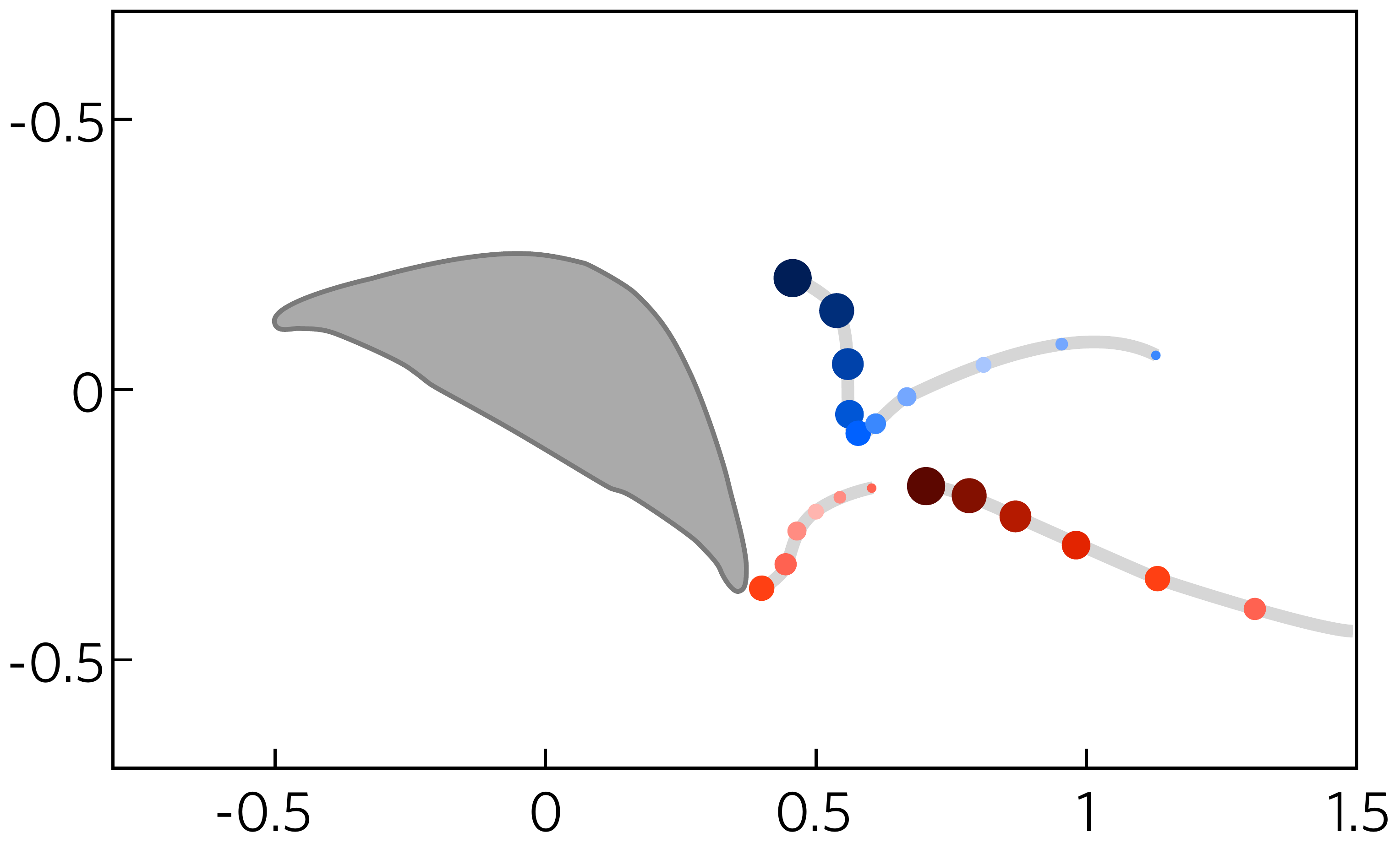}} \qquad
	\subfloat[$Re=1000$, {\aoa} $35^\circ$]{ 
	\includegraphics[width=0.4\textwidth]{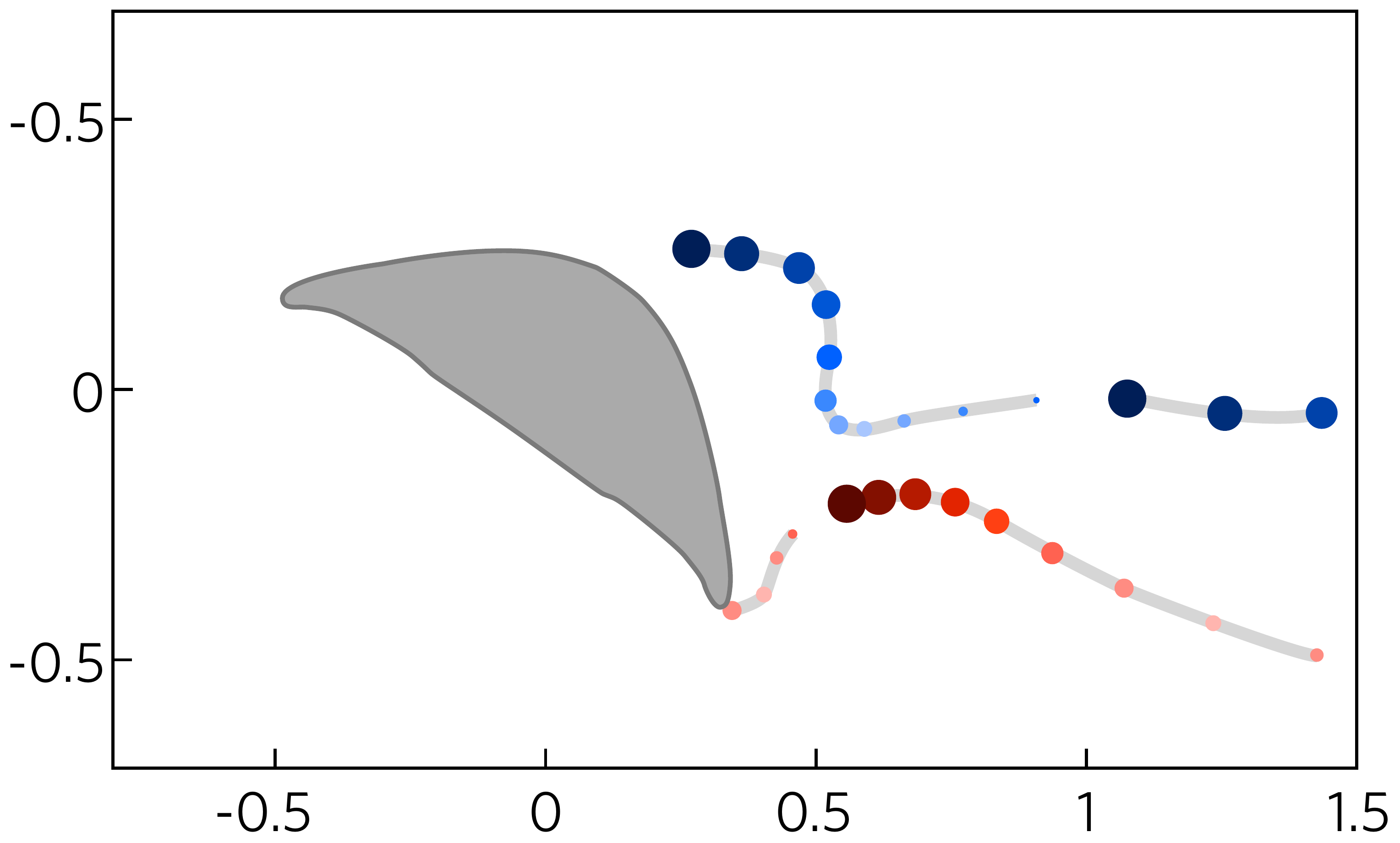}} \\ 
	\subfloat[$Re=2000$, {\aoa} $30^\circ$]{ 
	\includegraphics[width=0.4\textwidth]{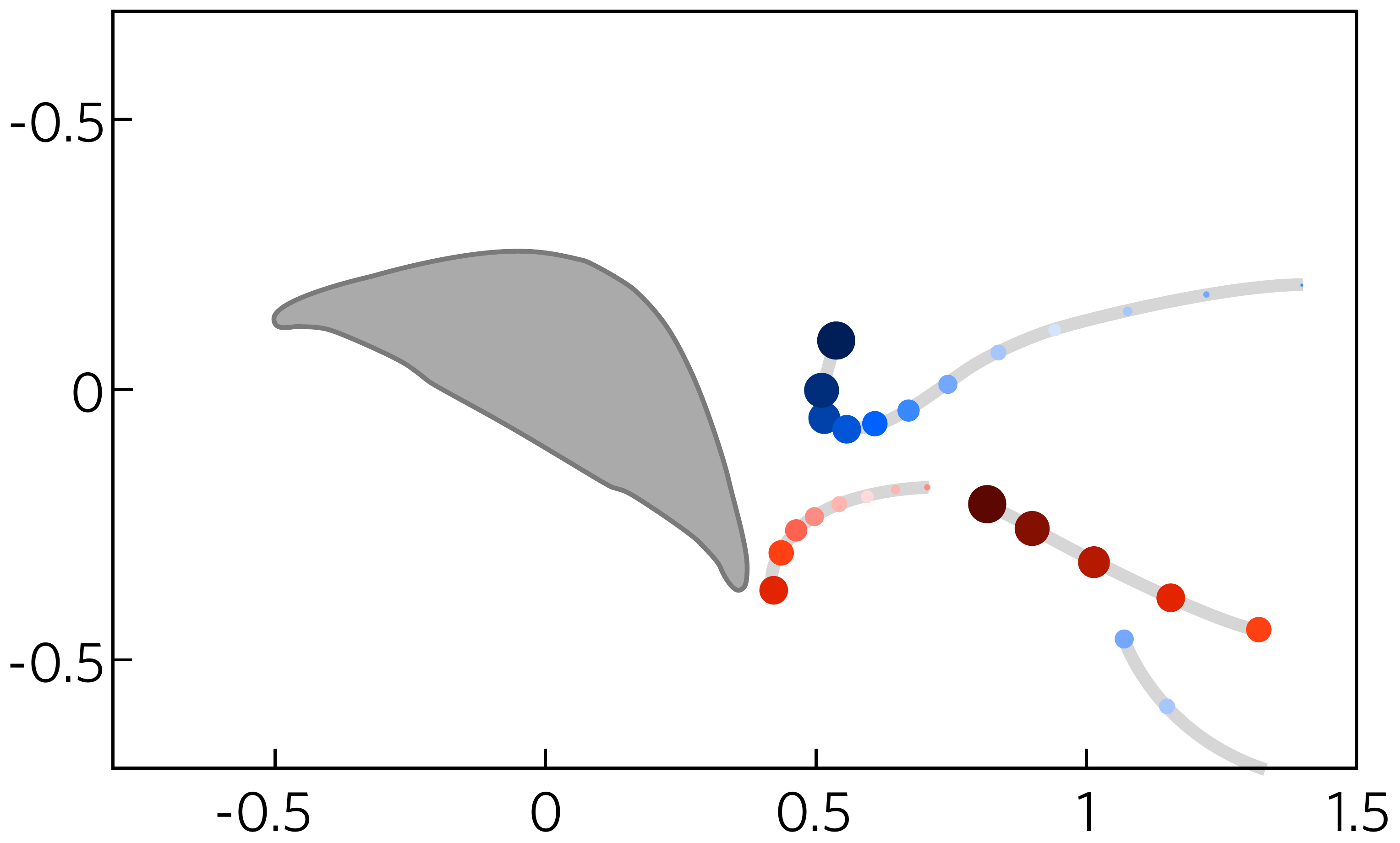} \label{subfig:trajectories2k30}} \qquad
	\subfloat[$Re=2000$, {\aoa} $35^\circ$]{ 
	\includegraphics[width=0.4\textwidth]{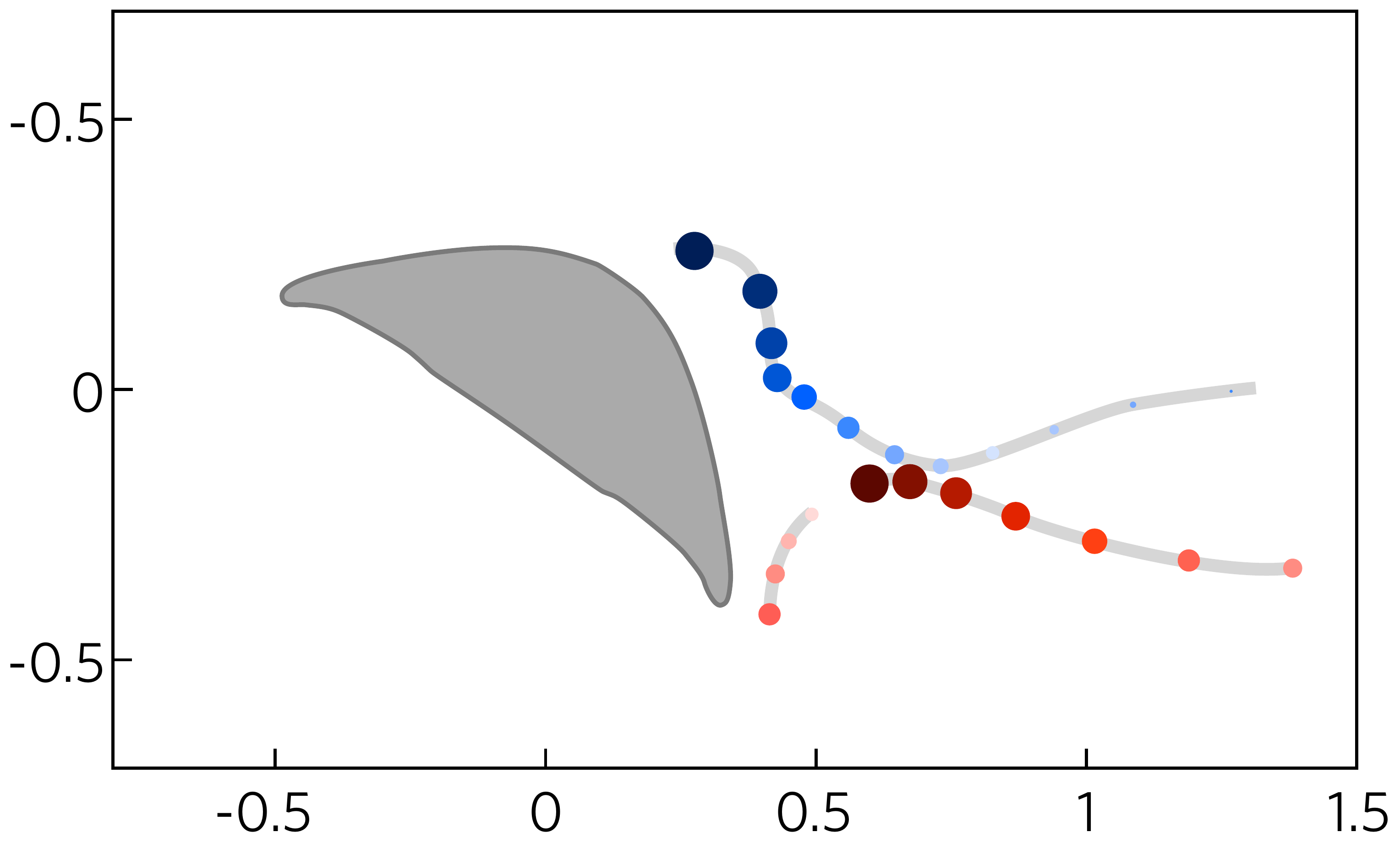}}
\caption{\small Pathlines of the centers of the coherent vortices in the wake. Clockwise vortices have been represented by blue dots and counter-clockwise vortices by red dots. Points of the same darkness and size represent the positions of the vortices at the same instant in time. Consecutive points are separated by the same periods of time (0.256 units). More than one point with the same color indicates the presence of more than one vortex at that instant. Note that the dorsal vortex stays closer to the body for a longer period of time at {\aoa} $35^\circ$ and $Re=2000$.}
\label{fig:trajectories}
\end{center}
\end{figure*}

The paths of vortices in the flow can provide a picture of how the vortices interact with the body and each other to produce lift. Figure \ref{fig:trajectories} shows the trajectories of the centers of the vortices in the flow for angles of attack $30^\circ$ and $35^\circ$ at Reynolds numbers 1000 and 2000. At $Re=1000$, the paths traced by the vortices at the two angles of attack are similar to each other. But at $Re=2000$, there is a marked difference. At $\aoa=30^\circ$, the newly forming trailing-edge vortex stretches and splits the vortex formed on the dorsal side. The trajectories of the split vortices are seen by the two tracks with blue circles in Figure \ref{subfig:trajectories2k30}. At $\aoa=35^\circ$, we can see that the dorsal vortex is formed nearer to the fore of the cross-section and follows a trajectory that stays closer to the body compared to the case at $\aoa=30^\circ$. The dorsal vortex is also stronger and does not split when it interacts with the trailing-edge vortex.

\subsection{Averaged surface pressure}

\begin{figure*}
\begin{center}
	\subfloat[$Re=1000$]{ \includegraphics[width=0.4\textwidth]{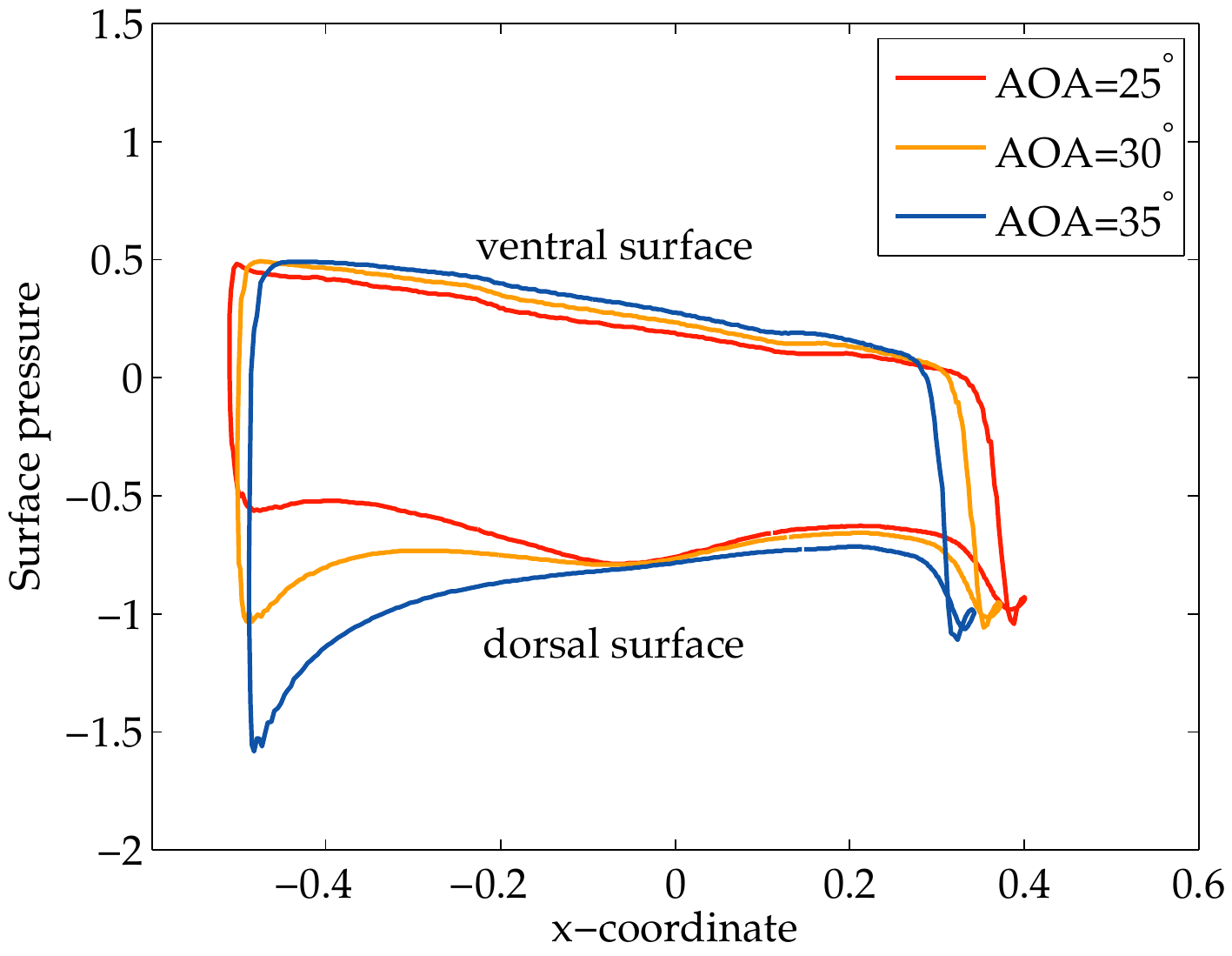} \label{subfig:surfP1k}}
	\subfloat[$Re=2000$]{ \includegraphics[width=0.4\textwidth]{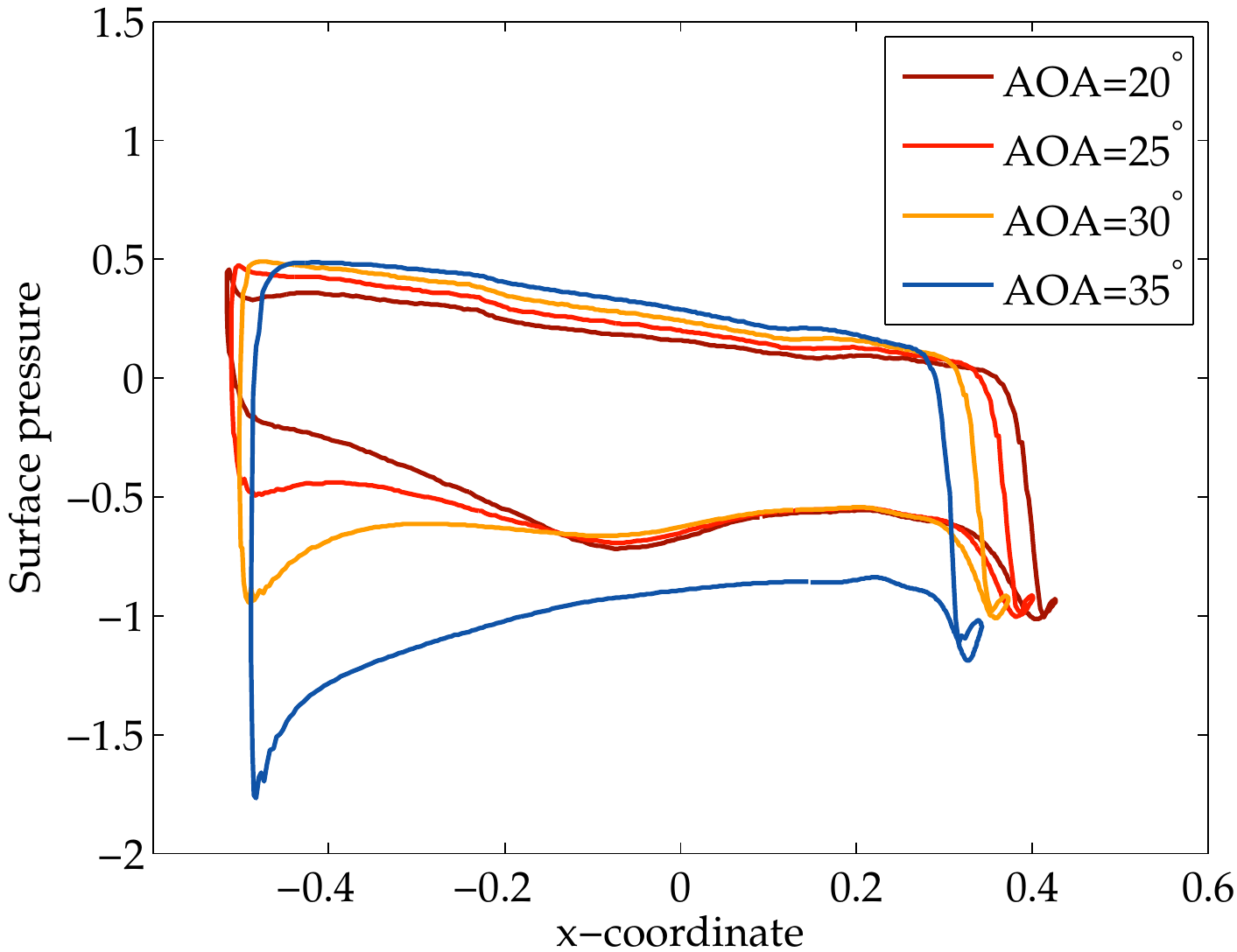} \label{subfig:surfP2k}}
\caption{\small Time-averaged surface pressure distributions on the snake profile at various angles of attack for flows with $Re=1000$ and $2000$. Pressure is plotted against the $x$-coordinate of the surface locations, so the area enclosed by each loop gives the pressure contribution to the lift force acting on the body. The top portion of the graph represents the pressure distribution on the ventral surface and the bottom portion of the graphs represent the pressure distributions on the dorsal surface (higher pressure must exist on the ventral surface for lift to be generated). See Figure 13 for orientation of the pressure fields relative to the cross-section. Data sets, plotting scripts and figures available online under CC-BY.\cite{KrishnanETal-share705890} }
\label{fig:surfPres}
\end{center}
\end{figure*}

The majority of the force experienced by a bluff body in a fluid flow is due to differences in the pressure along the surface of the body. The time-averaged surface pressure distribution is therefore directly related to the average lift and drag experienced by the body. Figure \ref{fig:surfPres} shows the surface distribution of the average pressure acting on the body at $Re=1000$ and $2000$ and at different angles of attacks before the stall regime. These have been plotted against the $x$-coordinate of the corresponding surface points so that the area under the graph gives the pressure contribution to the lift on the body. As expected, the leading-edge suction increases with an increase in angle of attack, associated with greater lift production.  But in addition, there is an extended region of low pressure over the rear part of the dorsal surface at {\aoa} $35^\circ$ (Figure \ref{subfig:surfP2k}). This is observed only at $Re=2000$, and the sudden pressure drop in this region is associated with the spike in the lift coefficient at this angle of attack.

\subsection{Swirling strength in the wake}

\begin{figure*}[p]
\begin{center}
	\subfloat[Time of $C_{L, max}$, ${\aoa}=20^\circ$]{ \includegraphics[width=0.41\textwidth]{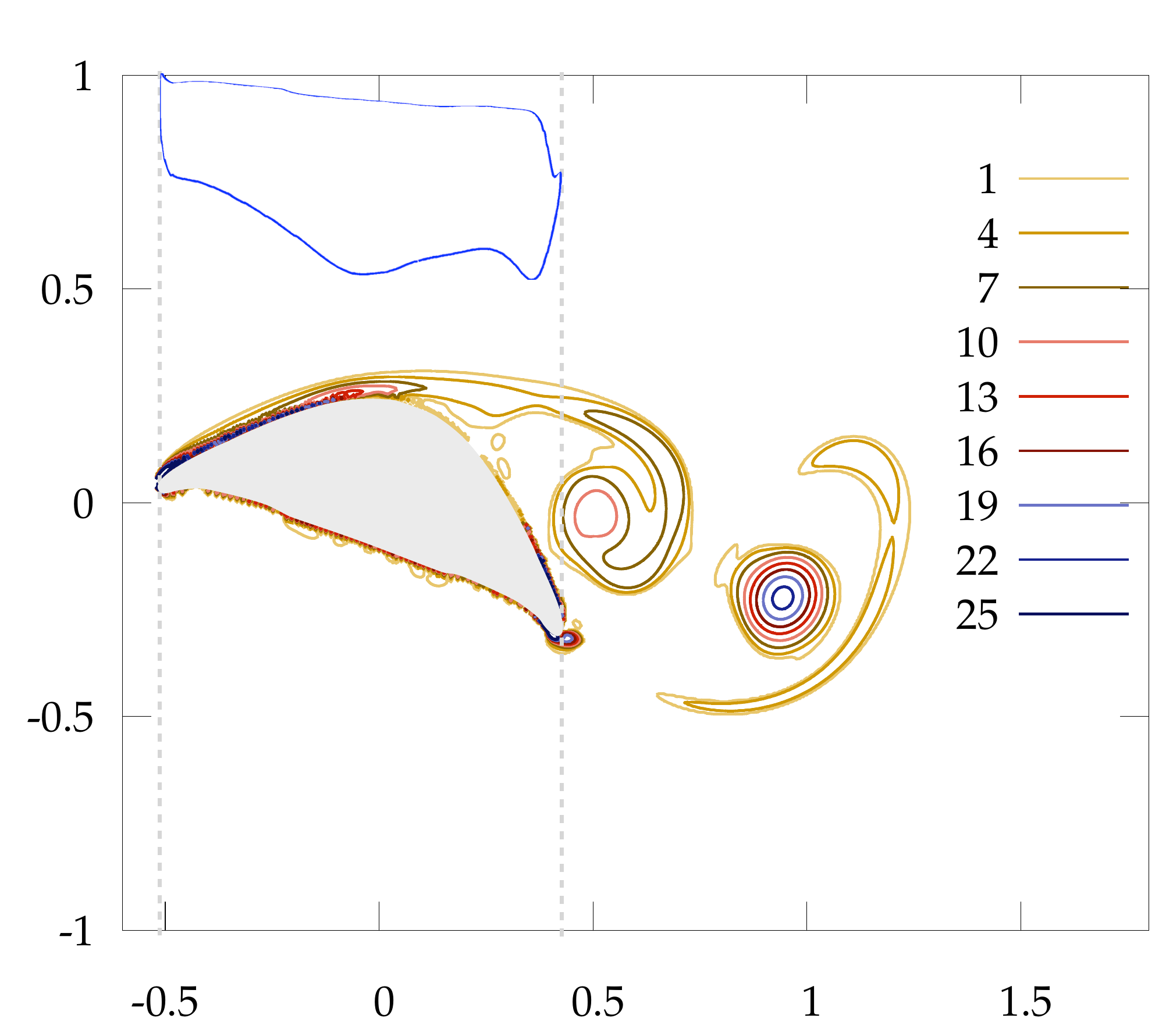} \label{subfig:q2k20}}
	\subfloat[Time of $C_{L, max}$, ${\aoa}=25^\circ$]{ \includegraphics[width=0.4\textwidth]{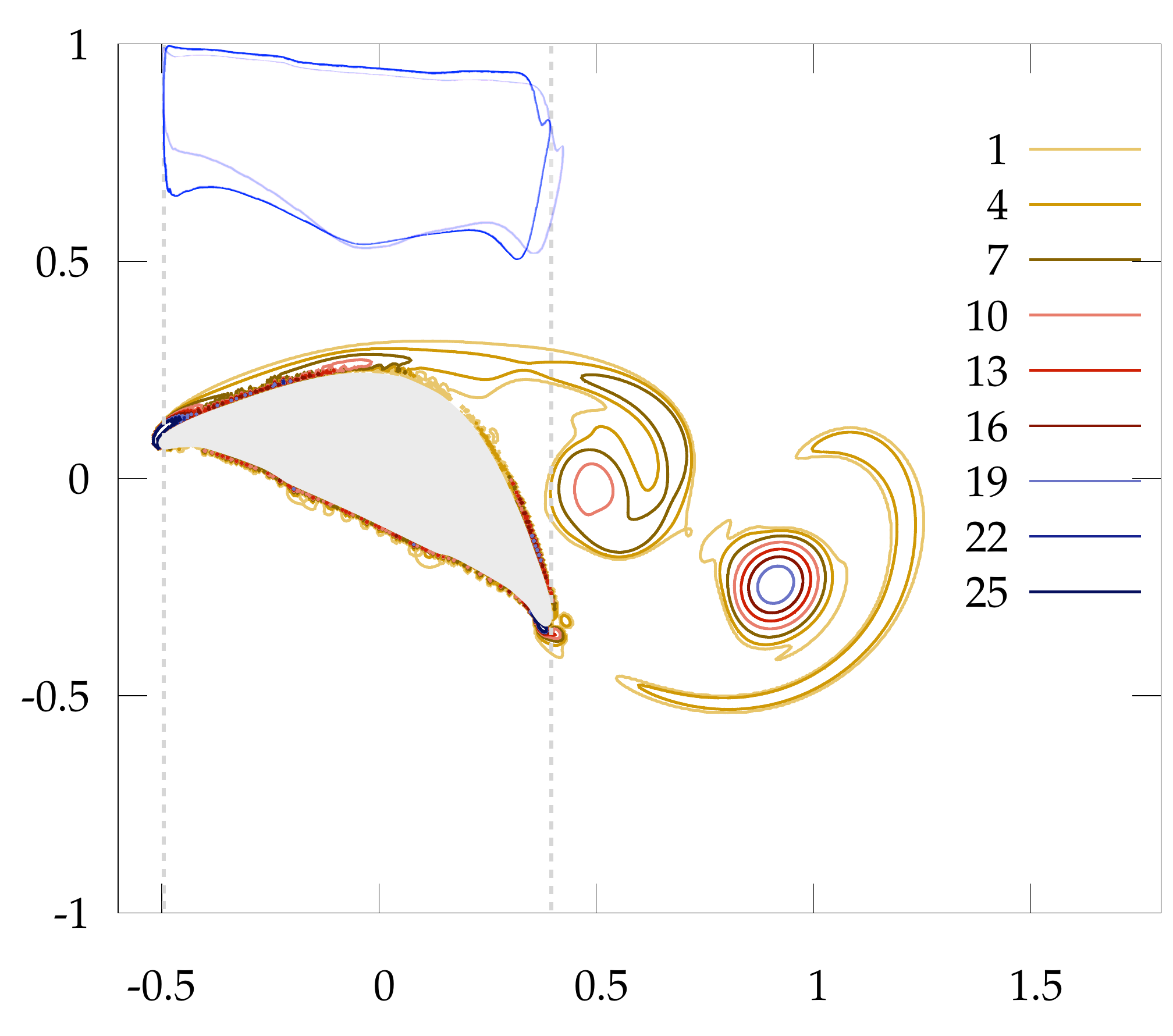} \label{subfig:q2k25}}\\
	\subfloat[Time of $C_{L, max}$, ${\aoa}=30^\circ$]{ \includegraphics[width=0.4\textwidth]{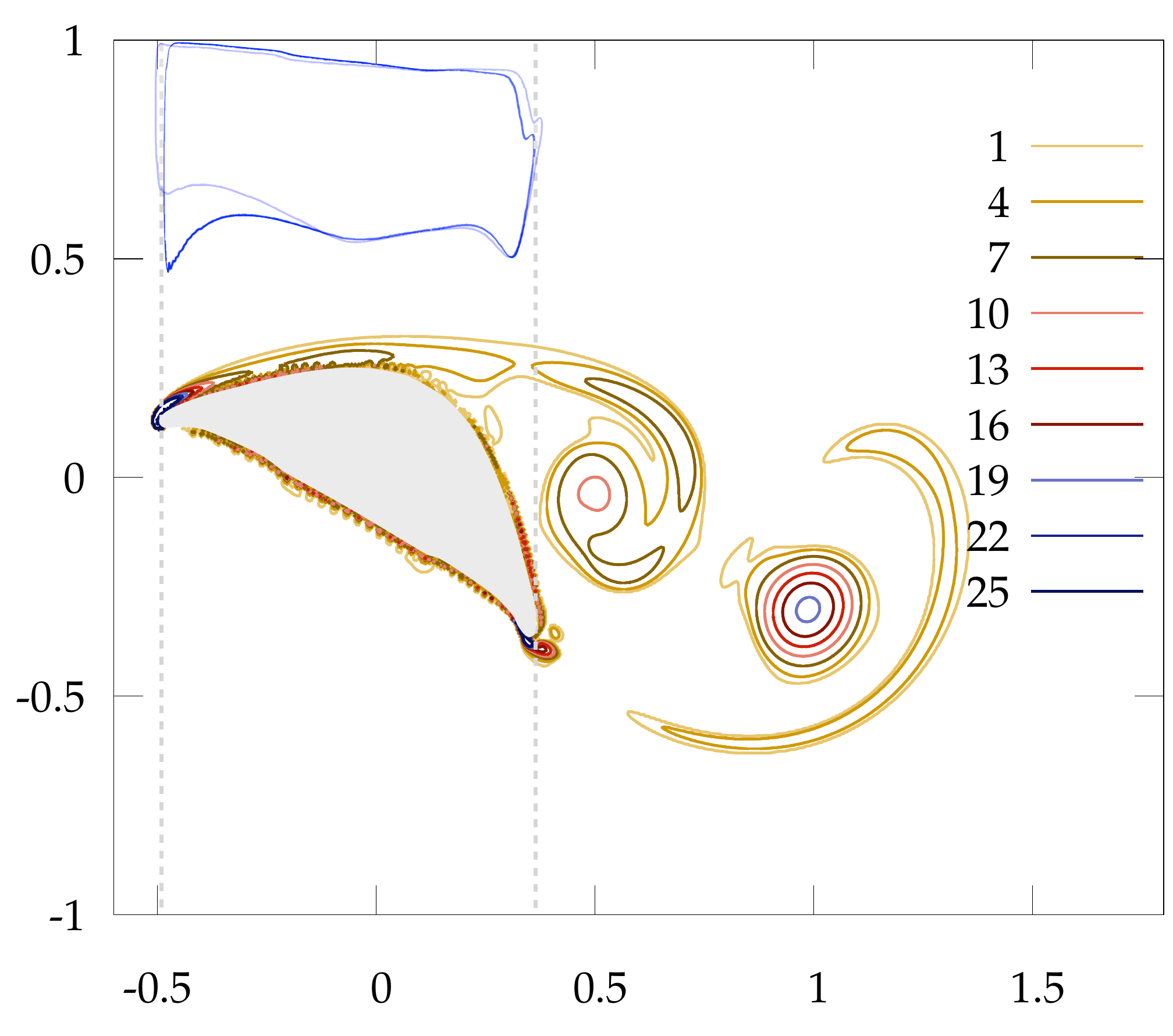} \label{subfig:q2k30}}
	\subfloat[Time of $C_{L, max}$, ${\aoa}=35^\circ$]{ \includegraphics[width=0.4\textwidth]{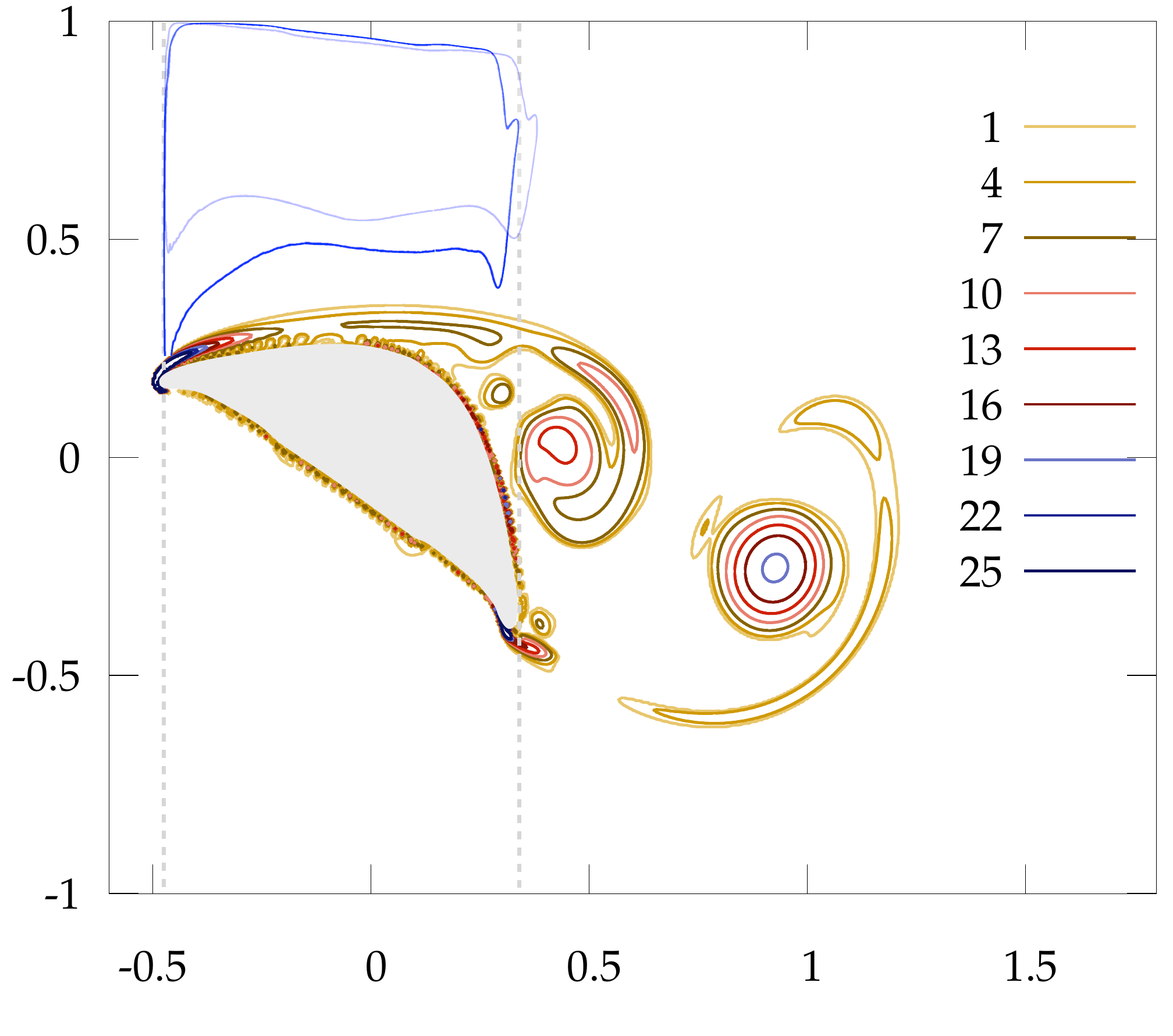} \label{subfig:q2k35}}\\
	\subfloat[Time of $C_{L, min}$, ${\aoa}=30^\circ$]{ \includegraphics[width=0.4\textwidth]{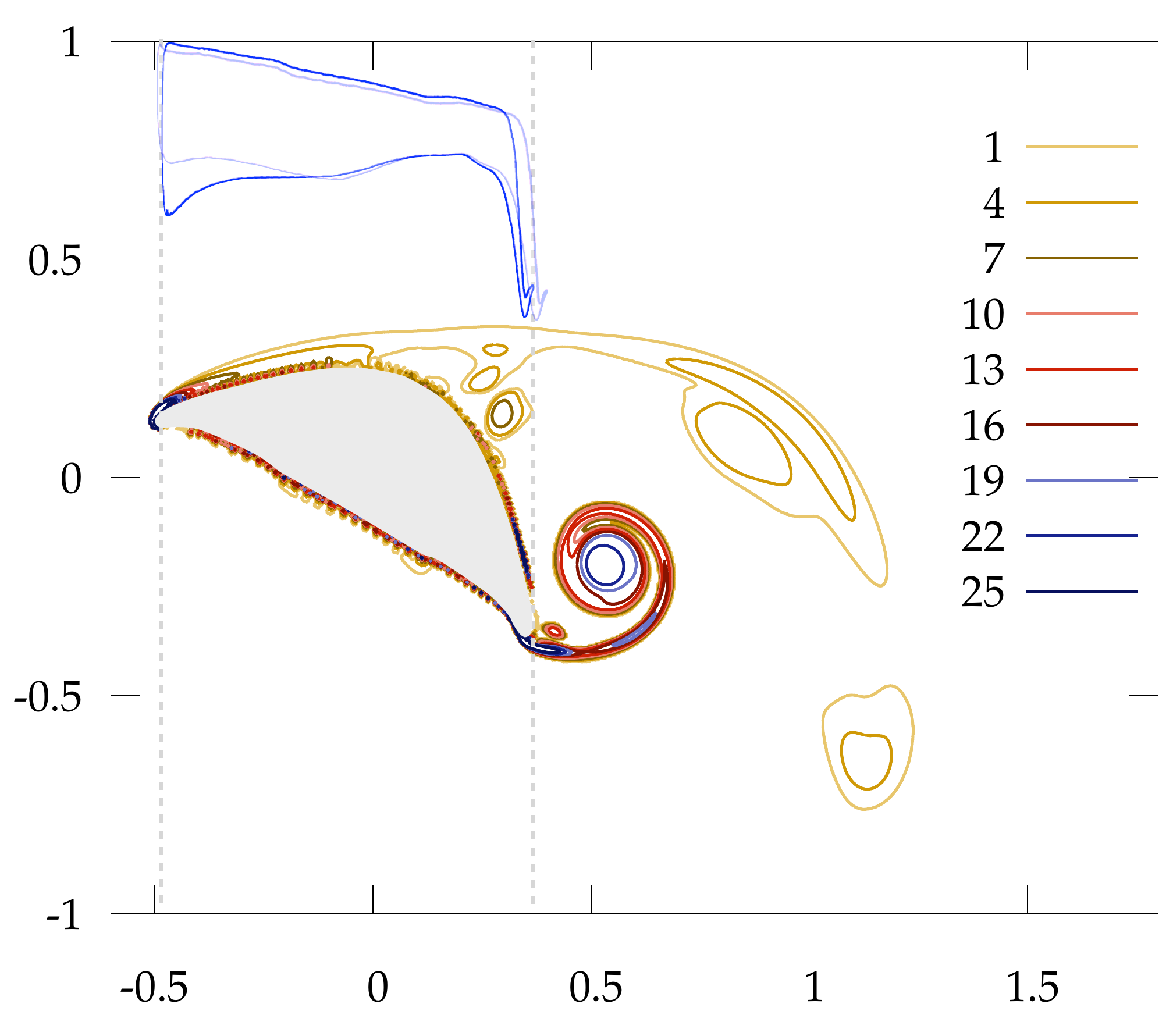} \label{subfig:q2k30m}}
	\subfloat[Time of $C_{L, min}$, ${\aoa}=35^\circ$]{ \includegraphics[width=0.4\textwidth]{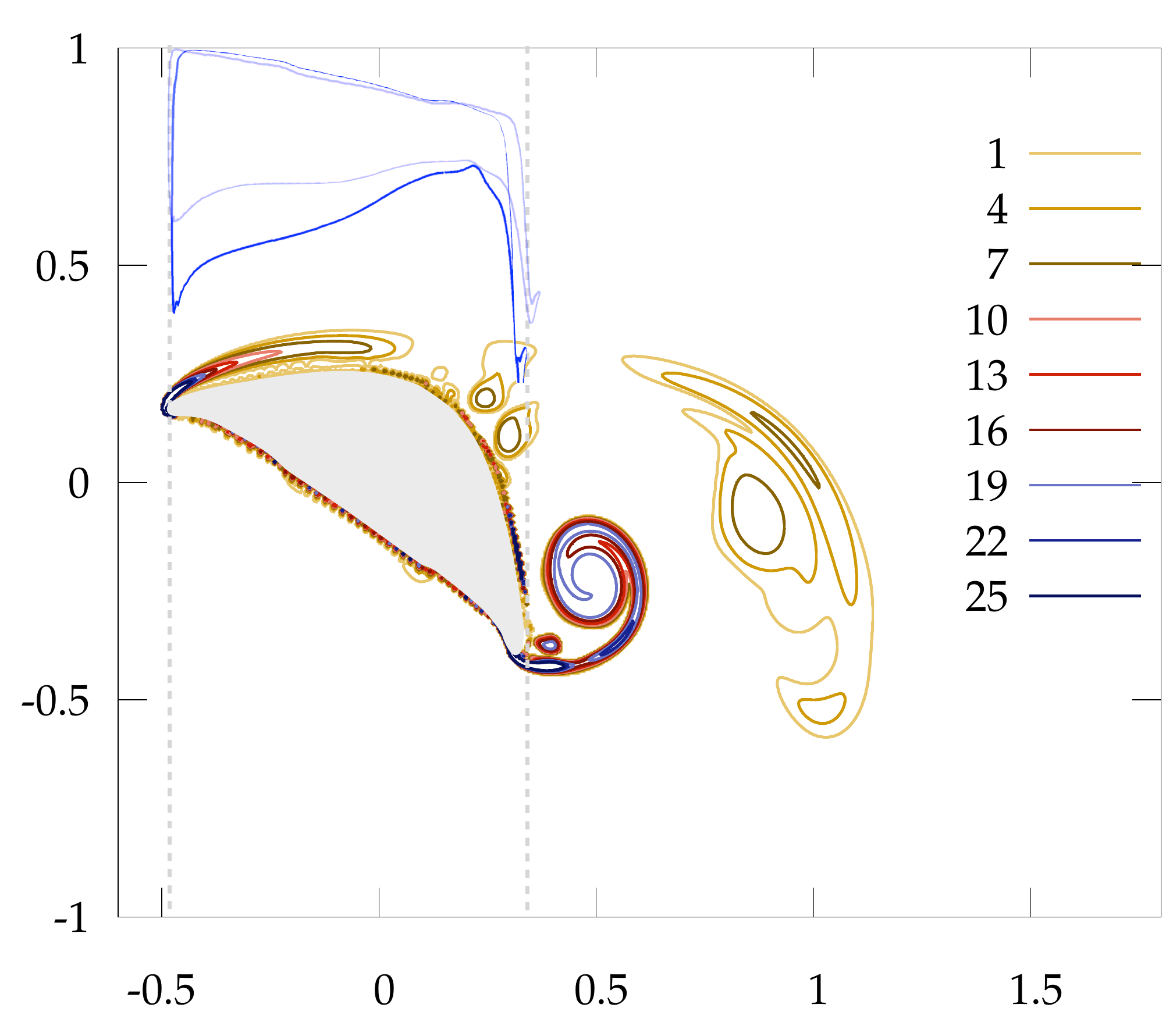} \label{subfig:q2k35m}}
\caption{\small $Re=2000$: Contours of the swirling strength of the flow at times of maximum (a--d) and minimum (e--f) lift, for various angles of attack. The maximum value of swirling strength in the dorsal vortex at the point of maximum lift is 10.7 at ${\aoa}=30^\circ$, and 14.3 at ${\aoa}=35^\circ$. The corresponding pressure distribution along the surface is also shown, along with the distribution at the previous angle of attack for comparison.}
\label{fig:qValues}
\end{center}
\end{figure*}

The swirling strength of the flow field identifies the regions of the flow that contain coherent vortical structures, as opposed to regions of vorticity that are dominated by shear.\cite{Zhou1999}
Figure \ref{fig:qValues} shows the contours of swirling strength  at $Re=2000$, at instants when the unsteady lift coefficient is maximum (first four frames) and minimum (last two frames).
The corresponding surface pressure distribution at these instants is also shown in each frame. 
Note that only the leading-edge suction changes up to {\aoa} $30^\circ$. But at ${\aoa}=35^\circ$, there is a decrease in the pressure along the whole dorsal surface, which is associated with a marked increase in the lift coefficient. 
At the instants of both maximum and minimum unsteady lift, strong vortices decrease pressure on a large portion of the dorsal surface. These features contribute to the observed enhanced time-averaged lift at ${\aoa}=35^\circ$.

\subsection{Near-body vorticity}

\begin{figure}
\begin{minipage}{0.47\linewidth}
\centering
	\subfloat[$t^*=44.544$]{ 
	\includegraphics[width=\textwidth]{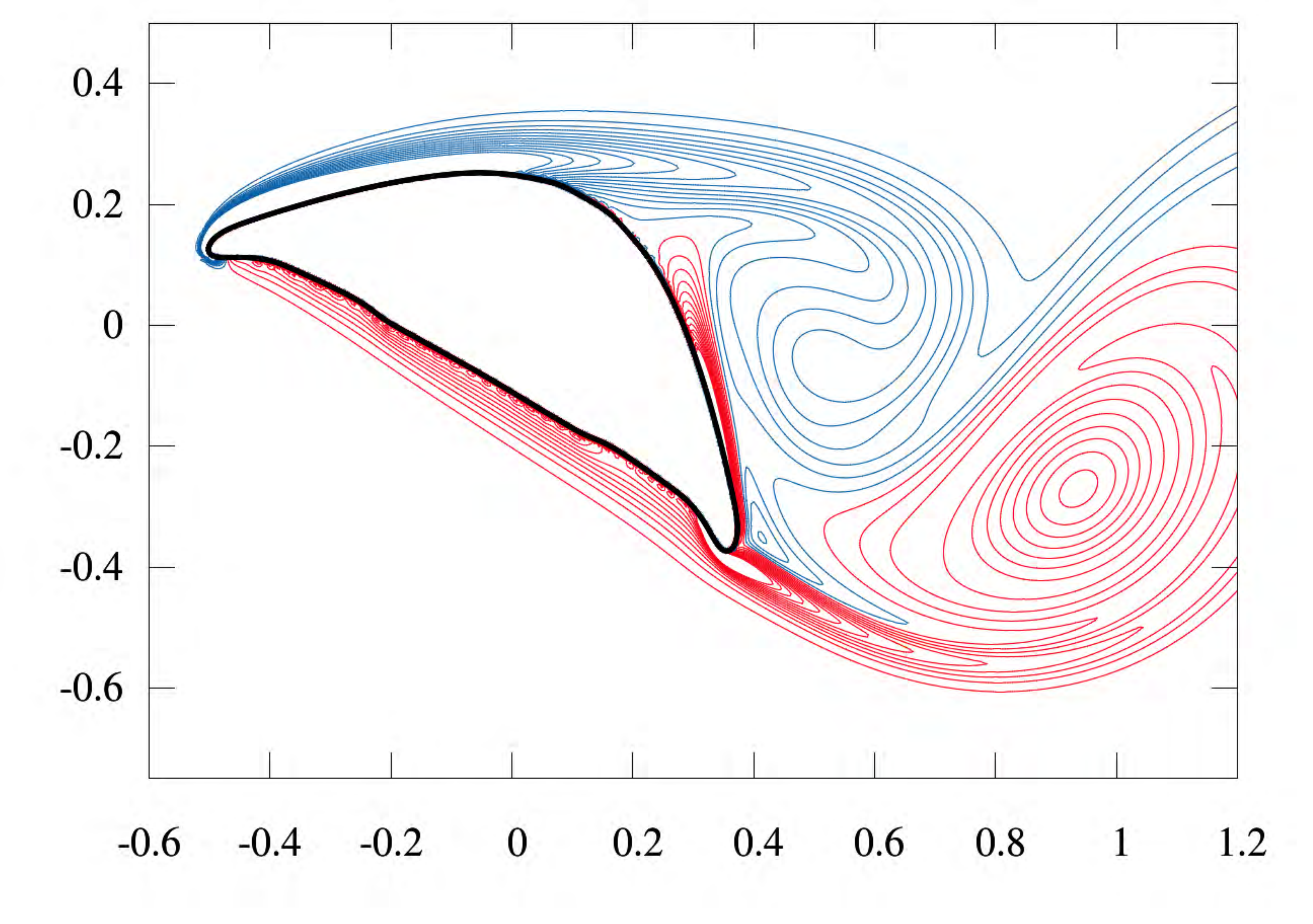} \label{subfig:vc1k30a}
	}\\
	\subfloat[$t^*=45.568$]{ 
	\includegraphics[width=\textwidth]{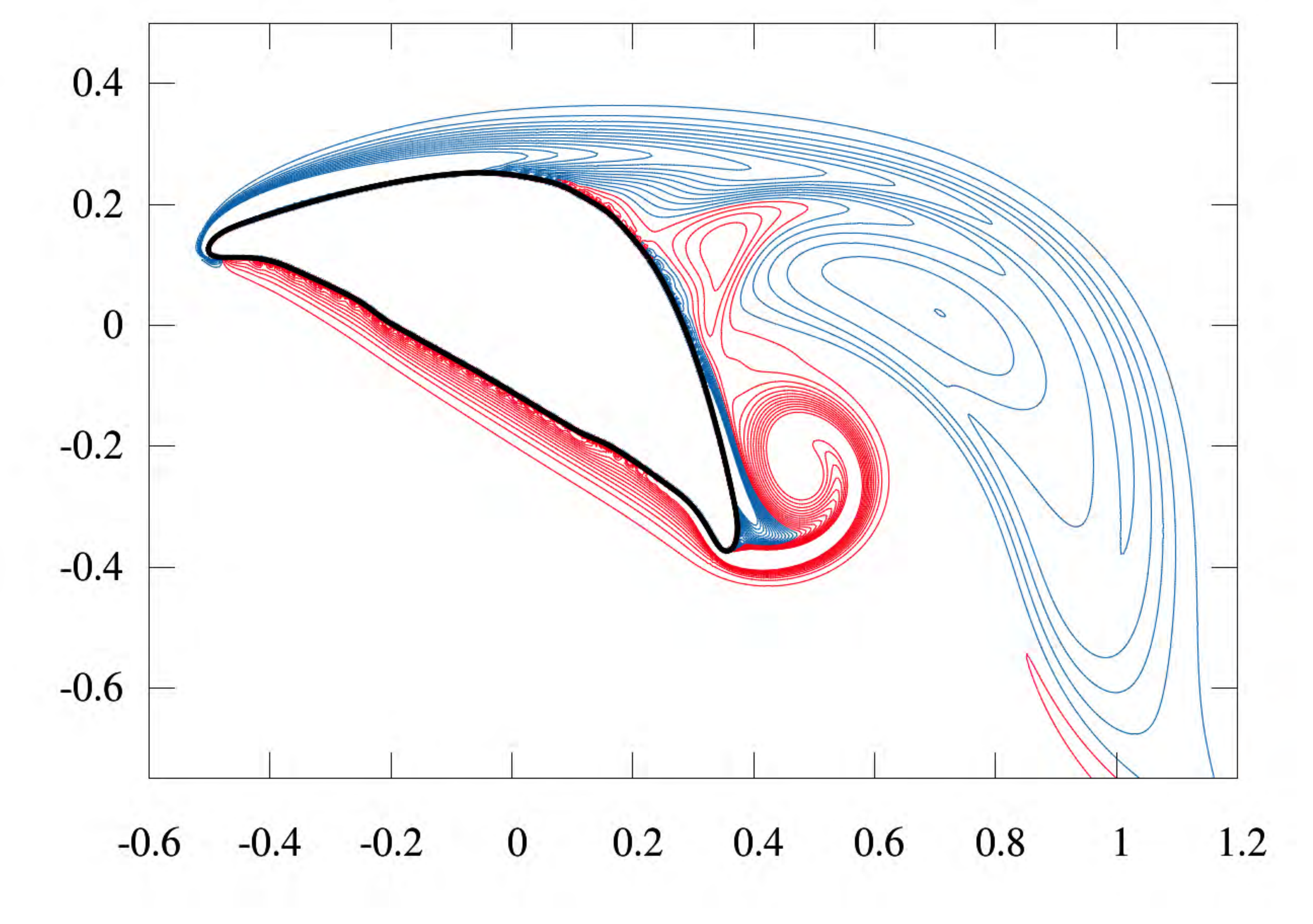} \label{subfig:vc1k30b}
	}\\
	\subfloat[$t^*=46.592$]{ 
	\includegraphics[width=\textwidth]{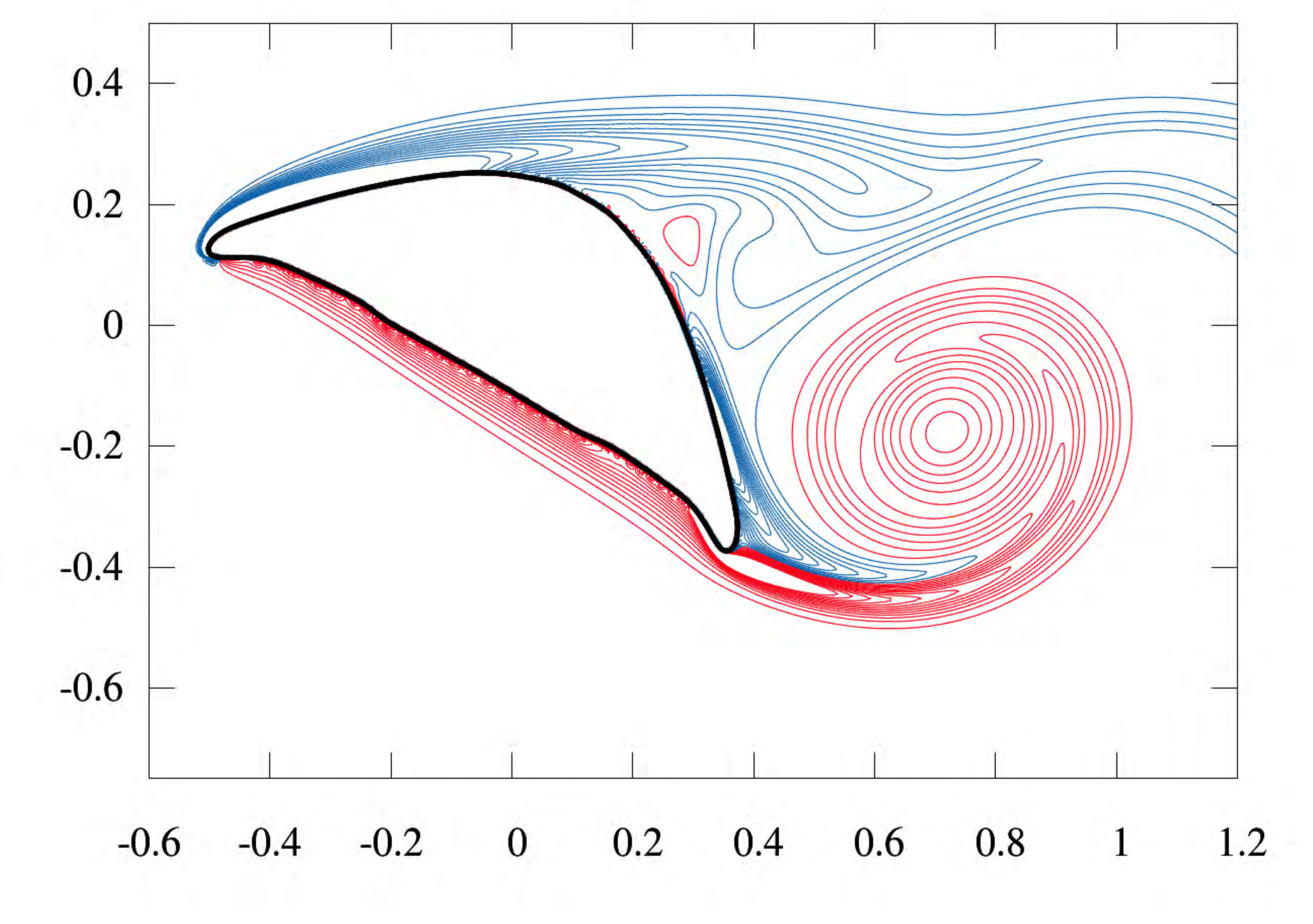} \label{subfig:vc1k30c}\
	}
\caption[justification=justified,singlelinecheck=false]{\small $Re=1000$, {\aoa} $30^\circ$: Vorticity contours from -25 to +25 in steps of 2. Contour lines of negative values of vorticity are blue and the contours of positive values are red.}
\label{fig:vConts1k30}
\end{minipage}
\qquad
\begin{minipage}{0.47\linewidth}
\centering
	\subfloat[$t^*=44.288$]{ 
	\includegraphics[width=\textwidth]{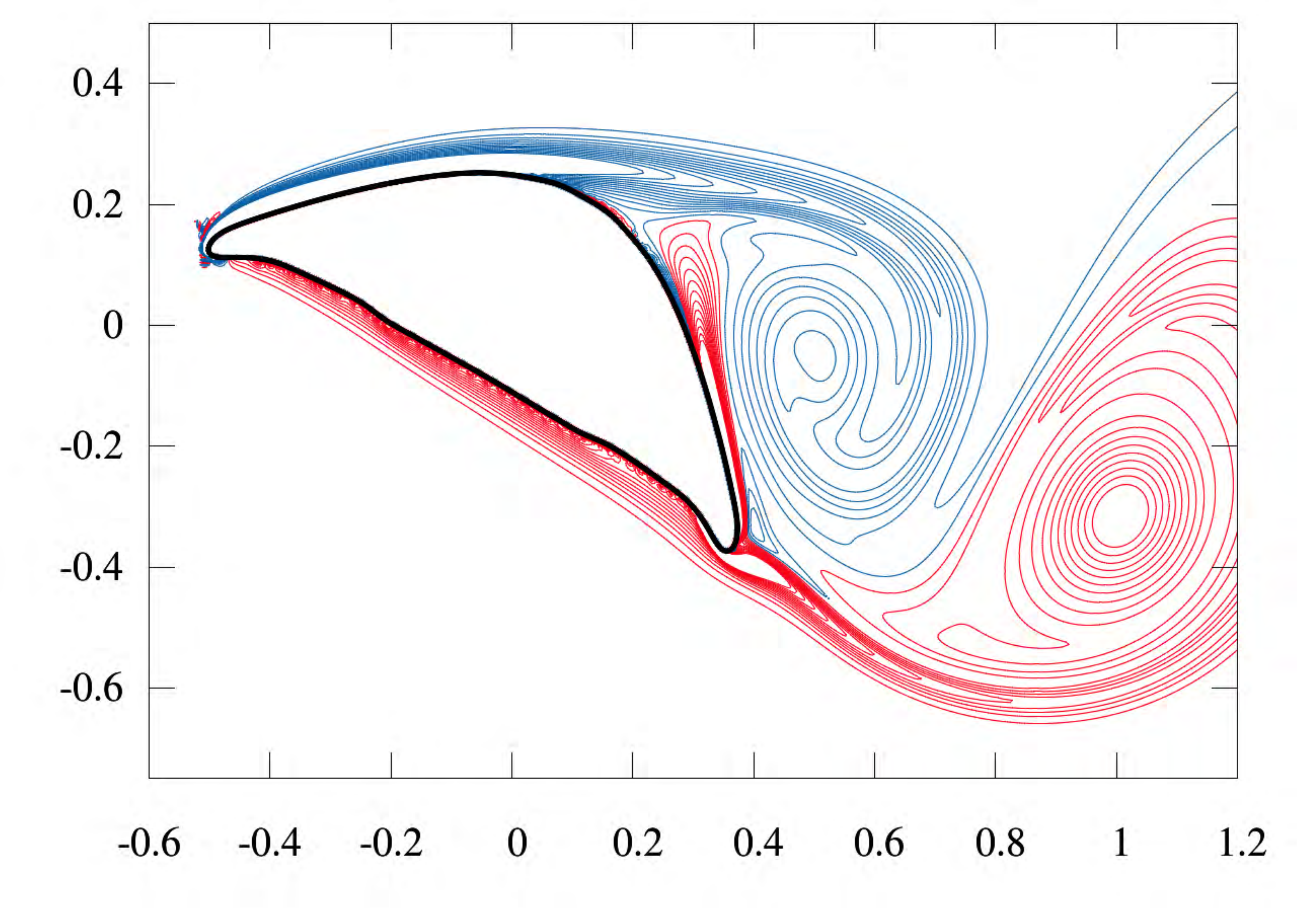} \label{subfig:vc2k30a}
	}\\
	\subfloat[$t^*=45.312$]{ 
	\includegraphics[width=\textwidth]{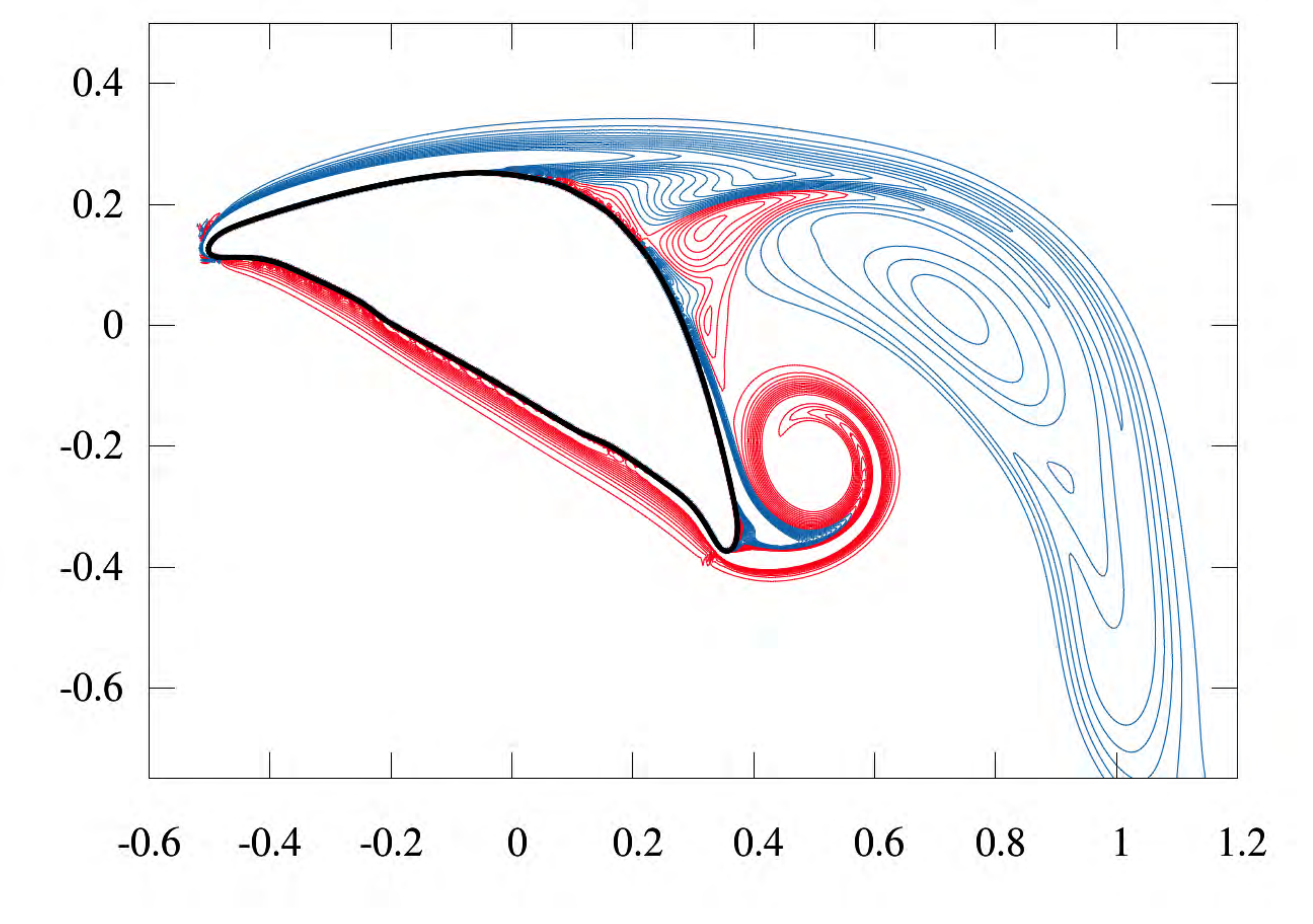} \label{subfig:vc2k30b}
	}\\
	\subfloat[$t^*=46.336$]{ 
	\includegraphics[width=\textwidth]{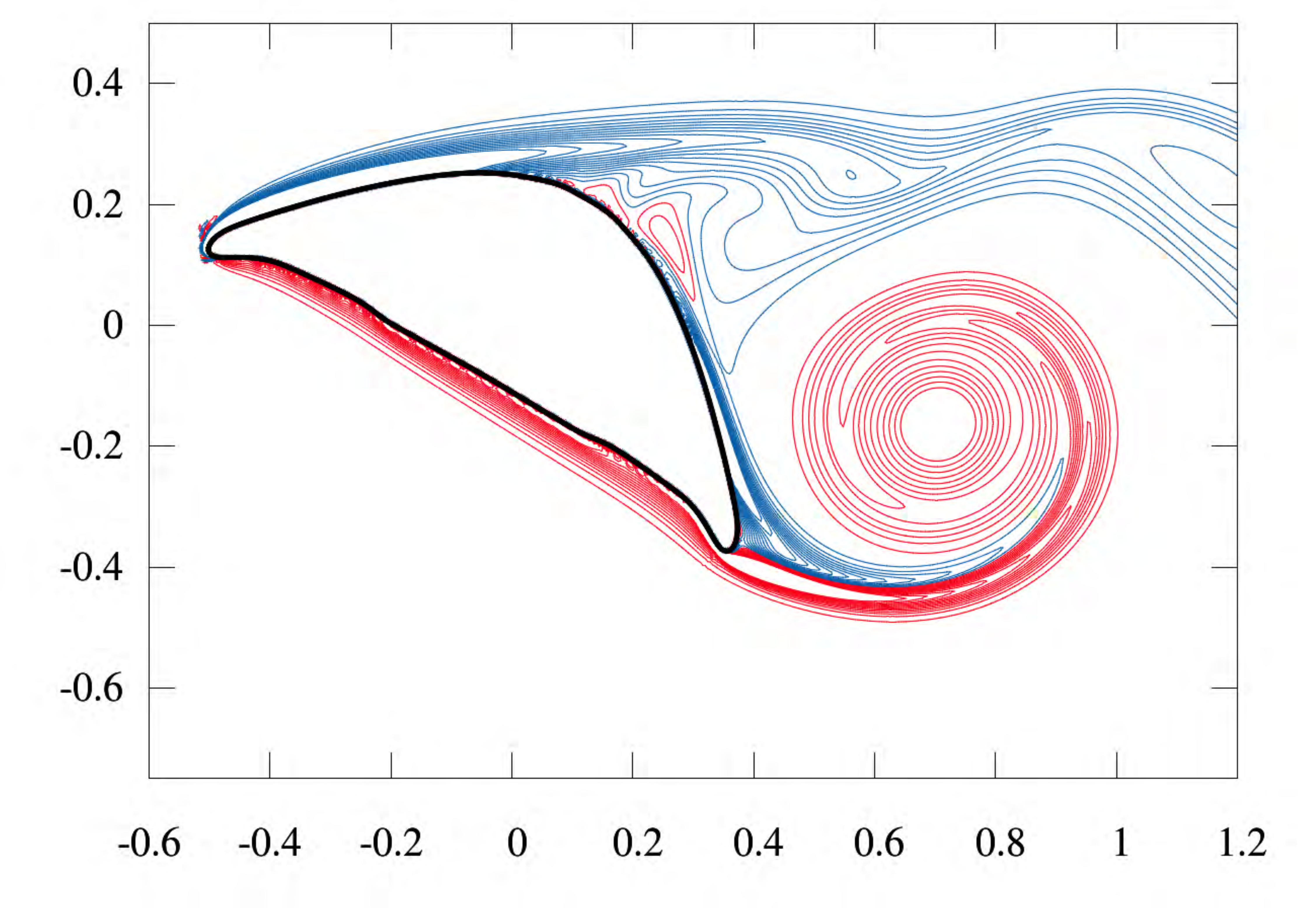} \label{subfig:vc2k30c}
	}
\caption[justification=justified,singlelinecheck=false]{\small  $Re=2000$, {\aoa} $30^\circ$: Vorticity contours from -25 to +25 in steps of 2. Contour lines of negative values of vorticity are blue and the contours of positive values are red.}
\label{fig:vConts2k30}
\end{minipage}
\end{figure}

\begin{figure*}[p]
\begin{center}
	\subfloat[$t^*=43.52$]{ \includegraphics[width=0.5\textwidth]{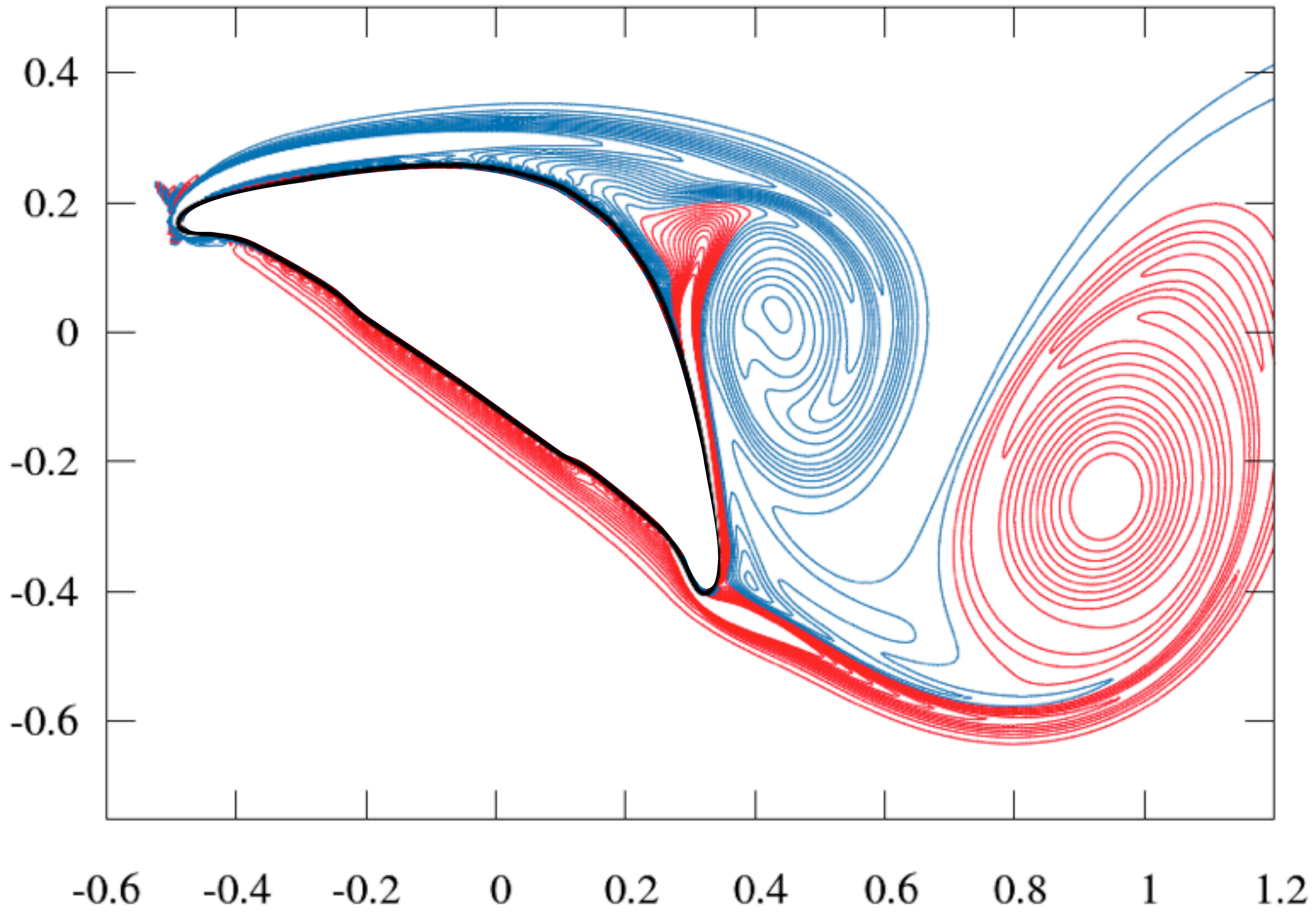} \label{subfig:vc2k35a}}
	\subfloat[$t^*=44.032$]{ \includegraphics[width=0.5\textwidth]{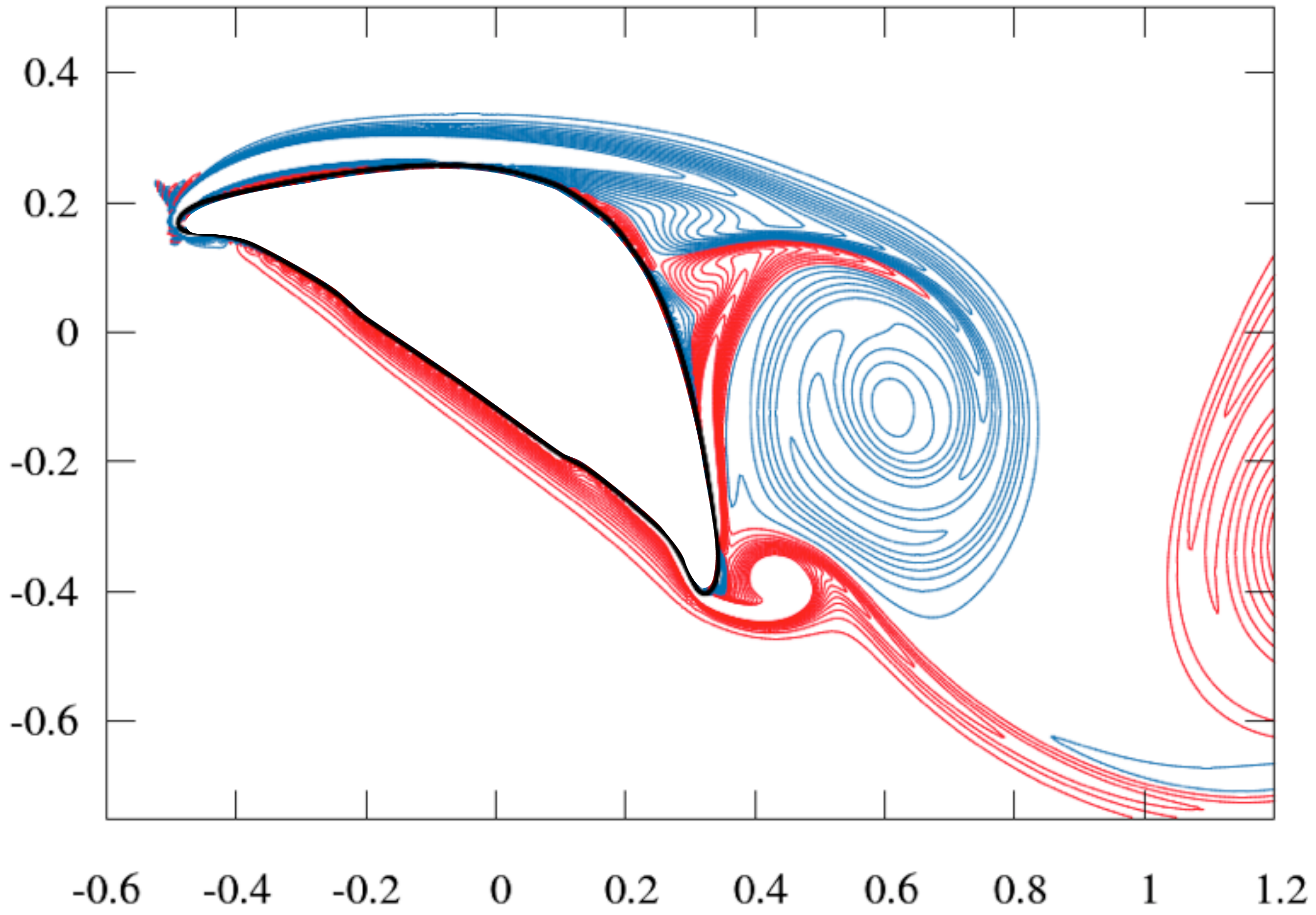} \label{subfig:vc2k35b}}  \\
	\subfloat[$t^*=44.544$]{ \includegraphics[width=0.5\textwidth]{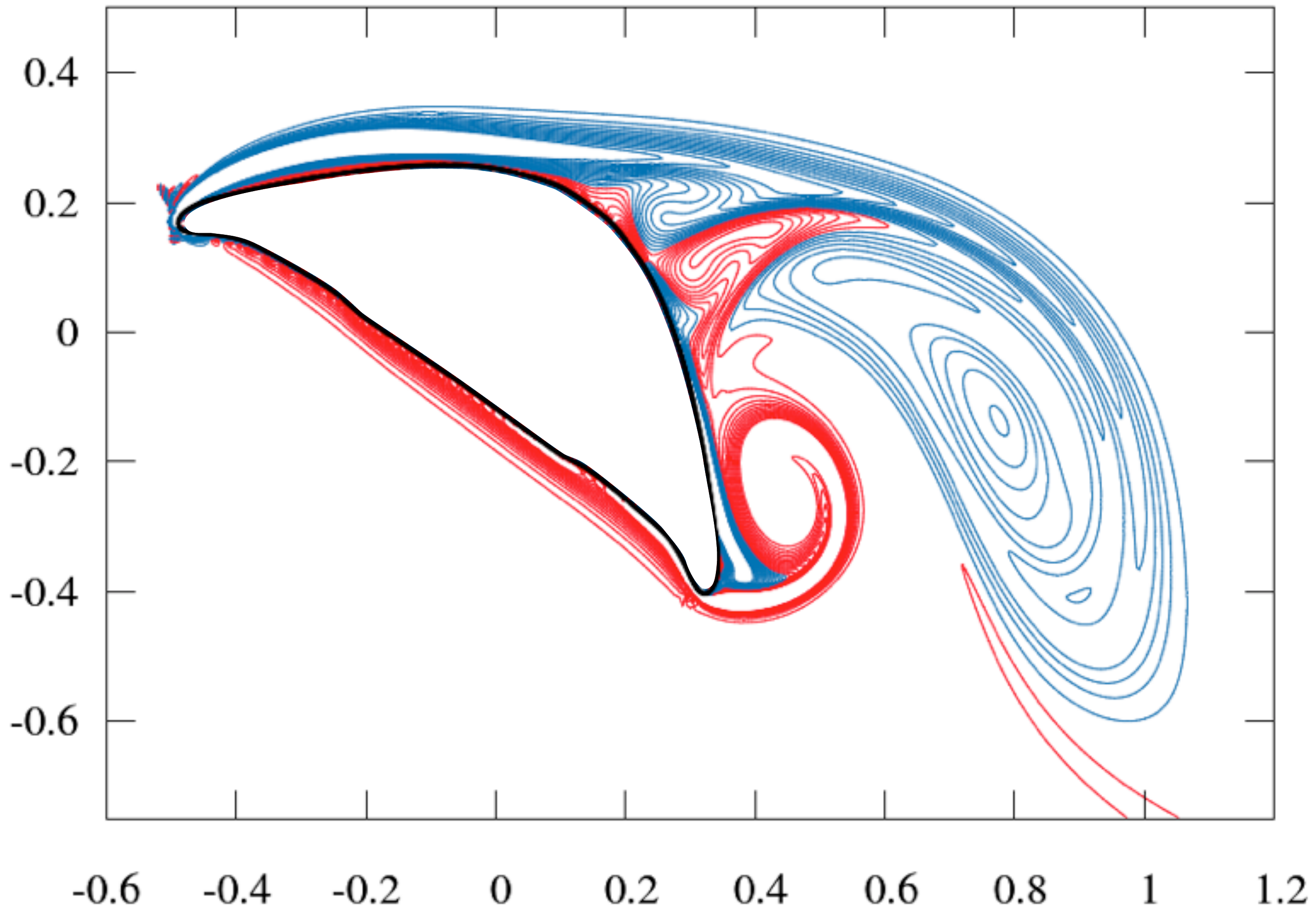} \label{subfig:vc2k35c}} 
	\subfloat[$t^*=45.056$]{ \includegraphics[width=0.5\textwidth]{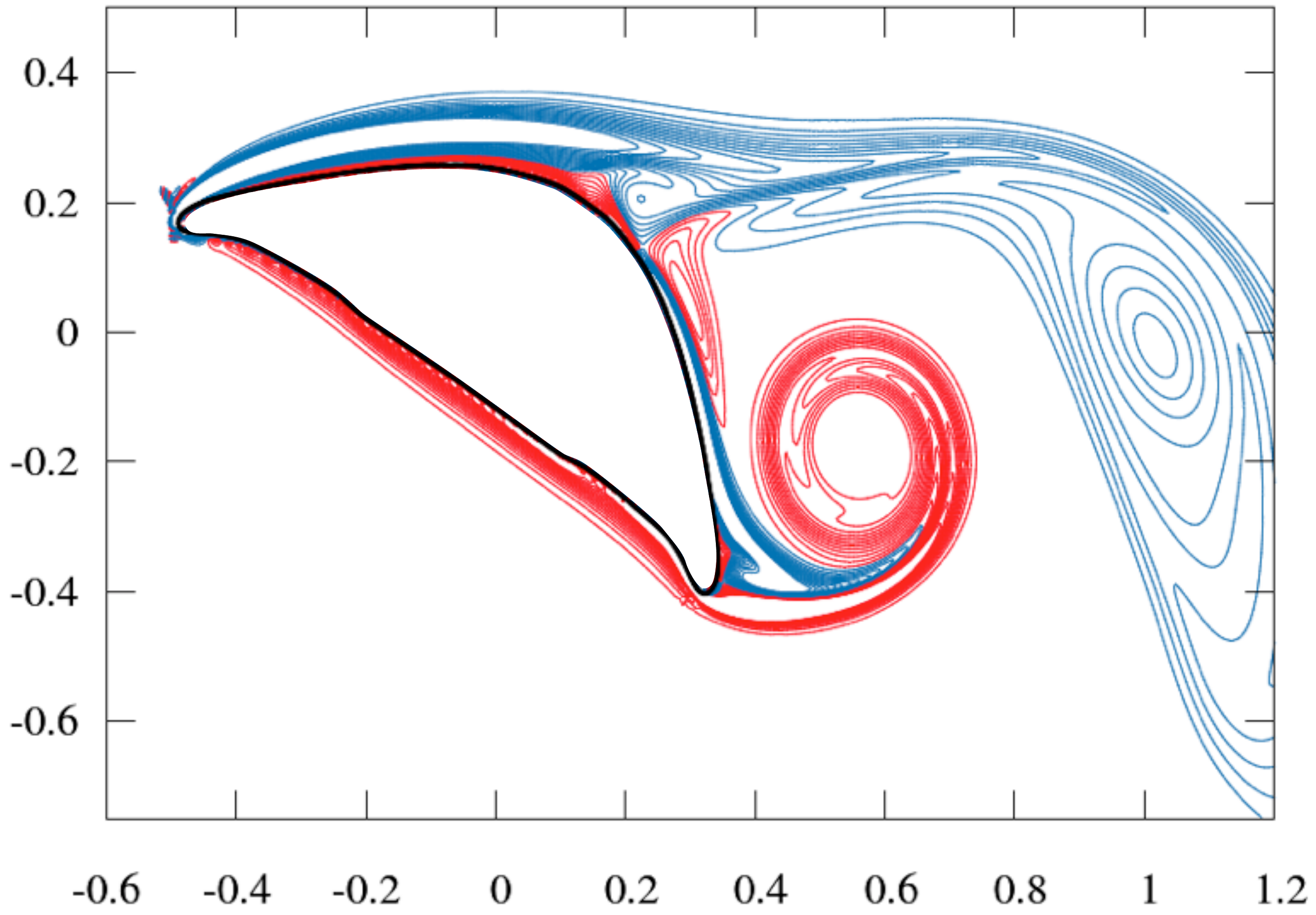} \label{subfig:vc2k35d}}  \\
	\subfloat[$t^*=45.568$]{ \includegraphics[width=0.5\textwidth]{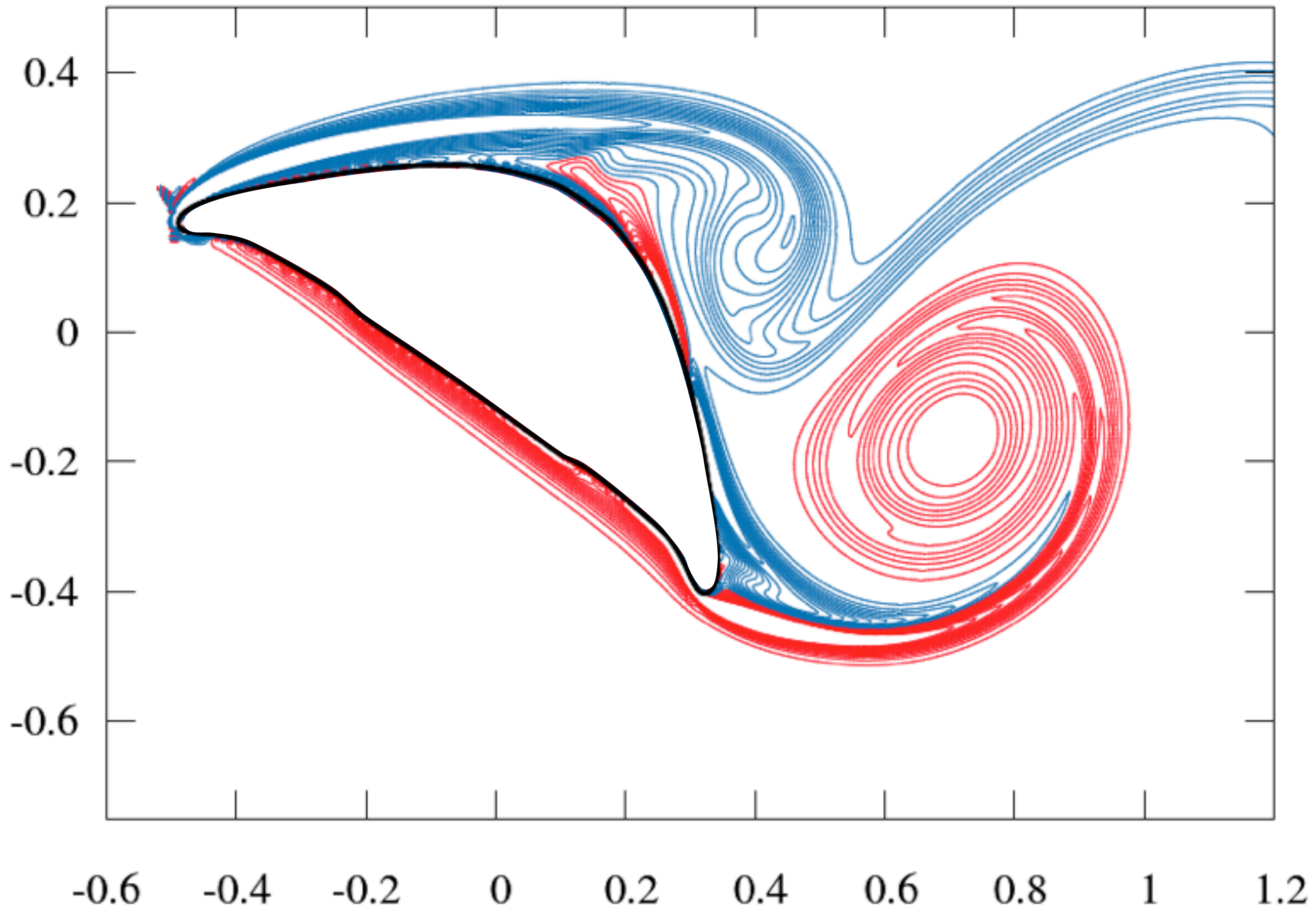} \label{subfig:vc2k35e}} 
	\subfloat[$t^*=46.08$]{ \includegraphics[width=0.5\textwidth]{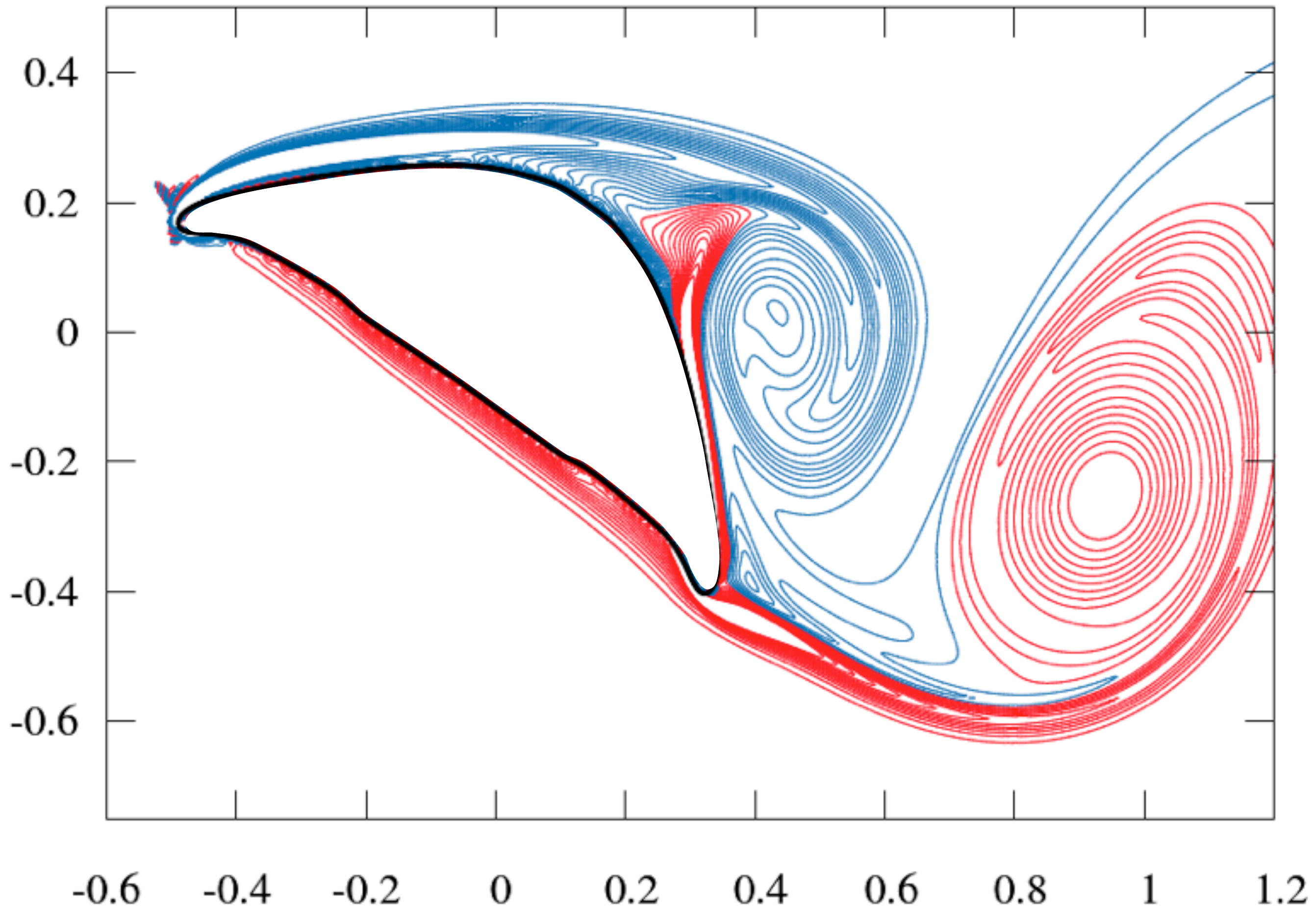} \label{subfig:vc2k35f}} 
\caption[justification=justified,singlelinecheck=false]{\small  $Re=2000$, {\aoa} $35^\circ$: Vorticity contours from -25 to +25 in steps of 2. Contour lines of negative values of vorticity are blue and the contours of positive values are red.}
\label{fig:vConts2k35}
\end{center}
\end{figure*}

The swirling strength helps to identify regions of vorticity that correspond to rotating flow, rather than strain-dominated regions. But we cannot identify the sign of vorticity, so we do not recognize primary vs.\ secondary vortices in these plots.
To examine near-body features, we plotted the contours of vorticity in the region close to the body. Figures \ref{fig:vConts1k30} and \ref{fig:vConts2k30} consist of three frames showing the vorticity plots of the flow at different points within one cycle of vortex shedding for flows at $Re=1000$ and $Re=2000$, respectively, when the angle of attack is $30^\circ$. Because the case with $\aoa=35^\circ$  and $Re=2000$ is of interest to us and we would like to examine it more carefully, Figure \ref{fig:vConts2k35} shows six frames of vorticity contours of this flow from one cycle of shedding.  These were plotted during the times when periodic vortex shedding had been established. A detailed description of the flow features and their effect on the force coefficients is saved for the following section.

\clearpage

\section{Discussion}
\label{discussion}

The first question that we aimed to answer with this study was whether an enhanced lift at a specific angle of attack would also appear in two-dimensional simulations of the flow around the cross-section of the snake, as it does in the experiments.
Our results show a pronounced peak in lift at angle of attack $35^{\circ}$ for Reynolds numbers 2000 and beyond. This Reynolds number at which the observed switch occurs is lower than in the two previous experimental studies, \cite{MiklaszETal2010, Holden2011} but the angle of attack at which the peak appears is the same in the simulations and the experiments. 

The snake's cross-section acts like a lifting bluff body, hence the pressure component accounts for most of the force acting on the body, with viscosity contributions being small. Figure \ref{fig:surfPres} shows the time-averaged pressure distribution along the surface of the body for flows with Reynolds numbers 1000 and 2000. The plots show an increased suction all along the dorsal surface of the snake for the case when ${\aoa}=35^\circ$ and $Re=2000$, which accounts for the enhanced lift. To more fully explain the mechanism of lift enhancement, it is necessary to consider the unsteady flow field.

\begin{figure}
\begin{center}
	{ \includegraphics[width=0.5\textwidth]{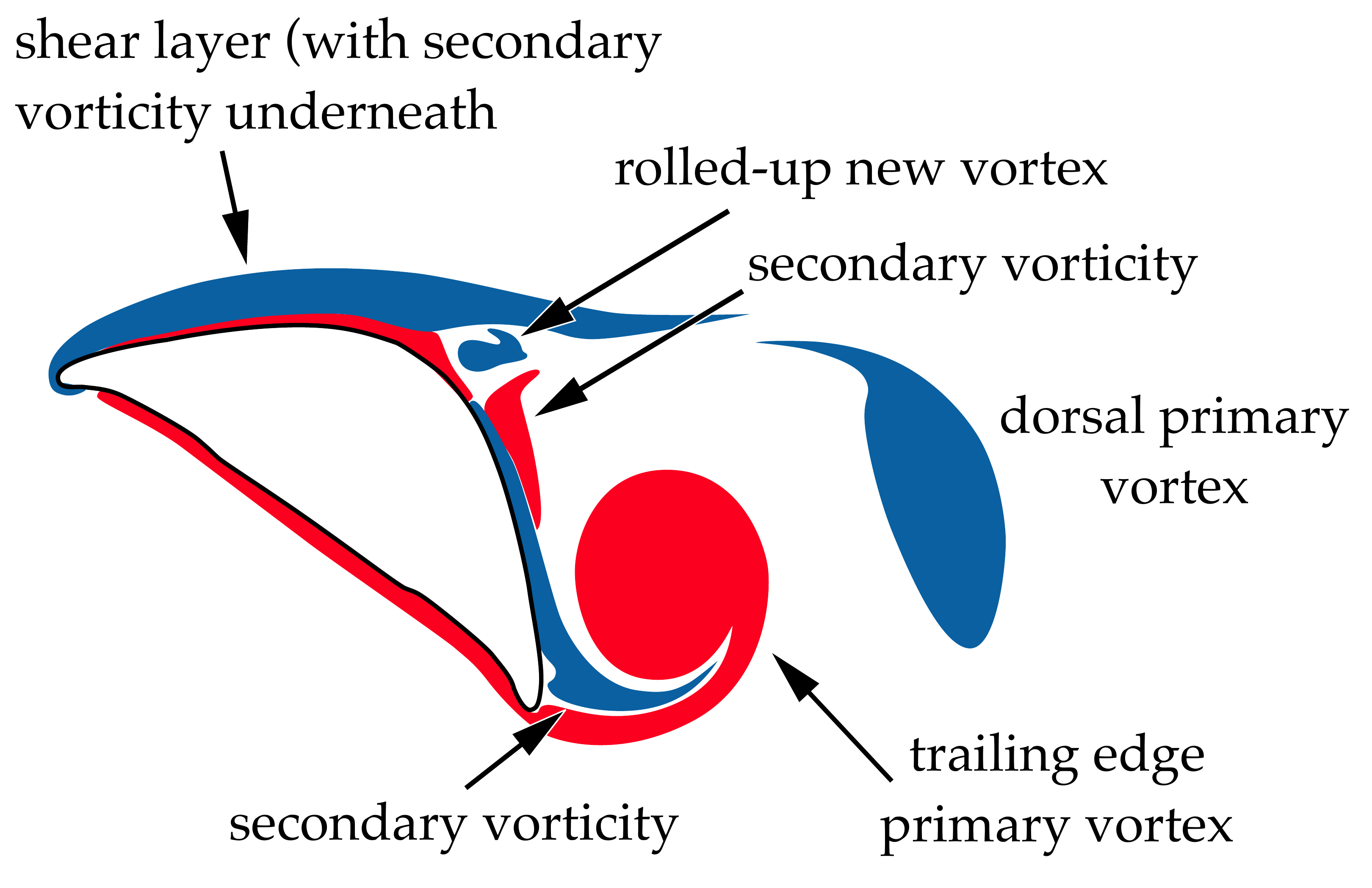} }
\caption{\small A summary of the major features found in the near-body wake.}
\label{fig:sketch}
\end{center}
\end{figure}

We compared the development of the vorticity in the wake of the cross-section of the snake with that of a two-dimensional circular cylinder. Koumoutsakos and Leonard \cite{koumoutsakos+leonard1995} studied the incompressible flow field of an impulsively started two-dimensional circular cylinder. They analyzed how the evolution of vorticity affected the drag coefficient at various Reynolds numbers between $40$ and $9500$, arriving at a description of the role of secondary vorticity and its interaction with the separating shear layer. Here, we present a similar analysis for the flow over the snake's cross-section at Reynolds numbers 1000 and 2000---the lift curve of the latter exhibits the spike at ${\aoa}=35^\circ$ and the former does not. Figure \ref{fig:sketch} shows a sketch of the near-wake region, indicating the terms that we will use to describe the flow.

Figure \ref{fig:vConts1k30} shows that at $Re=1000$ and \aoa\ $30^{\circ}$, the primary vortex generated on the dorsal side of the body induces a region of secondary vorticity. The boundary layer feeding the primary vortex at this Reynolds number is thick and the secondary vorticity is relatively weak.  The primary vortex generated at the trailing edge interacts with the dorsal primary vortex, straining it and weakening it. A similar flow pattern is observed at ${\aoa}=35^\circ$ for this Reynolds number (not shown).

At the higher Reynolds number of 2000, the vortices are stronger and more compact, as expected. But the flow fields at angles of attack $30^\circ$ and $35^\circ$ are qualitatively different from each other. 
At ${\aoa}=30^\circ$, the trailing-edge vortex is strong enough to weaken, stretch, and split the dorsal primary vortex into two (see Figure \ref{subfig:vort2k30}), resulting in a wake that consists of single vortices on top and pairs of vortices at the bottom of the vortex street. This wake pattern appears at this Reynolds number for angles of attack $30^{\circ}$ and lower.
But at ${\aoa}=35^\circ$, the stronger vortex on the dorsal side induces a stronger secondary vorticity (Figure \ref{subfig:vc2k35a}). The vortices along the dorsal surface are associated with an enhanced suction all along the upper side of the body, as seen in Figure \ref{subfig:surfP2k}. As the secondary vorticity infiltrates the primary vortex, a new second vortex of negative sign is formed by the shear layer on the dorsal surface (Figures \ref{subfig:vc2k35b} and \ref{subfig:vc2k35c}). The shear layer on the dorsal side separates near the leading edge, as evidenced by the positive vorticity near the surface in Figures \ref{subfig:vc2k35c} and \ref{subfig:vc2k35d}. The separated shear layer does not stall, but instead rolls up into the new vortex of negative sign due to the influence of the secondary vortex. This new second vortex forms a dipole with the secondary vorticity (Figure \ref{subfig:vc2k35c}) and the vortices are pushed towards the profile's surface. As the primary vortex convects away from the body, the secondary vorticity weakens (Figure \ref{subfig:vc2k35d}) and the second vortex formed due to the separated shear layer initiates a new primary vortex (Figure \ref{subfig:vc2k35e}) that remains close to the body and contributes to the enhanced suction on the dorsal surface. The trajectory plots in Figure \ref{fig:trajectories} show that the dorsal primary vortex for the case when $Re=2000$ and ${\aoa}=35^\circ$ indeed forms and remains closer to the body surface as compared to the other cases.

The swirling strength plots at $Re=2000$ (Figure \ref{fig:qValues}) are also consistent with these interpretations. At the instant when lift is minimum, the flow remains attached at the leading edge for the case when ${\aoa}=30^\circ$, but is separated when ${\aoa}=35^\circ$. Regions of high swirling strength  correspond to regions of low pressure in the flow, as on the front half of the dorsal surface. At the instant of maximum instantaneous lift, when ${\aoa}=35^\circ$ swirling strength is higher at the locations of the secondary vorticity and of the new second vortex, associated with a lowering of pressure across the dorsal surface and not just at the leading edge. Both the minima and maxima of the instantaneous lift increase sharply at ${\aoa}=35^\circ$ (see Figure \ref{fig:unsteadyLift}).

One phenomenon that occurs at this range of Reynolds number for flows over bluff bodies is the instability of the shear layer. For circular cylinders, Kelvin-Helmholtz instability of the shear layers is observed for flows with Reynolds numbers approximately 1200 and higher.\cite{Prasad1997} This is a two-dimensional phenomenon, and causes an increase in the 2-D Reynolds stresses, and subsequently increases base suction.\cite{Williamson1996} For the snake's cross-section at $Re=2000$ and ${\aoa}=35^\circ$, the separated shear layer on the dorsal side is also subject to this instability and can form small-scale vortices that eventually merge into the primary vortices. This could explain why the vortices are stronger and remain closer to the body. At lower angles of attack, the boundary layer is attached to the body and the instability is not manifested, which implies a reduced value of base suction, and subsequently lift and drag. 
\comment{[ak-should we retain the paragraph talking about the K-H instability?]}

As mentioned at the beginning of this section, the enhanced lift occurs at the same angle of attack in previous experiments with the cross-section of the snake, but in a different range of Reynolds number. This difference reflects the limitations of our study, which only computes two-dimensional flow. 
Experiments with bluff bodies show that beyond a certain Reynolds number ($\sim 190$ for circular cylinders), three-dimensional instabilities produce steam-wise vortices in the wake.\cite{Williamson1996} The formation and effect of these vortices are more pronounced for bluff bodies---they increase the Reynolds stresses in the wake of the cylinder, and wake vortices are formed further from the body surface, causing a decrease in the base suction when compared to two-dimensional simulations of the same flow. \cite{MittalBalachandar1995} Hence, 2-D simulations overestimate the unsteady lift and drag forces. This is also observed in our computed values of lift and drag for the snake model when compared to the experiments of Holden et al.\cite{Holden2011} Taking these facts into account, the simulations we report here can be considered a reduced model, one whose utility we could not assert \textit{a-priori}. It is perhaps a surprising result that a peak in lift is in fact obtained in the 2D simulations, as observed in the experiments. We surmise that the discrepancy in Reynolds number range where the enhanced lift appears can be ascribed to 3D disturbances, which inhibit the mechanism responsible for the peak in lift, so that it can only manifest itself at higher Reynolds number when the wake vortices are stronger and more compact.

A flying snake in the field exhibits complex three-dimensional motions and a number of other factors could affect its aerodynamics, some of which we briefly list here. The finite size of the sections generating lift means that wingtip vortices will be generated, causing a decrease in lift and an increase in drag. The $S$-shape of the snake in the air suggests that multiple segments of the snake body can generate lift simultaneously, and the wake of the section furthest upstream could have an effect on the downstream sections. A preliminary result by Miklasz et al.\cite{MiklaszETal2010} with two tandem models whose cross-sections approximated the snake geometry found that the value of the lift-to-drag ratio of the downstream section could increase by as much as 54\%, depending on its position in the wake. The undulatory motion of the snake in the air means that the body also moves laterally in addition to moving forward along the glide path. The sideways motion may generate spanwise flow, which has been known to stabilize leading-edge vortices \cite{Ellington1996} and increase the lift on the body. Another mechanism that we do not know if the snake makes use of is thrust generation due to heaving or pitching motions of its body. \cite{Freymuth1988} Fluid-structure interactions could also be present.

Without discounting the limitations of two-dimensional models, we have sought to characterize the wake mechanism that could explain the lift enhancement on the snake's cross-section in this work. The results give insight into the vortex structures that are involved in this mechanism and suggest new directions to interrogate the flow in three-dimensional studies. These constitute future work that we aim to carry out once the appropriate extensions to the code have been completed.

\section{Conclusion}
We studied computationally the aerodynamics of a species of flying snake, \emph{Chrysopelea paradisi}, by simulating two-dimensional incompressible flow over the cross-sectional shape of the snake during gliding flight. We obtained lift and drag characteristics of this shape in the Reynolds number range 500--3000. In flows with Reynolds numbers 2000 and beyond, the lift curve showed a sharp increase at an angle of attack of $35^\circ$, followed by a gradual stall. This behavior is similar to that observed previously in experimental studies of cylindrical sections with the same cross section, for Reynolds numbers 9000 and above. Our unsteady simulations reveal that in flows with Reynolds number 2000 and above and at angle of attack $35^\circ$, the 2-D flow separates at the leading edge but does not produce a stall. The free shear layer thus generated over the dorsal surface of the body can roll up and interact with secondary vorticity, resulting in vortices remaining closer to the surface and an associated increase in lift. Differences between the experiments and simulations in the magnitude of the force coefficients and the value of the threshold Reynolds number beyond which the phenomenon is observed may be attributed to the three-dimensional effects.

\medskip

The code, running scripts and data necessary to run the computations that produced the results of this paper can all be found online, and are shared openly. The \texttt{cuIBM} code is open source under the MIT license and can be obtained from its version-controlled repository at \href{https://bitbucket.org/anushk/cuibm/}{https://bitbucket.org/anushk/cuibm/}. Data sets, figure-plotting scripts, figures and supplementary animations have been deposited on \href{http://figshare.com/}{http://figshare.com/}.


\begin{acknowledgments}
LAB acknowledges partial support from NSF CAREER award ACI-1149784. She thanks the support from NVIDIA via an Academic Partnership award (2012) and the CUDA Fellows program. This research was also partially supported by NSF grant 1152304 to PPV and JJS.
\end{acknowledgments}


\section*{References}

\end{document}